\documentclass[journal]{IEEEtran}

\usepackage{tabularx} 
\usepackage{booktabs}
\usepackage{supertabular}
\usepackage{longtable}
\usepackage{makecell}
\usepackage{soul}
\usepackage{xcolor}
\usepackage{times}
\usepackage{bm}

\usepackage{cite}

\ifCLASSINFOpdf
  \usepackage[pdftex]{graphicx}
  \graphicspath{{./pdf/}{./jpeg/}{./png}}
  \DeclareGraphicsExtensions{.pdf,.jpeg,.png}
\else
  \usepackage[dvips]{graphicx}
  \graphicspath{{./eps/}}
\fi
\usepackage{amsmath}
\usepackage{array}

\ifCLASSOPTIONcompsoc
 \usepackage[caption=false,font=normalsize,labelfont=sf,textfont=sf]{subfig}
\else
 \usepackage[caption=false,font=footnotesize]{subfig}
\fi

\usepackage{stfloats}
\def\BibTeX{{\rm B\kern-.05em{\sc i\kern-.025em b}\kern-.08em
    T\kern-.1667em\lower.7ex\hbox{E}\kern-.125emX}}
\AtBeginDocument{\definecolor{ojcolor}{cmyk}{0.93,0.59,0.15,0.02}}

\begin{document}


%
\title{A Comparative study on THz Communication Systems: Photonics versus Electronics Approaches}
%
%
%

\author{Talha Rahman,~\IEEEmembership{Senior Member,~IEEE,}       
        and~Murat Uysal,~\IEEEmembership{Fellow,~IEEE}
\thanks{T. Rahman and M. Uysal are with the Division of Engineering, New York University Abu Dhabi (NYUAD), PO Box 129188, Abu Dhabi, UAE and also with the Research Institute NYUAD Wireless Center. \newline
This work is supported by Tamkeen under the Research Institute NYUAD grant CG017.}
\thanks{Manuscript dated April 28, 2026; revised xxx.}
\thanks{This work has been submitted to the IEEE Open Journal of the Communications Society for possible publication.}}


%
%
%
\markboth{A Comparative Study on THz Communication Systems:
Photonics versus Electronics Approaches}{Talha Rahman \textit{et al.}}
%
%
%
%
\maketitle
\begin{abstract}
Terahertz (THz) communication has emerged as a key enabler for sixth-generation (6G) networks, offering ultra-wide bandwidths to support data-intensive applications such as holographic telepresence and immersive extended reality. Recent advances have enabled both electronics-based and photonics-based THz front-ends, each with distinct advantages and hardware limitations. While electronics-based solutions leverage mature semiconductor platforms, they suffer from amplified oscillator phase noise, frequency offsets, and nonlinearities introduced by multiplier and amplifier chains. Photonics-based systems, in turn, enable highly tunable and spectrally pure carriers but are subject to laser intensity noise, amplified spontaneous emission, shot noise in photomixers, and thermal noise in RF mixers. This article provides a comprehensive review of experimental demonstrations in electronics-, photonics-, and hybrid-based THz links, highlighting their hardware architectures, performance metrics, and implementation trade-offs. We then survey theoretical modeling efforts, emphasizing how hardware impairments affect system reliability and identifying limitations in existing studies. Building on this, we develop comprehensive signal models for both approaches, derive analytical expressions for signal-to-noise ratio (SNR), and evaluate bit error rate (BER) performance under realistic system parameters. Comparative results demonstrate how distinct impairment mechanisms shape the overall link performance of electronics- versus photonics-based THz systems. The insights offered aim to guide the design of robust transceiver architectures and accelerate the integration of THz technologies into future 6G deployments.

\end{abstract}

%
\IEEEpeerreviewmaketitle

\section{Introduction}
%
%
%
%

\IEEEPARstart{S}{ixth-generation} (6G) cellular systems represent the next step in wireless communication, targeting terabit-per-second data rates, sub-millisecond latency, and deeply integrated artificial intelligence (AI). These networks are expected to enable transformative applications such as immersive extended reality, real-time holographic telepresence, ultra-reliable low-latency communications for remote operations, and massive digital twinning \cite{zugnoUseCasesTerahertz2025, jiangTerahertzCommunicationsSensing2024, alhajIntegrationHybridNetworks2024, wangRoad6GVisions2023, hong6GRDVision2022}.

Initial 6G deployments, expected in the early 2030s, will likely build upon existing spectrum allocations while expanding into new frequency ranges. In particular, the centimeter-wave band (7–24 GHz), also referred to as the upper midband or FR3, has emerged as a strong candidate due to its favorable balance between capacity, coverage, and mobility \cite{6GSpectrumUnleashing, keysightExploring6GSpectrum, 6GNextHorizon}. Beyond FR3, the exploration of sub-terahertz (100–300 GHz) and terahertz (300 GHz – 3 THz) bands is also gaining momentum, offering vast bandwidth potential to support extreme data rates and dense deployments in future 6G scenarios. These high-frequency bands are particularly critical for data-intensive and latency-sensitive applications such as real-time holographic telepresence. Delivering life-sized, interactive holograms in real time requires end-to-end data rates on the order of hundreds of gigabits per second to a few terabits per second, depending on resolution, frame rate, and the number of viewing angles. Similarly, fully immersive XR experiences, combining ultra-high-definition visuals, spatial audio, and low-latency feedback across multiple synchronized users, can exceed the capacity limits of conventional spectrum bands. The exceptionally wide bandwidths available in the THz spectrum are uniquely suited to meet these demanding requirements, making them essential for enabling such futuristic use cases in 6G \cite{6GSpectrumUnleashing, 6GNextHorizon}. In line with 6G development timelines, the THz spectrum is anticipated to be utilized during the second phase of deployment, around 2035.

Historically, the terahertz (THz) band was referred to as the “THz gap” due to the absence of efficient devices. However, recent breakthroughs in semiconductor technologies have enabled the development of reliable sources, detectors, and integrated components capable of operating at these high frequencies \cite{jiangTerahertzCommunicationsSensing2024}. As a result, THz system prototypes have been developed by several research groups around the world, demonstrating their potential for ultra-high-speed wireless links. This technological maturation now positions the THz band as a promising frontier for 6G networks. THz waves lie in the portion of the frequency spectrum between the realms of light waves and radio waves. This enables two primary methods for generating THz signals: electronics-based and photonics-based approaches 
\cite{thomasSurveyAdvancementsTHz2025, seedsTeraHertzPhotonicsWireless2015, liPhotonicsAidedTerahertzWaveWireless2022, songTerahertzCommunicationsChallenges2022, sungDesignConsiderationsPhotonic2021}. In the electronics-based approach, the output of a lower-frequency electronic oscillator is upconverted to the desired THz frequency using an amplifier and frequency multiplier chain. At the receiver side, a similar configuration is used, employing a lower-frequency local oscillator and multiplier chain, to downconvert the received THz signal to an intermediate frequency (IF) or baseband \cite{thomasSurveyAdvancementsTHz2025}. Standard baseband signal processing steps then follow at the digital receiver. In the photonics-based approach, two optical sources (i.e., lasers) with different wavelengths are fed into a photomixer, generating a THz signal at the frequency corresponding to the difference between the two optical carriers \cite{liPhotonicsAidedTerahertzWaveWireless2022}. At the receiver, a similar dual-laser configuration produces the local oscillator (LO) signal, which is used in an RF mixer to recover the baseband signal. In addition to these two approaches, hybrid systems have also been explored to leverage the complementary strengths of both technologies. In a typical hybrid implementation, photonics is used for THz signal generation at the transmitter—providing wide tunability and high spectral purity—while electronics handle signal detection and downconversion at the receiver, where electronic components are often more compact and power-efficient. 

Since electronics-based and photonics-based front-ends differ significantly, each approach introduces unique hardware constraints and performance bottlenecks. Capturing these effects in accurate signal models is therefore essential for meaningful system-level analysis. In communication theory, signal models were primarily developed for lower-frequency microwave and millimeterwave systems. Although hardware impairments such as phase noise, nonlinearities, and I/Q imbalance have been studied in those bands, their impact becomes far more pronounced at THz frequencies due to wider bandwidths, higher carrier frequencies, and stricter alignment requirements. At the same time, the photonics-based implementation approach introduces its own challenges, including laser-associated impairments such as relative intensity noise (RIN), amplified spontaneous emission (ASE) noise from optical amplifiers, compounded phase noise from multiple lasers, carrier frequency offsets due to wavelength drift, as well as shot noise in photomixers and thermal noise in RF mixers. Consequently, customized signal models are required for THz communication systems to capture the implementation-specific distortions. While some efforts in this direction have been made \cite{bicaisOptimizedSingleCarrier2020, a.TeraHertzTechnologyTHz2025}, the available signal models are typically fragmented. Most prior studies address only a single impairment or noise source, rather than providing a unified treatment. Furthermore, most studies focus on a specific implementation type and do not present a one-to-one comparison. 

In this article, we first present a thorough survey of the existing literature on THz communication systems with hardware impairments to highlight the aforementioned research gaps. Then, we develop comprehensive signal models for both electronics-based and photonics-based approaches.
\begin{itemize}
\item In the development of signal models for photonics-based THz systems, the characteristics of optical components play a central role. For example, the RIN of lasers and the ASE noise of optical amplifiers are key contributors to system degradation. ASE originates from spontaneous electron decay in the gain medium, while RIN arises from intensity fluctuations caused by the beating between stimulated and spontaneous emission photons inside the laser cavity. In addition, the photomixing process used for THz generation inherently produces shot noise. Although each of these noise sources can individually be approximated as additive white Gaussian noise (AWGN), their combined effect in the context of THz communication results in an aggregate noise process that cannot be accurately captured by a simple additive model. The limitations of an AWGN assumption become especially apparent when considering spectrally efficient, higher-order QAM formats. Furthermore, phase noise from different lasers employed in photonic THz transmitters and receivers results in a compounded phase noise effect at the receiver, which requires careful analysis in conjunction with the other noise processes. 
\item Electronics-based implementation approach faces a different set of challenges. The noise introduced by the base oscillator is amplified through the multiple frequency multiplication stages needed to generate THz carriers. This process not only increases the effective phase noise, but also aggravates carrier frequency offset. Furthermore, electrical and optical amplifiers, which are often employed in both architectures, contribute additional noise while simultaneously enhancing the noise already present. 
\end{itemize}

Based on the developed signal models, we derive analytical expressions for the signal-to-noise ratio (SNR) at both the transmitter and receiver. We then complement this analysis with Monte Carlo simulations, assessing the effect of effective additive noise on the bit error rate (BER) performance of square QAM formats. Our main contributions are summarized as follows:
\begin{itemize}
    \item \textbf{Literature overview on THz architectures and hardware impairments:}
We provide a structured overview of experimental demonstrations of electronics-based, photonics-based, and hybrid THz communication systems, highlighting their hardware architectures and related impairments. We also review prior theoretical studies that model hardware impairments in THz systems and identify limitations of existing analytical approaches. 

\item \textbf{Development of unified signal models with hardware impairments:}
We develop comprehensive signal models for both photonics-based and electronics-based THz transceivers that explicitly incorporate implementation-specific hardware impairments. In particular, the photonics-based model captures laser relative intensity noise (RIN), amplified spontaneous emission (ASE), photomixer shot noise, and mixer thermal noise, while the electronics-based model accounts for oscillator noise floor, thermal noise, and phase-noise amplification through frequency multiplier chains.

\item \textbf{Derivation of SNR expressions for photonics-based and electronics-based THz systems:}
Based on the proposed signal models, we derive analytical expressions for the SNR at both the transmitter and receiver, explicitly tracking how different hardware noise sources propagate through the transceiver chain.

\item \textbf{Evaluation of BER performance:}
Using the derived signal and noise models, we evaluate the BER performance of square QAM formats through Monte Carlo simulations under realistic system parameters.

\item \textbf{Comparative performance insights:}
Finally, we provide a comparative analysis of photonics-based and electronics-based THz architectures, identifying the dominant impairment mechanisms in each approach and quantifying their impact on system performance.

\end{itemize}

The rest of the paper is organized as follows. In Section \ref{sec:literatureReview}, we present a literature overview covering both experimental THz studies and theoretical works that analyze the impact of hardware impairments. In Section \ref{sec:photonics-basedTHz}, we describe the photonics-based coherent THz transmission systems, and introduce the related signal and noise models. In addition, the signal to noise ratio (SNR) expressions both at transmitter and receiver are derived. In Section \ref{sec:electronics-basedTHz}, we detail the electronics-based coherent THz transmission system, the related signal and noise models and SNR analysis. In Section \ref{sec:numericalResults}, we present numerical results and discussions based on the previously derived signal models for both photonics-based and electronics-based THz systems. We quantify the effect of hardware impairments and compare the performance of both implementation approaches. Finally, we present concluding remarks in Section \ref{sec:conclusions}.

\noindent\textbf{Notations:}
\begin{itemize}
    \item ${S_x}\left( f \right)$  is the power spectral density of variable $x\left( t \right)$.
    \item $\Re \left\{ . \right\}$ is the real operator.
    \item $\Im \left\{ . \right\}$ is the imaginary operator.
    \item $x^*(t)$ is the complex conjugate of $x(t)$.
    \item $N\left( {0;{\sigma ^2}} \right)$ is a zero-mean real valued Gaussian random variable with variance ${\sigma ^2}$.
    \item $CN\left( {0;{\sigma ^2}} \right)$ is a zero-mean complex symmetric Gaussian random variable with variance ${\sigma ^2}$.
    \item $\left\langle  \cdot  \right\rangle $ represents the time average.
    \item $E\left[ x \right]$ represents the expectation of $x$.
    \item $\chi_n^2$ is a real-valued variable with $n^{th}$ order chi-squared distribution.
    \item $C\chi _{2n}^2$ is a complex-valued variable defined by $C\chi _{2n}^2 = \chi _n^2 + j \cdot \chi _n^2$.
\end{itemize}
\section{Literature overview of THz communications}\label{sec:literatureReview}
In this section, we survey the state of the art in end-to-end THz transmission with particular emphasis on hardware architectures, implementation challenges, and their associated theoretical foundations. We classify the experimental literature according to the underlying front-end paradigm: electronics-based systems (Section \ref{sec:literatureReview-electronics}), photonics-based systems (Section \ref{sec:literatureReview-photonics}), and hybrid links that combine photonic transmitters with electronic receivers or vice versa (Section \ref{sec:literatureReview-hybrid}). To allow consistent comparison, we summarize each experiment in tables using common metrics such as carrier frequency, data rate, modulation scheme, transmission distance, transmitter/receiver approach, and distinctive architectural features. Finally, in Section \ref{sec:literatureReview-theoretical}, we summarize the theoretical works that model THz systems under hardware impairments.

 The experimental works included in this literature overview are selected based on the following criteria: (i) the work reports an end-to-end THz wireless transmission experiment, (ii) key system parameters such as carrier frequency, modulation format, data rate, and transmission distance are explicitly reported, and (iii) the implementation architecture can be clearly classified as electronics-based, photonics-based, or hybrid. Our survey aims to highlight representative experimental milestones and architectural trends rather than providing an exhaustive list of all reported experimental THz studies. It should also be noted that the reported experiments operate under different system assumptions, including antenna gains, EIRP levels, receiver noise figures, signal bandwidths, DSP techniques, and the use of forward error correction (FEC). Consequently, the reported data rate and transmission distance values should not be interpreted as strictly comparable performance metrics. Instead, these provide a qualitative overview of the capabilities and design trade-offs of different THz front-end architectures under their respective experimental conditions.

\begin{table*}[htb]
\setlength{\tabcolsep}{4pt}
\caption{Summary of Electronics-Based THz Experiments}
\label{tab:literatureReview-electronics}
\centering
\begin{tabularx}{\linewidth}{p{5mm}<{\raggedright} p{15mm} p{25mm}<{\raggedright} p{23mm}<{\raggedright} p{0.07\textwidth}<{\raggedright} X}
\toprule
\textbf{Ref.} & \textbf{Frequency} & \textbf{Data Rate} & \textbf{Modulation} & \textbf{Distance} & \textbf{Highlights} \\ 
\midrule
\mbox{\cite{moeller25GbitDuobinary2011}},\newline \mbox{\cite{shiodeGigabitWirelessCommunication2011}} & 625~GHz & 2.5~Gb/s & Duobinary & Few cm & SBD-based $\times$48 frequency multiplier; RTD receiver with higher sensitivity; horn antennas and focusing lenses enhance received power. \\
\mbox{\cite{wang034THzWirelessLink2014}} & 340~GHz & 3~Gb/s & 16QAM & 50~m & Custom Tx/Rx chains; SHM-based up/downconversion with $\times$2 multiplier; Cassegrain antennas (48~dBi combined gain) counter 117~dB path loss; FPGA-based DSP. \\
\mbox{\cite{danTerahertzWirelessCommunication2020}},\newline \mbox{\cite{danSuperheterodyne300GHz2020}} & 300~GHz & Up to 60~Gb/s & 16QAM (15~GBd) & 10~m & Experimental comparison of superheterodyne vs. zero-IF architectures; custom 35~nm InGaAs chipsets. \\
\mbox{\cite{senExperimentalDemonstrationUltrabroadband2019}} & 1~THz & 5–15~Gb/s & QPSK, OFDM & 13~cm & Performance limited by thermal and absorption noise; receiver noise statistics match Gaussian; BER measured across formats and distances. \\
\mbox{\cite{parisiModulationsTerahertzBand2024}} & 130~GHz, 225~GHz, 1.02~THz & Up to 20~Gb/s (analysis-based) & PSK, DSSS, OTFS, OFDM & Not specified & Phase noise (PN) and PAPR jointly modeled; PN mask adapted from 3GPP; QPSK/DSSS resilient under impairments, OFDM/OTFS more sensitive. \\
\mbox{\cite{wranaSensitivityAnalysis2803122022}} & $\sim$300~GHz & Up to 32~Gb/s & 32QAM–256QAM & Not specified & Integrated superheterodyne Tx/Rx chips (35~nm InGaAs, IEEE 802.15.3d); spectral efficiency up to 8~bits/Hz. \\
\mbox{\cite{songPrototypeKIOSKData2018}} & 300~GHz & Up to 20~Gb/s & OOK & 1~m & InP HEMT Tx/Rx chips; 10~dBm Tx power; 6~mm silicon lens ($\sim$20~dBi gain); extinction ratio 15~dB; real-time error-free post-FEC. \\
\mbox{\cite{chakrabortyCaseOFDMUltrabroadband2021}} & 140~GHz & $\sim$10~Gb/s (10~GHz BW) & OFDM & 10–50~cm & End-to-end custom converters; hardware-induced frequency-selective fading; CP length of 1/16 sufficient. \\
\mbox{\cite{hamada300GHz100GbInPHEMT2018}} & 300~GHz & 100~Gb/s (25~GBd 16QAM) & 16QAM & 2.22~m & First all-electronics 100~Gb/s THz link; InP-HEMT mixers (80~nm) with 15$\pm$2~dB conversion gain; 50~dBi antennas; BER $< 10^{-3}$ (lab). \\
\mbox{\cite{hara300GHzCMOSTransceiver2018}} & 300~GHz & 2~Gb/s (QPSK); 20~Gb/s (16QAM) & QPSK, 16QAM & 75~cm (QPSK); 10~cm (16QAM) & 40~nm CMOS transceiver; gate-pumped mixer-last Tx for high output power; mixer-first Rx with high conversion gain; DSP integration possible. \\
\mbox{\cite{castroLongrangeHighSpeedTHzWireless2020}} & 300~GHz & 76.8~Gb/s (16QAM, 500~m); 44.8~Gb/s (QPSK, 1~km) & 16QAM, QPSK & 500~m (16QAM); 1~km (QPSK) & Outdoor long-range THz link; MMIC-based IQ-conversion frontends with $\times$36 multiplier; 25~GHz Tx BW, 50~GHz Rx BW; 55~dBi parabolic antennas; offline DSP with pre-distortion/equalization; SD-FEC with rain-test validation. \\ 
\bottomrule
\end{tabularx}
\end{table*}

\subsection{Electronics-based THz transmission experiments}\label{sec:literatureReview-electronics}
In this section, we review key experimental milestones in electronics-based THz transmission, emphasizing system design choices, hardware configurations, and the resulting performance metrics.
In this section, we review major experimental milestones in electronics-based THz transmission, emphasizing system architectures, hardware configurations, and performance metrics.

In \cite{moeller25GbitDuobinary2011}, a 2.5 Gb/s duobinary transmission was demonstrated using an all-electronic setup. A Schottky barrier diode (SBD)–based $\times$48 frequency multiplier upconverted the input to 625 GHz, while a resonant tunneling diode (RTD) receiver, offering higher sensitivity than SBDs \cite{shiodeGigabitWirelessCommunication2011}, enabled detection. Horn antennas and focusing lenses improved received power.

In \cite{wang034THzWirelessLink2014}, a 16-QAM link at 340 GHz achieved 3 Gb/s over 50 m using custom TX/RX modules. The transmitter employed an amplifier–multiplier chain to generate 170 GHz carrier and a subharmonic mixer (SHM) with $\times$2 multiplication achieving 340 GHz, and a Cassegrain antenna providing directivity. The total antenna gain of 48 dBi was cruicial to compensate 117 dB path loss. The receiver used a similar chain, downconverting the signal to a 750 MHz IF for FPGA-based DSP. Superheterodyne and zero-IF architectures were compared in \cite{danTerahertzWirelessCommunication2020, danSuperheterodyne300GHz2020}, achieving 60 Gb/s (15 GBd 16QAM) over 10 m at 300 GHz using 35 nm InGaAs TX/RX chips. A 1 THz link was experimentally demonstrated in \cite{senExperimentalDemonstrationUltrabroadband2019}, limited mainly by thermal noise and absorption of water-vapor. Data rates of 5–15 Gb/s over 13 cm yielded BERs between $1.3\times10^{-4}$ and $2.7\times10^{-2}$.

In \cite{parisiModulationsTerahertzBand2024}, authors analyzed modulation robustness to phase noise (PN) and peak-to-average power ratio (PAPR) across 130 GHz, 225 GHz, and 1.02 THz frequencies. Using 3GPP-based PN masks matched to measurements, they found QPSK and DSSS resilient to PN/PAPR, while OFDM and OTFS were more sensitive.

In \cite{wranaSensitivityAnalysis2803122022}, 35 nm InGaAs superheterodyne TX/RX chips near 300 GHz achieved up to 32 Gb/s (32QAM) and supported 256QAM with 8 bits/Hz spectral efficiency, targeting IEEE 802.15.3d. A 300 GHz Kiosk prototype in \cite{songPrototypeKIOSKData2018} achieved error-free 20 Gb/s real-time OOK transmission using InP-HEMT TX/RX chips with 10 dBm output power and 20 dBi lens gain under line-of-sight conditions.

End-to-end THz OFDM transmission at 140 GHz was evaluated in \cite{chakrabortyCaseOFDMUltrabroadband2021}. Hardware-induced frequency selectivity justified OFDM use; CP lengths of 1/16$^{th}$ symbol proved sufficient, with no gain from longer CPs. A 100 Gb/s 300 GHz link using 80 nm InP-HEMT technology was presented in \cite{hamada300GHz100GbInPHEMT2018}. Mixers with 15$\pm$2 dB gain enabled 25 GBd 16QAM transmission over 2.22 m using 50 dBi antennas, achieving BER  $<10^{-3}$. A 40 nm CMOS 300 GHz transceiver in \cite{hara300GHzCMOSTransceiver2018} demonstrated 2 Gb/s QPSK over 75 cm and 20 Gb/s 16QAM over 10 cm. Gate-pumped-mixer-last (TX) and fundamental-mixer-first (RX) architectures were used to overcome PA/LNA bandwidth limits and seamless integration with DSP. 

Finally, \cite{castroLongrangeHighSpeedTHzWireless2020} reported 300 GHz outdoor links up to 1 km. Using direct IQ-conversion MMICs and 55 dBi parabolic antennas, 16QAM and QPSK achieved 76.8 Gb/s (500 m) and 44.8 Gb/s (1 km) with SD-FEC. Ten-hour rain tests showed ~1.5 dB Q-factor degradation.

Table \ref{tab:literatureReview-electronics} summarizes these electronics-based THz transmission experiments by frequency, data rate, distance, modulation, and key design features.

\subsection{Photonics-based THz transmission experiments}\label{sec:literatureReview-photonics}

\begin{table*}[!htb]
\setlength{\tabcolsep}{4pt}
\caption{Summary of Photonics-Based THz Experiments.}
\label{tab:literatureReview-photonics}
\centering
\begin{tabularx}{\textwidth}{p{5mm}<{\raggedright} p{12mm}<{\raggedright} p{24mm}<{\raggedright} p{20mm}<{\raggedright} p{20mm}<{\raggedright} X}
\toprule
\textbf{Ref.} & \textbf{Frequency} & \textbf{Data Rate} & \textbf{Modulation} & \textbf{Distance} & \textbf{Highlights} \\ 
\midrule
\mbox{\cite{nagatsumaTerahertzWirelessCommunications2013}} & 300–720 GHz & Up to 48~Gb/s & Dual-polarization & 0.5–1~m & UTC-PD-based photonics transmitter; comparison of SBD and harmonic mixer receivers; first error-free 40~Gb/s link without FEC; optical comb source for high-order modulations. \\ 
\mbox{\cite{li400GOpticalWireless2013}} & 37.5 \& 100~GHz & 4$\times$100~Gb/s & QPSK, 16QAM & 0.7–1.5~m & 80~km SMF followed by THz wireless; offline DSP used to correct polarization crosstalk. \\ 
\mbox{\cite{moonCostEffectivePhotonicsBasedTHz2021}} & 300~GHz & PAM4/8 up to 30~GBd & PAM4, PAM8 & Up to 1.4~m & UTC-PD transmitter and SBD receiver; limited by MZM; ISI mitigated using FFE (101~taps) and DFE (5~taps). \\ 
\mbox{\cite{datTransparentFiberMillimeter2022}},\newline
\mbox{\cite{horstTransparentOpticalTHzOpticalLink2022}},\newline
\mbox{\cite{ummethalaTHztoopticalConversionWireless2019}} & 101–288 GHz & Up to 240~Gb/s gross (164~Gb/s net); 50~Gb/s & 32QAM, 64QAM, OFDM & 5–115~m wireless + up to 20~km fiber & First optical–THz–optical links; thin-film LiNbO$_3$ and plasmonic MZMs for direct THz-to-optical conversion; 240~Gb/s gross over 5~m; 192~Gb/s gross over 115~m + fiber; 50~Gb/s at 288~GHz over 16~m. \\ 
\mbox{\cite{maekawaSinglecarrier220GbitSubTHz2024}} & 300~GHz & Up to 220~Gb/s (single-carrier) & 16QAM, 32QAM, 64QAM, 128QAM & 214~m (extrapolated to 500~m) & First beyond-200~Gb/s single-carrier THz link; exploited Fresnel region (2–3~dB FSPL); UTC-PD source, SHM-based downconversion, 60~dBi Cassegrain antennas; 220~Gb/s with 44~GBd 32QAM. \\ 
\mbox{\cite{maekawa300GHzbandWirelessLink2023a}} & 300~GHz & 80~Gb/s (16QAM), 60~Gb/s (32QAM) & 16QAM, 32QAM & Not specified & Photonics-based Tx/Rx; SBS-assisted low-noise lasers for THz generation; improved phase/amplitude stability; performance depends on laser CNR; 15~GBd 32QAM tested. \\ 
\mbox{\cite{nagatsumaTerahertzCommunicationsTechnologies2012}} & 300~GHz & 24~Gb/s (photonic), 2.5~Gb/s (electronic) & OOK & 1~m (photonic), few cm (electronic) & First direct comparison of photonic vs. electronic transmitter/receiver architectures. \\ 
\bottomrule
\end{tabularx}
\end{table*}
Photonics-based THz transmitters offer a compelling alternative to electronic systems by enabling ultra-high carrier frequencies, broad bandwidths, and superior frequency tunability. Unlike electronics—limited by device linearity and frequency-multiplication efficiency—photonic systems exploit photomixing and optical frequency combs to generate spectrally pure, phase-stable THz carriers. This facilitates complex modulation formats and higher aggregate data rates, making photonic architectures promising for next-generation ultra-broadband THz links.

Signal generation typically employs uni-traveling carrier (UTC) or PIN photodiodes, while detection uses either direct or coherent schemes. Direct detection relies on SBDs or RTDs, whereas coherent receivers use photomixing-based local oscillators (LOs) and subharmonic mixers (SHMs). Recent fully photonic receivers remodulate the incoming THz signal onto an optical carrier and downconvert it via coherent optical detection—simplifying the remote antenna unit and preserving wide bandwidth.

In \cite{nagatsumaTerahertzWirelessCommunications2013}, envelope and coherent detection were compared for photonics-based THz links. A phase-stabilized optical frequency comb enabled dual-polarization transmission achieving 48 Gb/s.
In \cite{li400GOpticalWireless2013}, four 100 Gb/s polarization-multiplexed channels (QPSK and 16QAM) were transmitted through 80 km SMF and short wireless links (0.7 and 1.5 m), with offline DSP mitigating polarization crosstalk.
Using a UTC-PD transmitter at 300 GHz, \cite{moonCostEffectivePhotonicsBasedTHz2021} demonstrated 4-PAM/8-PAM up to 30 GBd with 47~$\mu$W output power; equalization with FFE/DFE mitigated intersymbol interference.

A recent trend involves optical–THz–optical (OTO) links, which directly remodulate the received THz signal onto an optical carrier—eliminating electronic downconversion. The first OTO demonstration at 101 GHz \cite{datTransparentFiberMillimeter2022} used a thin-film LiNbO$_3$ Mach–Zehnder modulator (MZM) collocated with the receive antenna, achieving 71 Gb/s (single carrier) over 5 m + 10 km fiber and 54 Gb/s (OFDM) over 20 m + 10 km fiber.
Building on this, \cite{horstTransparentOpticalTHzOpticalLink2022} employed a UTC-PD source and plasmonic MZM to realize 230 GHz OTO links with 240 Gb/s (5 m) and 150 Gb/s (115 m) gross bit-rates, later enhanced to 192 Gb/s using THz amplification.
Similarly, \cite{ummethalaTHztoopticalConversionWireless2019} achieved 50 Gb/s over 16 m at 288 GHz using a plasmonic MZM with 360 GHz modulation bandwidth, demonstrating efficient THz-to-optical conversion.

At longer distances, \cite{maekawaSinglecarrier220GbitSubTHz2024} achieved 220 Gb/s over 214 m using 300 GHz UTC-PD transmitters, Cassegrain antennas (60 dBi), and a photomixed LO for downconversion. 16QAM–128QAM formats were tested, with 32QAM (44 GBd) yielding the best performance.
In \cite{maekawa300GHzbandWirelessLink2023a}, SBS-assisted low-noise lasers improved phase stability at 300 GHz, enabling 80 Gb/s (32QAM).
Finally, \cite{nagatsumaTerahertzCommunicationsTechnologies2012} directly compared photonic and electronic THz systems at 300 GHz, showing 24 Gb/s OOK over 1 m (photonic, UTC-PD + SBD) versus 2.5 Gb/s (electronic, RTD) over a few centimeters—illustrating the inherent bandwidth advantage of photonics. Table \ref{tab:literatureReview-photonics} provides a summary of photonics-based THz transmission experiments.

\begin{table*}
\setlength{\tabcolsep}{4pt}
\caption{Summary of hybrid THz experiments.}
\label{tab:literatureReview-hybrid}
\centering
\begin{tabularx}{\textwidth}{p{5mm} p{0.1\linewidth} p{0.1\linewidth}<{\raggedright} p{0.15\linewidth}<{\raggedright} p{0.1\textwidth}<{\raggedright} X}
\hline
\textbf{Ref.} & \textbf{Frequency} & \textbf{Data Rate} & \textbf{Modulation} & \textbf{Distance} & \textbf{Highlights} \\ \hline
\mbox{\cite{ducournauUltrawideBandwidthSingleChannel04THz2014}}  & 400~GHz & 40~Gb/s & OOK & $\sim$2 m & Photonic Tx + electronic Rx; horn antennas (40/52~dBi). \\ 
\mbox{\cite{koenig100GbitWireless2013}} & 237.5~GHz & 100~Gb/s & Up to 16QAM & 20~m & UTC-PD Tx (photonic) + MMIC Rx (electronic, 35~nm HEMT); path loss 106~dB; oscilloscope + offline DSP. \\ 
\mbox{\cite{chinniSinglechannel100Gbit2018}} & 280~GHz & 100~Gb/s & 16QAM & 0.5~m & UTC-PD Tx (photonic), SHM-based Rx (electronic, $\times$6 LO multiplier); Tx power –10~dBm, Rx SNR $\approx$ 25~dB. \\ 
\mbox{\cite{jia04THzPhotonicWireless2018}} & 425~GHz & 128~Gb/s & 16QAM (32~GBd) & 0.5~m & UTC-PD Tx (photonic), SBD-mixer Rx (electronic); optical comb for stability; advanced DSP equalization. \\ 
\mbox{\cite{castro32GBd16QAM2019}} & 300~GHz & 64~Gb/s and 128~Gb/s & QPSK, 16QAM (32~GBd) & 0.5~m & PIN-PD source (photonic Tx), electronic Rx; Tx/Rx antennas (21/26~dBi); $\sim$8~dB path/misalignment loss. \\ 
\mbox{\cite{li120GbWireless2019}} & 375–500~GHz & 120~Gb/s (6$\times$20~Gb/s) & DP-QPSK (WDM) & 10 km fiber + 1.42~m THz & Hybrid fiber–wireless; photonic Tx and electronic Rx + 2$\times$2 MIMO DSP (CMA, 87~taps). \\ 
\mbox{\cite{jia2300Gbit2020}} & 320–380~GHz & 2$\times$300~Gb/s & PS-64QAM, OFDM & 2.8~m & Dual-polarization OFDM (photonic Tx and electronic Rx); Volterra equalization; high spectral efficiency. \\ 
\mbox{\cite{uemura600GHzBandHeterodyneReceiver2020}} & 600~GHz & 15~Gb/s & OOK & Short & PIN-PD based source (photonic Tx); novel Rx with IF leakage suppression (electronic Rx). \\ 
\mbox{\cite{moon6GIndoorNetwork2022}} & 300~GHz & 100~Gb/s DL, 25~Gb/s UL & 16QAM/QPSK & $\leq$2.5~m + 10~km fiber & Indoor photonics–electronics hybrid network (OHU + RNs); coherent SHM Rx; bidirectional architecture. \\ 
\mbox{\cite{jiaIntegratedDuallaserPhotonic2022}} & 408~GHz & 131~Gb/s & 16QAM OFDM & 10.7~m & Dual-DFB PIC with injection locking; UTC-PD mixer (photonic Tx); advanced DSP (linear, PN, nonlinear equalization); electronic Rx. \\ 
\mbox{\cite{nellenCoherentWirelessLink2022}} & 300~GHz & Up to 160~Gb/s & QPSK–64QAM & 0.5~m & PIN-PD generation (photonic Tx); SHM-based coherent Rx (electronic Rx). \\ 
\mbox{\cite{liPhotonicsassisted320GHz2023}} & 320~GHz & 50~Gb/s & QPSK and 16QAM & 850~m & Long-range outdoor link; horn + lens antennas (113.5~dBi total gain); Volterra nonlinear equalizer; photonic Tx and electronic Rx. \\ 
\mbox{\cite{dingHighSpeedLongDistancePhotonicsAided2023}} & 335–339~GHz & 32–124.8~Gb/s & 16QAM–PS-256QAM & Up to 400~m & UTC-PD Tx (photonic) + SHM Rx (electronic); Tx amplifier + custom 70~dBi lenses; advanced DSP with probabilistic shaping. \\ 
\mbox{\cite{tong200mPhotonicsAidedTerahertz2024}} & 300~GHz & 253~Gb/s & DP-OFDM 16QAM (46~GBd) & 20~km fiber + 200~m THz & UTC-PD Tx (photonic) + SHM Rx ($\times$24 LO, electronic); DSP with novel 4$\times$4 memory polynomial equalizer. \\ 
\mbox{\cite{liPhotonicTerahertzWireless2024}} & 320~GHz & 50~Gb/s & QPSK and 16QAM & 850~m & UTC-PD Tx (–8~dBm, photonic); horn (48.5~dBi) + lens (65~dBi) antennas; electronic Rx; advanced 2$\times$2 Volterra DSP. \\ 
\mbox{\cite{rhaNovelPhaseCFO2022}} & 300~GHz & 30~GBd 16QAM & 16QAM & Not specified & CFO/PN estimation algorithm for free-running lasers; BER improved from 8.8$\times$10$^{-3}$ to 3.6$\times$10$^{-3}$; photonic Tx and electronic Rx. \\ 
\mbox{\cite{zhou938Gb51502024}} & 5–150~GHz & 938~Gb/s & OFDM + bit loading & 12~cm & Seamless multi-band backhaul; 5–75~GHz via DACs (electronic Tx), 75–150~GHz via photonic mixing (photonic Tx), electronic Rx. \\ \hline
\end{tabularx}
\end{table*}
\subsection{Hybrid THz transmission experiments}\label{sec:literatureReview-hybrid}
Hybrid THz systems that combine photonic transmitters with electronic receivers—or integrate both domains within one link—have emerged as a promising alternative combining the spectral purity of photonics and the compact, power-efficient design of electronics. Photonic generation enables tunable, high-purity THz carriers and higher bandwidth, while electronics provide low-cost, integrated front-end, and DSP-friendly receivers. By leveraging these complementary advantages, hybrid links achieve higher data rates, longer distances, and improved spectral efficiency compared with purely electronic or photonic systems.

In \cite{ducournauUltrawideBandwidthSingleChannel04THz2014}, a 40 Gb/s link at 400 GHz over 2 m was demonstrated using a photonic transmitter and electronic receiver with 40 dBi/52 dBi antenna gains.
A 100 Gb/s, 237.5 GHz link over 20 m was shown in \cite{koenig100GbitWireless2013} using a UTC-PD transmitter supporting 16QAM and an MMIC receiver fabricated in 35 nm HEMT technology; total antenna gain was 86 dBi for a 106 dB path loss.
At 280 GHz, \cite{chinniSinglechannel100Gbit2018} reported 100 Gb/s 16QAM transmission using a UTC-PD source and SHM-based receiver; the received power was -19 dBm with 25 dB SNR. Similarly, \cite{jia04THzPhotonicWireless2018} demonstrated 128 Gb/s 16QAM at 425 GHz using a UTC-PD and SBD mixer, with optical-comb stabilization and DSP equalization mitigating mixer filtering.
A 300 GHz link in \cite{castro32GBd16QAM2019} employed a PIN-PD transmitter and achieved 32 GBd 16QAM over 0.5 m with 21 dBi/26 dBi antenna gains.

WDM integration was shown in \cite{li120GbWireless2019}, where 6$\times$25 GHz-spaced channels (375–500 GHz) carried 20 Gb/s DP-QPSK signals, reaching an aggregate 120 Gb/s after 10 km SMF + 142 cm THz transmission. A fractionally spaced 2$\times$2 MIMO CMA equalizer is used to decouple polarization crosstalk. High-capacity polarization-multiplexed 2$\times$300 Gb/s 64QAM OFDM transmission over 2.8 m was reported in \cite{jia2300Gbit2020} with Volterra-based nonlinear compensation.
At 600 GHz, \cite{uemura600GHzBandHeterodyneReceiver2020} demonstrated OOK up to 15 Gb/s using a heterodyne receiver that removed IF leakage.

Hybrid THz–fiber networks were proposed in \cite{moon6GIndoorNetwork2022}, showing a 300 GHz, 100 Gb/s 16QAM downlink and 25 Gb/s QPSK uplink over 10 km fiber + 2.5 m wireless, enabled by photomixing, SHM downconversion, and DSP-based dispersion compensation.
A dual-DFB PIC for 408 GHz carrier generation via injection-locked lasers was presented in \cite{jiaIntegratedDuallaserPhotonic2022}; using 16QAM OFDM, the UTC-PD-based system achieved 131 Gb/s over 10.7 m with coherent SHM reception and nonlinear equalization.
At 300 GHz, \cite{nellenCoherentWirelessLink2022} demonstrated QPSK–64QAM formats up to 160 Gb/s (32QAM) over 50 cm using a PIN-PD transmitter and coherent SHM receiver.

Long-range demonstrations include \cite{liPhotonicsassisted320GHz2023}, achieving 50 Gb/s over 850 m at 320 GHz with 113.5 dBi antenna gain and Volterra equalization; and \cite{dingHighSpeedLongDistancePhotonicsAided2023}, achieving 124.8 Gb/s (100 m), 56 Gb/s (200 m), and 32 Gb/s (400 m) using probabilistically shaped 256QAM–16QAM, THz amplification, and lens-aided antennas (total gain$\approx$ 122 dBi). The configuration in \cite{liPhotonicTerahertzWireless2024} achieved a similar 50 Gb/s over 850 m at 320 GHz, with UTC-PD power of -8 dBm and 2$\times$2 MIMO second order Volterra equalization compensating amplitude and IQ skew.

\begin{table*}[!htb]
\setlength{\tabcolsep}{4pt}
\centering
\caption{Summary of Hardware Impairment Modeling Approaches in Photonics- and Electronics-Based Systems}
\label{tab:literatureReview-theoretical}
\begin{tabularx}{\textwidth}{p{0.04\linewidth} p{0.15\linewidth}<{\raggedright} X X X}
\toprule
\textbf{Ref.} & \textbf{Underlying Architecture} & \textbf{Hardware Impairment} & \textbf{Modeling Approach} & \textbf{Performance Metric} \\
\hline
\addlinespace
\cite{a.TeraHertzTechnologyTHz2025} & Electronics-based & Linear (bandwidth, I/Q imbalance), Nonlinear PA, Oscillator phase noise & 
Bandwidth limitation modeled as a low-pass filter; static I/Q imbalance; PA nonlinearity modeled by a memory polynomial; phase noise PSD modeled by a multi-pole multi-zero approach & 
BER vs. symbol rate \\
\addlinespace
\cite{bjornsonNewLookDualHop2013} & Electronics-based & Abstract hardware modeling & Complex Gaussian noise approximation & Outage probability, ergodic capacity \\
\cite{boulogeorgosAnalyticalPerformanceAssessment2019} & Electronics-based & Abstract hardware modeling & Extension of the Gaussian model with channel effects & Outage probability, ergodic capacity \\
\cite{antesPerformanceEstimationBroadband2015} & Electronics-based & I/Q imbalance, phase noise, DC offset, oscillator self-mixing & 
Static I/Q amplitude and phase imbalance; oscillator phase noise modeled as Gaussian noise & 
EVM, error floor vs. SNR \\
\cite{bicaisPhaseNoiseModel2019} & Electronics-based & Oscillator phase noise & Correlated vs. uncorrelated Gaussian models & Complexity vs. accuracy tradeoff \\
\cite{kokkoniemiImpactBeamMisalignment2020} & Electronics-based & Antenna misalignment & 1D/2D Gaussian motion models & Effective antenna gain degradation \\
\cite{saadMIMOTechniquesWireless2020} & Electronics-based & Oscillator phase noise & Gaussian approximation & Throughput, power consumption \\
\cite{neshaastegaranEffectOscillatorPhase2020} & Electronics-based & Oscillator phase noise & Near-carrier + white noise floor components & SINR vs. SNR, BER vs. SNR \\
\cite{tarboushSingleMulticarrierTerahertzBand2022} & Electronics-based & Additive noise and oscillator phase noise & Gaussian additive noise and Gaussian phase noise & BER vs. SNR, PAPR, computational complexity \\
\cite{oikonomouDesignSuperConstellations2025} & Electronics-based & Additive noise and oscillator phase noise & Gaussian additive noise and Gaussian phase noise & Symbol error probability, detection complexity \\
\cite{belloAnalysisGaussianPhase2024} & Electronics-based & Additive noise and oscillator phase noise & Gaussian additive noise and Gaussian phase noise model & BER vs. SNR \\
\cite{desombreContinuousPhaseModulation2024} & Photonics-Tx and Electronic-Rx & Photodiode nonlinearities & Nonlinear models for AM–AM and AM–PM transfer functions & BER vs. distance, energy efficiency vs. spectral efficiency \\
\cite{feketeEffectLaserCharacteristics2015} & Photonics-Tx \& No Rx & Laser RIN and phase noise; RF oscillator phase noise & 
Phase noise modeled as a Wiener process; laser models with RIN and phase noise derived from laser rate equations & 
Carrier-to-noise ratio of the generated signal \\
This work & Both electronics- and photonics-based approaches & 
\textbf{Electronics-based:} Thermal noise of mixers; noise floor of base oscillators; near-carrier (correlated) and white (uncorrelated) phase noise. 
\newline
\textbf{Photonics-based:} Laser RIN; optical amplifier ASE; photomixer shot noise; mixer thermal noise; laser phase noise &
\textbf{Electronics-based:} Thermal noise and oscillator noise floor modeled as AWGN; near-carrier phase noise modeled as fractional Brownian motion; white phase noise as Gaussian process.
\newline
\textbf{Photonics-based:} RIN, ASE, and shot noise modeled as AWGN; laser phase noise modeled as Wiener process &
SNR limitations at transmitter and receiver; BER vs. SNR for single-carrier QAM with realistic parameters; performance comparison of photonics- and electronics-based THz systems \\
\bottomrule
\end{tabularx}
\end{table*}

Algorithmic advances also support hybrid links: \cite{rhaNovelPhaseCFO2022} proposed a phase/CFO estimator improving BER from $8.8×10^{-3} \rightarrow 3.6×10^{-3}$ for 30 GBd 16QAM at 0.3 THz, tolerant to $\pm$5 GHz CFO.
In \cite{tong200mPhotonicsAidedTerahertz2024}, a 20 km fiber + 200 m wireless link achieved 253 Gb/s DP-OFDM 16QAM at 300 GHz, using high-gain lens antennas (97 dBi total) and a 4×4 trigonometric memory-polynomial equalizer.
Finally, \cite{zhou938Gb51502024} demonstrated hybrid signal generation across 5–150 GHz, combining electronic DAC-based (5–75 GHz) and photonics-based (75–150 GHz) signal generation to achieve 938 Gb/s OFDM with adaptive bit-loading. Table \ref{tab:literatureReview-hybrid} summarized the hybrid THz transmission experiments.
\color{black}

\subsection{Theoretical modeling of THz systems under hardware impairments}\label{sec:literatureReview-theoretical}
Experimental results show that hardware impairments strongly affect THz link performance. Compared with sub-6 GHz or mmWave systems, oscillator instabilities, amplifier nonlinearities, and laser or detector noise are far more pronounced at THz frequencies. Several studies have modeled these effects to capture their underlying mechanisms and system impact.

Most theoretical efforts address electronics-based transceivers, where key impairments include bandwidth limits, I/Q imbalance, PA nonlinearities, and oscillator-induced phase noise. The ETSI report \cite{a.TeraHertzTechnologyTHz2025} summarizes their impact on high-symbol-rate QAM, while simulation frameworks such as \cite{eckhardtModularLinkLevel2022} model linear distortion via FIR filters, PA nonlinearities via memory-polynomial, and phase noise using the multi-pole, multi-zero model from 3GPP \cite{3GPPTR38808}, with realistic parameters from \cite{scheyttUltraLowPhaseNoise2020,liEightElement136147GHz2022,johnBroadband300GHzPower2020}. Many works approximate cumulative hardware effects as complex Gaussian noise for analytical tractability \cite{bjornsonNewLookDualHop2013,boulogeorgosAnalyticalPerformanceAssessment2019,antesPerformanceEstimationBroadband2015}. The model of \cite{boulogeorgosAnalyticalPerformanceAssessment2019}, includes antenna misalignment (adapted from FSO \cite{faridOutageCapacityOptimization2007}), fading, and atmospheric absorption; assessing outage and ergodic capacity.

Other studies refine specific mechanisms. \cite{antesPerformanceEstimationBroadband2015} developed an EVM model including I/Q imbalance, DC offset, and self-mixing, identifying phase noise as the dominant limiting factor. \cite{bicaisPhaseNoiseModel2019} compared correlated and uncorrelated Gaussian phase-noise models, noting that the latter provides adequate accuracy for wideband systems. Beam-misalignment effects were analyzed in \cite{kokkoniemiImpactBeamMisalignment2020}, showing strong gain degradation under realistic motion models. Spatial-multiplexing strategies under Gaussian phase noise were evaluated in \cite{saadMIMOTechniquesWireless2020}, where generalized spatial multiplexing MIMO outperformed conventional schemes. OFDM impairments from oscillator noise were modeled in \cite{neshaastegaranEffectOscillatorPhase2020}, while \cite{tarboushSingleMulticarrierTerahertzBand2022} compared single- and multi-carrier modulations under correlated and uncorrelated phase noise. New constellations such as Super-APSK (SAPSK) were proposed in \cite{oikonomouDesignSuperConstellations2025} with closed-form SEP analysis, and \cite{belloAnalysisGaussianPhase2024} showed that DFT-s-OFDM even with suboptimal detection outperforms OFDM under strong Gaussian phase noise.

In contrast, photonics-based modeling remains limited. \cite{desombreContinuousPhaseModulation2024} introduced Continuous Phase Modulation (CPM) as a robust alternative to QAM/QPSK under photodiode nonlinearities, explicitly modeling AM–AM and AM–PM transfer characteristics. CPM proved markedly more resilient, achieving error-free 7.5 GBd transmission over 1 m at 107.5 GHz without FEC. Similarly, \cite{feketeEffectLaserCharacteristics2015} compared OPLL- and optical-comb-based heterodyning for 100 GHz signal generation, analyzing RIN and phase-noise trade-offs between setups.

Overall, theoretical modeling of THz systems remains sparse. Most studies treat hardware effects collectively as Gaussian noise, overlooking additive-noise contributions from mixers and oscillators. In this work, we explicitly model these additive noise sources and trace their propagation through the transmitter and receiver chains, deriving aggregate noise statistics that depart from the classical AWGN assumption. The resulting signal models—incorporating phase noise for both electronics- and photonics-based links—yield analytical SNR predictions for QAM formats under realistic hardware constraints, enabling a unified comparison of both system types.

Beyond advanced physical layer designs, future THz networks are also expected to rely on intelligent monitoring and control mechanisms operating at the network level. Recent advances in graphical modeling and real-time analysis of dynamic network data provide promising tools for tracking evolving network states, detecting anomalies, and enabling adaptive management of large-scale communication systems \cite{kayaNewHybridApproach2025,heIntelligentTerahertzMedium2023}. While such approaches are outside the scope of the present work, they may complement the impairment-aware physical-layer models developed here when designing intelligent THz network infrastructures.

\vspace{-5pt}
\section{Photonics-based THz systems}\label{sec:photonics-basedTHz}
\subsection{Signal Models for Transmission and Reception}

\begin{figure}  
    \centering
    \subfloat[]{\includegraphics[trim={0.1cm 0 0 0}, clip, scale=0.9]{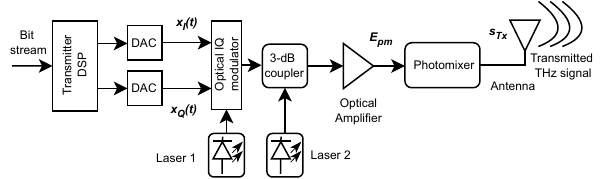}}\\
    \subfloat[]{\includegraphics[trim={0.25cm 0 0 0}, clip, scale=0.9]{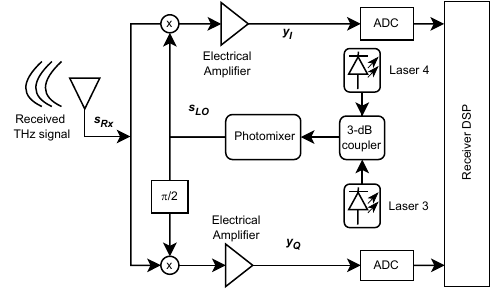}}
    \caption{Photonics based THz system (a) Transmitter (b) Receiver.}\label{fig:photonBlkDiagram}
\end{figure}

\begin{figure}  
    \centering
    \subfloat[][]{\includegraphics[trim={0.1cm 0 0 0}, clip, scale=0.9]{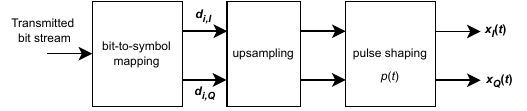}}\\
    \subfloat[][]{\includegraphics[trim={0.1cm 0 0 0}, clip, scale=0.7]{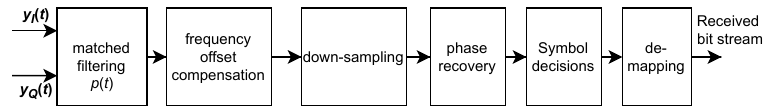}}
    \caption{Sub-blocks of DSP. (a) Transmitter DSP block diagram (b) Receiver DSP block diagram.}\label{fig:DSPBlkDiagram}
    \vspace{-10pt}
\end{figure}

The block diagrams of the photonics-based THz coherent transmitter and receiver are shown in Fig. \ref{fig:photonBlkDiagram}. The sub-blocks of baseband digital signal processing (DSP) are further illustrated in Fig. \ref{fig:DSPBlkDiagram}.

\noindent\textbf{Transmitted Signal Model:} The input bit stream is mapped to quadrature amplitude modulation (QAM) symbols, upsampled and subsequently filtered by a digital pulse shaping filter. The pulse shaped signals are converted to analog waveforms by digital to analog converters (DACs). The continuous time waveforms generated by the DACs corresponding to the in-phase (I) and quadrature-phase (Q) channels are given by ${x_I}\left( t \right) = \sum\nolimits_i {p(t - i{T_M}){d_{i,I}}}$, \textit{i=}1,2… and ${x_Q}\left( t \right) = \sum\nolimits_i {p(t - i{T_M}){d_{i,Q}}}$, \textit{i=}1,2… where $p\left( t \right)$ is the pulse shape, ${T_M}$ is the symbol period, and ${d_{i,I}}$, ${d_{i,Q}}$ are the real and imaginary components of the complex baseband QAM modulation symbol during the ${i^{th}}$ symbol period. The complex baseband waveform is defined as $x\left( t \right) = {x_I}\left( t \right) + j \cdot {x_Q}\left( t \right)$. The mean power of baseband signal is assumed to be unity, i.e., $E\left[ {{{\left| {x\left( t \right)} \right|}^2}} \right] = 1$.

The output of DACs corresponding to I- and Q-channels drive an IQ Mach-Zehnder-modulator (MZM) which is fed by a laser with a frequency of ${f_1}$ in the infrared band. Another infrared laser with a frequency ${f_2}$ is coupled with the modulated signal. These frequencies are selected such that the difference between the two frequencies ${f_c} = \left| {{f_1} - {f_2}} \right|$ yields the desired THz frequency. The coupled signals undergo optical amplification and then impinge on a photomixer which takes signal input at infrared wavelengths and is capable of radiating in THz regime. Mathematically speaking, the signal field at the input of photomixer is given by
\begin{flalign}\label{eq:1}
{E_{pm}} &= \frac{1}{{\sqrt 2 }}\left[ {\sqrt {{P_1}} x\left( t \right)\exp \left\{ { - j\left( {2\pi {f_1}t + {\phi _1}\left( t \right)} \right)} \right\}} \right.&&\\\nonumber
&\left. { + \sqrt {{P_2}} \exp \left\{ { - j\left( {2\pi {f_2}t + {\phi _2}\left( t \right)} \right)} \right\} + {{\tilde i}_{Opt}}\left( t \right)} \right]&&
\end{flalign}
where ${P_1}$ is the transmit signal power, ${P_2}$ is the power of the second laser, and ${\phi _1}(t)$ and ${\phi _2}(t)$ are the phase offsets of the first and second lasers. In \eqref{eq:1}, ${\tilde i_{Opt}}\left( t \right)$ is the optical noise term which incorporates ASE noise generated by the optical amplifier and the RIN of the lasers. The optical noise is modelled as a symmetric complex Gaussian distribution i.e., ${\tilde i_{Opt}}\left( t \right) \sim CN\left( {0;\sigma _{Opt}^2} \right) = CN\left( {0;\sigma _{ASE}^2 + \sigma _{RIN}^2} \right)$ where $\sigma _{ASE}^2$ and $\sigma _{RIN}^2$ are noise powers due to ASE and RIN, respectively.

The photomixer detects the intensity of the impinging electrical field, and the corresponding output current is given as
\begin{flalign}\label{eq:2}
    {s_{Tx}} = R{\left| {{E_{pm}}} \right|^2} + {i_{sh,Tx}}(t)&&
\end{flalign}
where   is the responsivity of the photomixer, and   is the shot noise generated by the photomixing process. Replacing \eqref{eq:1} in \eqref{eq:2} and expanding the resulting expression, we obtain

\begin{flalign}\label{eq:3}    
{s_{Tx}} &= \frac{R}{2}\left[ {{P_1} \cdot {{\left| {x\left( t \right)} \right|}^2} + {P_2} + {{\left| {{{\tilde i}_{Opt}}\left( t \right)} \right|}^2}} \right.&&\\\nonumber
 &+ 2\Re \left[ {\sqrt {{P_2}} \sqrt {{P_1}} x\left( t \right)\exp \left\{ { - j\left( {2\pi {f_c}t + {\phi _c}\left( t \right)} \right)} \right\}} \right.&&\\\nonumber
&\left. {\left. { + \sqrt {{P_2}}  \cdot {{\mathord{\buildrel{\lower3pt\hbox{$\scriptscriptstyle\frown$}} 
\over i} }_{Opt}}\left( t \right) + \sqrt {{P_1}} x\left( t \right) \cdot {{\hat i}_{Opt}}\left( t \right)} \right]} \right] + {i_{sh,Tx}}(t)&&
\end{flalign}
where ${\mathord{\buildrel{\lower3pt\hbox{$\scriptscriptstyle\frown$}} 
\over i} _{Opt}}\left( t \right) = {\tilde i_{Opt}}\left( t \right)\exp \left\{ { - j\left( {2\pi {f_2}t + {\phi _2}\left( t \right)} \right)} \right\}$, ${\hat i_{Opt}}\left( t \right) = {\tilde i_{Opt}}\left( t \right) \cdot \exp \left\{ { - j\left( {2\pi {f_1}t + {\phi _1}\left( t \right)} \right)} \right\}$, ${f_c} = \left| {{f_1} - {f_2}} \right|$ and ${\phi _c}(t) = {\phi _1}(t) - {\phi _2}(t)$ is the resulting phase of the generated THz carrier.

The first term in \eqref{eq:3} represents the signal-signal beat interference (SSBI) which occupies a bandwidth double of the baseband signal bandwidth around the zero frequency. The second term is the direct current (DC) offset. The photomixer is designed to radiate around the THz band of frequencies, while the SSBI and DC terms are far below its operational range. Hence, the SSBI and DC terms can be sufficiently suppressed by design. For example, a bandpass filter centered at ${f_c}$ will be sufficient to filter out unwanted DC and SSBI terms before transmission. The third term is optical beat noise generated due to the mixing of optical noise with itself, during the photomixing process. Optical beat noise occupies a broad spectrum and cannot be filtered out like the first two components. After removing the filtered-out components, \eqref{eq:3} reduces to
\begin{flalign}\label{eq:4}
{s_{Tx}} &= R\frac{{{{\left| {{{\tilde i}_{Opt}}\left( t \right)} \right|}^2}}}{2} + R \cdot \sqrt {{P_2}} \sqrt {{P_1}} \left\{ {{x_I}\left( t \right)\cos \left( {2\pi {f_c}t} \right.} \right.&&\\\nonumber
&\left. {\left. { + {\phi _c}\left( t \right)} \right) + {x_Q}\left( t \right)\sin \left( {2\pi {f_c}t + {\phi _c}\left( t \right)} \right)} \right\}&&\\\nonumber
& + R\sqrt {{P_2}} \Re \left\{ {{{\mathord{\buildrel{\lower3pt\hbox{$\scriptscriptstyle\frown$}} 
\over i} }_{Opt}}\left( t \right)} \right\} + R\sqrt {{P_1}} {x_I}\left( t \right) \cdot \Re \left\{ {{{\hat i}_{Opt}}\left( t \right)} \right\}&&\\\nonumber
 &- R\sqrt {{P_1}} {x_Q}\left( t \right) \cdot \Im \left\{ {{{\hat i}_{Opt}}\left( t \right)} \right\} + {i_{sh,Tx}}\left( t \right)&&
\end{flalign}

Note that the second term in \eqref{eq:4} is the signal of interest and other components represent the unwanted noise. Rearranging \eqref{eq:4}, we can write 
\begin{flalign}\label{eq:5}
{s_{Tx}} &= {G_1} \cdot \left\{ {{x_I}\left( t \right)\cos \left( {2\pi {f_c}t + {\phi _c}\left( t \right)} \right)} \right.&&\\\nonumber
&\left. { + {x_Q}\left( t \right)\sin \left( {2\pi {f_c}t + {\phi _c}\left( t \right)} \right)} \right\} + {n_{Tx}}\left( t \right)&&
\end{flalign}
where we define ${G_1} = R \cdot \sqrt {{P_2}{P_1}}$. Here, the noise term ${n_{Tx}}\left( t \right)$ is defined by
\begin{flalign}\label{eq:6}
{n_{Tx}}\left( t \right) &= R\frac{{{{\left| {{{\tilde i}_{Opt}}\left( t \right)} \right|}^2}}}{2} + R\sqrt {{P_2}} \Re \left\{ {{{\mathord{\buildrel{\lower3pt\hbox{$\scriptscriptstyle\frown$}} 
\over i} }_{Opt}}\left( t \right)} \right\}&&\\\nonumber
& + R\sqrt {{P_1}} {x_I}\left( t \right) \cdot \Re \left\{ {{{\hat i}_{Opt}}\left( t \right)} \right\}&&\\\nonumber
& - R\sqrt {{P_1}} {x_Q}\left( t \right) \cdot \Im \left\{ {{{\hat i}_{Opt}}\left( t \right)} \right\} + {i_{sh,Tx}}\left( t \right)&&
\end{flalign}

It should be noted that ${\tilde i_{Opt}}\left( t \right)$ is modeled a symmetric complex Gaussian random variable, therefore its statistics will not change by multiplication with a complex exponential, i.e., ${\tilde i_{Opt}}\left( t \right) \sim {\mathord{\buildrel{\lower3pt\hbox{$\scriptscriptstyle\frown$}} 
\over i} _{Opt}}\left( t \right) \sim {\hat i_{Opt}}\left( t \right) \sim CN\left( {0;\sigma _{Opt}^2} \right)$. The statistical model of ${n_{Tx}}\left( t \right)$ is given as
\begin{flalign}\label{eq:7}
{n_{Tx}}\left( t \right) &\sim \frac{{R}}{2}{\left| {CN\left( {0;\sigma _{Opt}^2} \right)} \right|^2} + R\sqrt {{P_2}} N\left( {0;{{\sigma _{Opt}^2} \mathord{\left/
 {\vphantom {{\sigma _{Opt}^2} 2}} \right.
 \kern-\nulldelimiterspace} 2}} \right)&&\\\nonumber
& + R\sqrt {{P_1}} {x_I}\left( t \right) \cdot N\left( {0;{{\sigma _{Opt}^2} \mathord{\left/
 {\vphantom {{\sigma _{Opt}^2} 2}} \right.
 \kern-\nulldelimiterspace} 2}} \right)&&\\\nonumber
& - R\sqrt {{P_1}} {x_Q}\left( t \right) \cdot N\left( {0;{{\sigma _{Opt}^2} \mathord{\left/
 {\vphantom {{\sigma _{Opt}^2} 2}} \right.
 \kern-\nulldelimiterspace} 2}} \right) + N\left( {0;\sigma _{sh,Tx}^2} \right)&&
\end{flalign}

Expressing ${\left| {CN\left( {0;\sigma _{Opt}^2} \right)} \right|^2}$ in terms of real and imaginary parts, \eqref{eq:7} can be rewritten as
\begin{flalign}\label{eq:8}
{n_{Tx}}\left( t \right) &\sim \frac{{R}}{2}\left[ {\left( {{N_I}{{\left( {0;{{\sigma _{Opt}^2} \mathord{\left/
 {\vphantom {{\sigma _{Opt}^2} 2}} \right.
 \kern-\nulldelimiterspace} 2}} \right)}^2} + {N_Q}{{\left( {0;{{\sigma _{Opt}^2} \mathord{\left/
 {\vphantom {{\sigma _{Opt}^2} 2}} \right.
 \kern-\nulldelimiterspace} 2}} \right)}^2}} \right)} \right]&&\\\nonumber
& + N\left( {0;{{{R^2}{P_2}\sigma _{Opt}^2} \mathord{\left/
 {\vphantom {{{R^2}{P_2}\sigma _{Opt}^2} 2}} \right.
 \kern-\nulldelimiterspace} 2}} \right) + {x_I}\left( t \right) \cdot N\left( {0;{R^2}{P_1}{{\sigma _{Opt}^2} \mathord{\left/
 {\vphantom {{\sigma _{Opt}^2} 2}} \right.
 \kern-\nulldelimiterspace} 2}} \right)&&\\\nonumber
& - {x_Q}\left( t \right) \cdot N\left( {0;{{{R^2}{P_1}\sigma _{Opt}^2} \mathord{\left/
 {\vphantom {{{R^2}{P_1}\sigma _{Opt}^2} 2}} \right.
 \kern-\nulldelimiterspace} 2}} \right) + N\left( {0;\sigma _{sh,Tx}^2} \right)&&
\end{flalign}

Note that ${N_I}\left( {0;{{\sigma _{Opt}^2} \mathord{\left/
 {\vphantom {{\sigma _{Opt}^2} 2}} \right.
 \kern-\nulldelimiterspace} 2}} \right) = \left( {{{{\sigma _{Opt}}} \mathord{\left/
 {\vphantom {{{\sigma _{Opt}}} {\sqrt 2 }}} \right.
 \kern-\nulldelimiterspace} {\sqrt 2 }}} \right) \cdot {N_I}\left( {0;1} \right)$ and ${N_Q}\left( {0;{{\sigma _{Opt}^2} \mathord{\left/
 {\vphantom {{\sigma _{Opt}^2} 2}} \right.
 \kern-\nulldelimiterspace} 2}} \right) = \left( {{{{\sigma _{Opt}}} \mathord{\left/
 {\vphantom {{{\sigma _{Opt}}} {\sqrt 2 }}} \right.
 \kern-\nulldelimiterspace} {\sqrt 2 }}} \right) \cdot {N_Q}\left( {0;1} \right)$, hence, \eqref{eq:8} can be simplified as
\begin{flalign}\label{eq:9}
    {n_{Tx}}\left( t \right)\sim{n_{Tx1}}\left( t \right) + {n_{Tx2}}\left( t \right) + {n_{Tx3}}\left( t \right) - {n_{Tx4}}\left( t \right)&&
\end{flalign}
where we define ${n_{Tx1}}\left( t \right) \sim \left( {{{R\sigma _{Opt}^2} \mathord{\left/
 {\vphantom {{R\sigma _{Opt}^2} 4}} \right.
 \kern-\nulldelimiterspace} 4}} \right) \cdot \chi _2^2$, \\${n_{Tx2}}\left( t \right) \sim N\left( {0;{{\sigma _{sh,Tx}^2 + {R^2}{P_2}\sigma _{Opt}^2} \mathord{\left/
 {\vphantom {{\sigma _{sh,Tx}^2 + {R^2}{P_2}\sigma _{Opt}^2} 2}} \right.
 \kern-\nulldelimiterspace} 2}} \right)$,\\ ${n_{Tx3}}\left( t \right) \sim {x_I}\left( t \right) \cdot N\left( {0;{R^2}{P_1}{{\sigma _{Opt}^2} \mathord{\left/
 {\vphantom {{\sigma _{Opt}^2} 2}} \right.
 \kern-\nulldelimiterspace} 2}} \right)$, \\and ${n_{Tx4}}\left( t \right) \sim {x_Q}\left( t \right) \cdot N\left( {0;{{{R^2}{P_1}\sigma _{Opt}^2} \mathord{\left/
 {\vphantom {{{R^2}{P_1}\sigma _{Opt}^2} 2}} \right.
 \kern-\nulldelimiterspace} 2}} \right)$. 

\noindent\textbf{Received signal model:} The complex equivalent representation of transmit signal in \eqref{eq:5} is given as ${s_{Tx}} = {G_1} \cdot x\left( t \right)\exp \left( { - j\left( {2\pi {f_c}t + {\phi _c}\left( t \right)} \right)} \right) + {\tilde n_{Tx}}\left( t \right)$ where ${\tilde n_{Tx}}\left( t \right)$ is the complex equivalent representation of ${n_{Tx}}\left( t \right)$. Its statistical description is given as ${\tilde n_{Tx}}\left( t \right) \sim \left( {{{R\sigma _{Opt}^2} \mathord{\left/
 {\vphantom {{R\sigma _{Opt}^2} 4}} \right.
 \kern-\nulldelimiterspace} 4}} \right)\left[ {C\chi _2^2} \right] + CN\left( {0;{{\sigma _{sh,Tx}^2 + {R^2}{P_2}\sigma _{Opt}^2} \mathord{\left/
 {\vphantom {{\sigma _{sh,Tx}^2 + {R^2}{P_2}\sigma _{Opt}^2} 2}} \right.
 \kern-\nulldelimiterspace} 2}} \right) + x\left( t \right) \cdot CN\left( {0;{R^2}{P_1}\sigma _{Opt}^2} \right)$.

The transmit signal is launched into the propagation medium through a horn antenna followed by a collimating lens. Assuming that we have a line of sight (LOS) link and there are no reflected waves, the received signal is given as
\begin{flalign}\label{eq:10}
{s_{Rx}} &= \alpha  \cdot \left[ {{G_1} \cdot x\left( t \right)\exp \left\{ { - j\left( {2\pi {f_c}t + {\phi _c}\left( t \right)} \right)} \right\}} \right.&&\\\nonumber
&\left. { + {{\tilde n}_{Tx}}\left( t \right)} \right]&&
\end{flalign}

where $\alpha$ is the channel coefficient representing the overall propagation gain between the transmitter and receiver. In practice, $\alpha$ accounts for the combined effects of transmit and receive antenna gains,  geometrical loss, atmospheric absorption, and additional implementation-dependent losses such as antenna misalignment or coupling loss. Mathematically speaking, the corresponding power-domain gain is given by
\begin{equation}
|\alpha|^2=\frac{G_{\mathrm{Tx}}G_{\mathrm{Rx}}}{L_{\mathrm{FS}}(d,f_c)\,L_{\mathrm{atm}}(d,f_c)\,L_{\mathrm{align}}(d)},
\end{equation}
where $G_{\mathrm{Tx}}$ and $G_{\mathrm{Rx}}$ denote the transmit and receive antenna gains, $L_{\mathrm{FS}}(d,f_c)$ is the geometrical-loss at carrier frequency $f_c$ and distance $d$, $L_{\mathrm{atm}}(d,f_c)$ denotes the atmospheric attenuation factor, and $L_{\mathrm{align}}(d)$ accounts for antenna misalignment loss.

The geometrical loss is modeled by the Friis free-space propagation model as ${L_{FS}}\left( {d,{f_c}} \right) = {\left( {{{4\pi d{f_c}} \mathord{\left/
 {\vphantom {{4\pi d{f_c}} c}} \right.
 \kern-\nulldelimiterspace} c}} \right)^2}$  where $f_c$ is the carrier frequency and $c$ is the speed of light in vacuum. The atmospheric loss is modeled by the Beer-Lambart law as ${L_{atm}}\left( {{f_c},d} \right) = \exp \left( {\kappa \left( {{f_c}} \right)d} \right)$ where $\kappa \left( {{f_c}} \right)$ is the absorption-coefficient of the propagation medium at $f_c$. Under the assumption of a Gaussian intensity profile, the alignment loss is formulated as ${L_{align}}\left( d \right) = \exp \left( {{{{r^2}} \mathord{\left/
 {\vphantom {{{r^2}} {w{{\left( d \right)}^2}}}} \right.
 \kern-\nulldelimiterspace} {w{{\left( d \right)}^2}}}} \right)$ where $r$ is the lateral beam displacement at the receiver plane and $w(d) = {\theta _{div}}d$ is the beam diameter at the receiver for a divergence angle $\theta_{div}$. 

As an example, consider a carrier frequency of $f_c=300$~GHz and a transmission distance of $d=1$~m, the geometrical loss is approximately $L_{FS}=82$~dB. A beam diameter of $w\left( d \right) = 10$~mm and a lateral displacement of $r = 2$~mm result in an alignment loss of approximately ${L_{align}} = 0.35$~dB. For clear weather condition, the absorption coefficient is  $\kappa \left( {{f_c}} \right) \approx 0.058$~m$^{-1}$ (equivalently, an attenuation of 0.5~dB/m) and the resulting atmospheric loss is ${L_{atm}} = 0.5$~dB.
Assuming antenna gains of $G_{Tx}=G_{Rx}=40$~dBi, the resulting link budget yields a total loss of approximately 2.85~dB. This corresponds to an effective propagation gain of $\alpha\approx0.72$. Under the same assumptions, if higher-gain antennas are employed such that the combined antenna gain is $\sim$91~dBi, the effective propagation gain increases to $\alpha\approx2.5$.

In our study, our primary focus is on analyzing the impact of hardware impairments rather than propagation effects. Therefore, when deriving and analyzing the noise statistics, we normalize the channel gain by setting $\alpha=1$ in order to isolate the contribution of hardware-induced noise sources. In the BER performance evaluation, however, $\alpha$ is varied to sweep the received signal power and evaluate system performance over a range of operating conditions. Under this formulation, propagation losses are implicitly reflected through the corresponding value of $\alpha$.

At the receiver side, the required THz local oscillator (LO) signal is generated using the same principle as described for the transmitter. The only difference here is that both the coupled lasers input to the photomixer are unmodulated. The LO signal at the output of photomixer can be written as 
\begin{flalign}\label{eq:11}
{s_{LO}} &= R \cdot \sqrt {{P_3}{P_4}} \cos \left( {2\pi {f_{LO}}t + {\phi _{LO}}\left( t \right)} \right)&&\\\nonumber
& + {n_{LO}}\left( t \right)&&
\end{flalign}
where ${P_3}$, and ${P_4}$ are the powers of the third and fourth lasers, respectively, ${n_{LO}}\left( t \right)$ is the noise generated by the photomixer which generates the LO, ${f_{LO}} = \left| {{f_3} - {f_4}} \right|$ and ${\phi _{LO}}(t) = {\phi _3}(t) - {\phi _4}(t)$ are the frequency and phase of the generated LO, respectively. The noise term ${n_{LO}}\left( t \right)$ is given as
\begin{flalign}\label{eq:12}
{n_{LO}}\left( t \right) &= R\frac{{{{\left| {{{\tilde i}_{Opt,LO}}\left( t \right)} \right|}^2}}}{2} + R\sqrt {{P_4}} \Re \left\{ {{{\mathord{\buildrel{\lower3pt\hbox{$\scriptscriptstyle\frown$}} 
\over i} }_{Opt,LO}}\left( t \right)} \right\}&&\\\nonumber
& + R\sqrt {{P_3}} \Re \left\{ {{{\hat i}_{Opt,LO}}\left( t \right)} \right\} + {i_{sh,LO}}\left( t \right)&&
\end{flalign}
where ${\tilde i_{Opt,LO}}\left( t \right) \sim CN\left( {0;\sigma _{Opt,LO}^2} \right) = CN\left( {0;\sigma _{ASE,LO}^2 + \sigma _{RIN,LO}^2} \right)$ is the optical noise injected to the photomixer generating the LO, and ${i_{sh,LO}}\left( t \right)$ is the shot noise added by the photomixer generating the LO. It is worth mentioning that due to high power of both lasers generating the LO, an optical amplifier may not be needed. In that case, $\sigma _{ASE,LO}^2 = 0$ and the optical noise is solely due to RIN of lasers, i.e., ${\tilde i_{Opt,LO}}\left( t \right) \sim CN\left( {0;\sigma _{RIN,LO}^2} \right)$. 

The LO can be written in the complex form as
\begin{flalign}\label{eq:13}
{\tilde s_{LO}} = {G_2}\exp \left( {j\left( {2\pi {f_{LO}}t + {\phi _{LO}}\left( t \right)} \right)} \right) + {\tilde n_{LO}}\left( t \right)&&
\end{flalign}
where ${\tilde n_{LO}}\left( t \right)$ is the complex representation of LO noise current and ${G_2} = R \cdot \sqrt {{P_3}{P_4}}$. Accordingly, the power of LO is given as ${P_{THz,LO}} = G_2^2 = {R^2}{P_3}{P_4}$. The noise term ${\tilde n_{LO}}\left( t \right)$ is statistically described by
\begin{flalign}\label{eq:14}
{{\tilde n}_{LO}}\left( t \right) &\sim \frac{{R\sigma _{Opt,LO}^2}}{4}\left[ {C\chi _2^2} \right] + CN\left( {0;{{{R^2}{P_4}\sigma _{Opt,LO}^2} \mathord{\left/
 {\vphantom {{{R^2}{P_4}\sigma _{Opt,LO}^2} 2}} \right.
 \kern-\nulldelimiterspace} 2}} \right)&&\\\nonumber
& + CN\left( {0;{{{R^2}{P_3}\sigma _{Opt,LO}^2} \mathord{\left/
 {\vphantom {{{R^2}{P_3}\sigma _{Opt,LO}^2} 2}} \right.
 \kern-\nulldelimiterspace} 2}} \right) + CN\left( {0;\sigma _{sh,LO}^2} \right)&&\\\nonumber
& \sim \frac{{R\sigma _{Opt,LO}^2}}{4}\left[ {C\chi _2^2} \right]&&\\\nonumber
& + CN\left( {0;\sigma _{sh,LO}^2 + {{\left( {{P_3} + {P_4}} \right){R^2}\sigma _{Opt,LO}^2} \mathord{\left/
 {\vphantom {{\left( {{P_3} + {P_4}} \right){R^2}\sigma _{Opt,LO}^2} 2}} \right.
 \kern-\nulldelimiterspace} 2}} \right)&&
\end{flalign}

The LO and received signal are mixed, i.e.,
\begin{flalign}\label{eq:15}
y\left( t \right) = {\tilde s_{LO}} \cdot {\tilde s_{Rx}} + {\tilde i_{th}}\left( t \right)&&
\end{flalign}
where ${\tilde i_{th}}\left( t \right)$ is the thermal noise added by the THz mixer and the subsequent electrical amplifiers. Similar to the shot noise added by the photomixers, thermal noise is modeled as a symmetric complex Gaussian distribution, i.e., ${\tilde i_{th}}\left( t \right) \sim CN\left( {0;\sigma _{th}^2} \right)$. Replacing \eqref{eq:10} and \eqref{eq:13} in \eqref{eq:15} and expanding the resulting expression, we obtain
\begin{flalign}\label{eq:16}
y\left( t \right) &= \alpha  \cdot \left[ {{G_1}{G_2} \cdot x\left( t \right)\exp \left( { - j\left( {2\pi \Delta ft + \Delta \phi \left( t \right)} \right)} \right)} \right.&&\\\nonumber
& + {G_2}\exp \left( {j\left( {2\pi {f_{LO}}t + {\phi _{LO}}\left( t \right)} \right)} \right) \cdot {{\tilde n}_{Tx}}\left( t \right)&&\\\nonumber
& + {G_1}x\left( t \right)\exp \left( { - j\left( {2\pi {f_c}t + {\phi _c}\left( t \right)} \right)} \right) \cdot {{\tilde n}_{LO}}\left( t \right)&&\\\nonumber
&\left. { + {{\tilde n}_{LO}}\left( t \right) \cdot {{\tilde n}_{Tx}}\left( t \right)} \right] + {{\tilde i}_{th}}\left( t \right)&&
\end{flalign}
where we define $\Delta f = \left| {{f_c} - {f_{LO}}} \right|$ and $\Delta \phi \left( t \right) = {\phi _c}\left( t \right) - {\phi _{LO}}\left( t \right)$ as the resultant frequency and phase of the mixed signal. This can be written in more compact form as
\begin{flalign}\label{eq:17}
y\left( t \right) &= \alpha {G_1}{G_2} \cdot x\left( t \right)\exp \left( { - j\left( {2\pi \Delta ft + \Delta \phi \left( t \right)} \right)} \right)&&\\\nonumber
& + \tilde n\left( t \right)&&
\end{flalign}
where $\tilde n\left( t \right)$ is the complex noise term which is defined as 
\begin{flalign}\label{eq:18}
\tilde n\left( t \right) &= \alpha \left[ {{G_2}\exp \left( {j\left( {2\pi {f_{LO}}t + {\phi _{LO}}\left( t \right)} \right)} \right) \cdot {{\tilde n}_{Tx}}\left( t \right)} \right.&&\\\nonumber
& + {G_1}x\left( t \right)\exp \left( { - j\left( {2\pi {f_c}t + {\phi _c}\left( t \right)} \right)} \right) \cdot {{\tilde n}_{LO}}\left( t \right)&&\\\nonumber
&\left. { + {{\tilde n}_{LO}}\left( t \right) \cdot {{\tilde n}_{Tx}}\left( t \right)} \right] + {{\tilde i}_{th}}\left( t \right)&&
\end{flalign}

By defining ${\mathord{\buildrel{\lower3pt\hbox{$\scriptscriptstyle\frown$}} 
\over n} _{Tx}}\left( t \right) = \exp \left( {j\left( {2\pi {f_{LO}}t + {\phi _{LO}}\left( t \right)} \right)} \right) \cdot {\tilde n_{Tx}}\left( t \right)$ and ${\mathord{\buildrel{\lower3pt\hbox{$\scriptscriptstyle\frown$}} 
\over n} _{LO}}\left( t \right) = \exp \left( { - j\left( {2\pi {f_c}t + {\phi _c}\left( t \right)} \right)} \right) \cdot {\tilde n_{LO}}\left( t \right)$, \eqref{eq:18} can be rewritten as
\begin{flalign}\label{eq:19}
\tilde n\left( t \right) &= \alpha \left[ {{G_2} \cdot {{\mathord{\buildrel{\lower3pt\hbox{$\scriptscriptstyle\frown$}} 
\over n} }_{Tx}}\left( t \right) + {G_1}x\left( t \right) \cdot {{\mathord{\buildrel{\lower3pt\hbox{$\scriptscriptstyle\frown$}} 
\over n} }_{LO}}\left( t \right)} \right.&&\\\nonumber
&\left. { + {{\tilde n}_{LO}}\left( t \right) \cdot {{\tilde n}_{Tx}}\left( t \right)} \right] + {{\tilde i}_{th}}\left( t \right)&&
\end{flalign}

Assuming that ${\mathord{\buildrel{\lower3pt\hbox{$\scriptscriptstyle\frown$}} 
\over n} _{Tx}}\left( t \right)$ and ${\mathord{\buildrel{\lower3pt\hbox{$\scriptscriptstyle\frown$}} 
\over n} _{LO}}\left( t \right)$ are statistically equivalent to ${\tilde n_{Tx}}\left( t \right)$ and ${\tilde n_{LO}}\left( t \right)$, respectively, the statistical description of $\tilde n\left( t \right)$ is given as 
\begin{flalign}\label{eq:20}
\tilde n\left( t \right) &\sim \alpha \left[ {{G_2} \cdot {{\tilde n}_{Tx}}\left( t \right) + {G_1}x\left( t \right) \cdot {{\tilde n}_{LO}}\left( t \right)} \right.&&\\\nonumber
&\left. { + {{\tilde n}_{LO}}\left( t \right) \cdot {{\tilde n}_{Tx}}\left( t \right)} \right] + CN\left( {0;\sigma _{th}^2} \right)&&
\end{flalign}

The constituents of the aggregate noise term can be defined as ${n_1}\left( t \right) = \alpha {G_2} \cdot {\tilde n_{Tx}}\left( t \right)$, ${n_2}\left( t \right) = \alpha {G_1}x\left( t \right) \cdot {\tilde n_{LO}}\left( t \right)$, ${n_3}\left( t \right) = \alpha {\tilde n_{LO}}\left( t \right) \cdot {\tilde n_{Tx}}\left( t \right)$, and ${n_4}\left( t \right) = CN\left( {0;\sigma _{th}^2} \right)$. The aggregate noise term can be equivalently written as
\begin{flalign}\label{eq:21}
\tilde n\left( t \right) \sim {n_1}\left( t \right) + {n_2}\left( t \right) + {n_3}\left( t \right) + {n_4}\left( t \right)&&
\end{flalign}

The real and imaginary parts of $y\left( t \right)$ are written as
\begin{flalign}\label{eq:22}
{y_I}\left( t \right) &= \Re \left\{ {y\left( t \right)} \right\} &&\\\nonumber
&= \alpha {G_1}{G_2} \cdot \left[ {{x_I}\left( t \right)\cos \left( {2\pi \Delta ft + \Delta \phi \left( t \right)} \right)} \right.&&\\\nonumber
&\left. { + {x_Q}\left( t \right)\sin \left( {2\pi \Delta ft + \Delta \phi \left( t \right)} \right)} \right] + {n_I}\left( t \right)&&
\end{flalign}
\begin{flalign}\label{eq:23}
{y_Q}\left( t \right) &= \Im \left\{ {y\left( t \right)} \right\} &&\\\nonumber
&= \alpha {G_1}{G_2} \cdot \left[ {{x_I}\left( t \right)\sin \left( {2\pi \Delta ft + \Delta \phi \left( t \right)} \right)} \right.&&\\\nonumber
& \left. { + {x_Q}\left( t \right)\cos \left( {2\pi \Delta ft + \Delta \phi \left( t \right)} \right)} \right] + {n_Q}\left( t \right)&&
\end{flalign}
where ${n_I}\left( t \right) = \Re \left\{ {\tilde n\left( t \right)} \right\}$ and ${n_Q}\left( t \right) = \Im \left\{ {\tilde n\left( t \right)} \right\}$. Under the assumption of ideal homodyne detection, we have ${f_{LO}} = {f_c}$ and $\Delta f = 0$. In addition, if there is no phase offset, i.e., $\Delta \phi \left( t \right) = 0$, $y\left( t \right)$ reduces to ${y_I}\left( t \right) = \alpha {G_1}{G_2} \cdot {x_I}\left( t \right) + {n_I}\left( t \right)$ and ${y_Q}\left( t \right) = \alpha {G_1}{G_2} \cdot {x_Q}\left( t \right) + {n_Q}\left( t \right)$. Here we observe that in the absence of frequency and phase offset, i.e.,$\Delta f = 0$, and $\Delta \phi \left( t \right) = 0$, ${y_I}\left( t \right)$ and ${y_Q}\left( t \right)$ are scaled and noisy version of the transmitted waveforms ${x_I}\left( t \right)$ and ${x_Q}\left( t \right)$, respectively. 

At the receiver, the signal is sampled by two independent analog-to-digital convertors (ADCs); one for ${y_I}\left( t \right)$ and the other for ${y_Q}\left( t \right)$. Fig. \ref{fig:DSPBlkDiagram}b shows the sub-blocks of receiver DSP. First step in receiver DSP is the matched filtering of the received signal. The matched filter is another root-raised cosine filter with the same roll-off factor as the pulse shaping filter. After matched filtering, frequency offset estimation is performed by analyzing the spectrum of signal raised to the 4$th$ power \cite{farukDigitalSignalProcessing2017}. The estimated frequency offset is compensated and the resulting signal is down-sampled to 1 sample per symbol and a phase locked loop (PLL) is employed for phase recovery. We assume that the channel coefficient is perfectly estimated and made available to the maximum likelihood (ML) detector which takes the form of
\begin{flalign}\label{eq:24}
\hat d = \arg \mathop {\min }\limits_i {\left\| {z - \alpha  \cdot {d_i}} \right\|^2}&&
\end{flalign}
where $z$ is the output of phase recovery block, and $\hat d$ is the estimate of the transmitted data symbol. A QAM symbol-to-bit de-mapper is then employed to convert detected modulation symbols $\hat d$ into a bit stream. 

\subsection{Effect of Hardware Impairments in Photonics-based THz Systems}\label{sec:photon-hardwareImpairments}
At the transmitter, the lasers deployed to generate the THz signal generate RIN, which is then amplified by the subsequent erbium-doped fiber amplifier (EDFA). The EDFA also adds ASE noise to the optical signal before it is converted into the THz domain via a photomixer. Additionally, the photomixers at both the transmitter and receiver produce additive shot noise currents, further degrading the signal. The RF mixer introduces thermal noise during up- and down-conversion. Since these noises are originated at the transmitter, the SNR of a THz communication systems becomes limited even before transmission, unlike the conventional lower-frequency systems. This SNR is further degraded by the noise contributions at the receiver, ultimately affecting overall system performance.

Other critical hardware impairments include frequency mismatch and phase noise. In practice, the transmitter’s carrier frequency and the receiver’s local oscillator frequency are never perfectly aligned, resulting in a non-zero $\Delta f$. Additionally, the lasers’ instantaneous frequencies exhibit slow drift around their central values, causing $\Delta f$ to vary over time. However, since this drift occurs at a much slower rate than the symbol rate, $\Delta f$ can be treated as constant over a single data block, provided the receiver’s DSP uses an appropriately designed block length. Moreover, $\Delta f$ is typically much smaller than the signal bandwidth, placing the system in the intradyne regime of coherent detection. Beyond frequency variations, the phase of each laser source undergoes a random walk process, introducing time-dependent phase rotation. Both frequency and phase impairments must be actively compensated in the receiver’s DSP prior to detection.
\subsubsection{Effect of noise}\hfill\newline
\noindent\textbf{Signal to noise ratio at transmitter output: }The SNR at the transmitter is defined by
\begin{flalign}\label{eq:25}
SN{R_{Tx}} = \frac{{{P_{THz}}}}{{{P_{noise,Tx}}}}&&
\end{flalign}
where ${P_{THz}}$ is the power of generated THz signal and ${P_{noise,Tx}} = E\left| {{n_{Tx}}{{\left( t \right)}^2}} \right|$ is the power of transmitter noise. Based on the transmitted signal model in \eqref{eq:5}, the power of generated THz signal is given by
\begin{flalign}\label{eq:26}
{P_{THz}} = G_1^2 = {R^2}{P_2}{P_1}&&
\end{flalign}

Based on \eqref{eq:9}, we can calculate the noise power present at the transmitter as
\begin{flalign}\label{eq:27}
{P_{noise,Tx}} &= E\left[ {{n_{Tx}}{{\left( t \right)}^2}} \right]&&\\\nonumber
& = E\left[ {n_{Tx1}^2} \right] + E\left[ {n_{Tx2}^2} \right] + E\left[ {n_{Tx3}^2} \right] + E\left[ {n_{Tx4}^2} \right]&&
\end{flalign}

Note that, ${n_{Tx1}}$ is a product of a scalar with a second order chi-squared distribution. Its variance is therefore given as $E\left[ {n_{Tx1}^2} \right] = {{{R^2}\sigma _{Opt}^4} \mathord{\left/
 {\vphantom {{{R^2}\sigma _{Opt}^4} 4}} \right.
 \kern-\nulldelimiterspace} 4}$ where we use the fact that the variance of a $k^{th}$ order Chi-squared distribution is given as $2k$. The second noise term is a normal distribution with variance ${{\sigma _{sh,Tx}^2 + {R^2}{P_2}\sigma _{Opt}^2} \mathord{\left/
 {\vphantom {{\sigma _{sh,Tx}^2 + {R^2}{P_2}\sigma _{Opt}^2} 2}} \right.
 \kern-\nulldelimiterspace} 2}$. The third and the fourth noise terms are products of signal I- and Q-components, respectively with a normal distribution. Since ${x_I}\left( t \right)$, ${x_Q}\left( t \right)$, and $N\left( {0;{{{R^2}{P_1}\sigma _{Opt}^2} \mathord{\left/
 {\vphantom {{{R^2}{P_1}\sigma _{Opt}^2} 2}} \right.
 \kern-\nulldelimiterspace} 2}} \right)$ are independent of each other and they have zero mean, the variance of their product can be replaced by the product of their variances. Hence, inserting the constituent values in \eqref{eq:27}, we obtain
\begin{flalign}\label{eq:28}
{P_{noise,Tx}} = \sigma _{sh,Tx}^2 + \frac{{{R^2}\sigma _{Opt}^2}}{4} \cdot \left[ {\sigma _{Opt}^2 + 2{P_2} + 2{P_1}} \right]&&
\end{flalign}

Replacing \eqref{eq:26} and \eqref{eq:28} in \eqref{eq:25}, the SNR of the transmitted signal is given by 
\begin{flalign}\label{eq:29}
SN{R_{Tx}} = \frac{{4{R^2}{P_2}{P_1}}}{{4\sigma _{sh,Tx}^2 + {R^2}\sigma _{Opt}^4 + 2{R^2}\sigma _{Opt}^2\left( {{P_2} + {P_1}} \right)}}&&
\end{flalign}

This indicates that the $SN{R_{Tx}}$ is limited due to shot noise of photomixer as well as the optical noise from the lasers and optical amplifier.\\ 
\textbf{Signal to noise ratio at the receiver:} Now we can calculate the signal to noise ratio after coherent detection of THz signal. It is given by
\begin{flalign}\label{eq:30}
SN{R_{Rx}} = \frac{{{P_{sig}}}}{{{P_{noise}}}}&&
\end{flalign}
where ${P_{sig}}$ and ${P_{noise}}$ are the signal and noise powers at the receiver, respectively. Based on \eqref{eq:17}, the received signal power is given as
\begin{flalign}\label{eq:31}
    {P_{sig}} = {\left( {\alpha {G_1}{G_2}} \right)^2}&&
\end{flalign}

Based on \eqref{eq:19}, we can calculate the noise power present at the receiver. This yields
\begin{flalign}\label{eq:32}
{P_{noise}} &= E\left[ {{{\left| {\tilde n\left( t \right)} \right|}^2}} \right]&&\\\nonumber
& = {\alpha ^2}\left[ {G_2^2 \cdot E\left[ {{{\left| {{{\tilde n}_{Tx}}\left( t \right)} \right|}^2}} \right]} \right. + G_1^2 \cdot E\left[ {{{\left| {{{\tilde n}_{LO}}\left( t \right)} \right|}^2}} \right]&&\\\nonumber
&\left. { + E\left[ {{{\left| {{{\tilde n}_{LO}}\left( t \right)} \right|}^2}} \right] \cdot E\left[ {{{\left| {{{\tilde n}_{Tx}}\left( t \right)} \right|}^2}} \right]} \right] + E\left[ {{{\left| {{{\tilde i}_{th}}\left( t \right)} \right|}^2}} \right]&&
\end{flalign}

Following the same steps that yield \eqref{eq:28}, the LO noise power $E\left[ {{{\left| {{{\tilde n}_{LO}}\left( t \right)} \right|}^2}} \right]$ is obtained as
\begin{flalign}\label{eq:33}
    E\left[ {{{\left| {{{\tilde n}_{LO}}\left( t \right)} \right|}^2}} \right] &= \sigma _{sh,LO}^2 &&\\\nonumber
    &+\frac{{{R^2}\sigma _{Opt,LO}^2}}{4}\left( {\sigma _{Opt,LO}^2 + 2{P_3} + 2{P_4}} \right)
\end{flalign}
where $\sigma _{sh,LO}^2$ is the shot noise power generated by the photomixer generating the LO. Replacing \eqref{eq:28} and \eqref{eq:33} in \eqref{eq:32}, we obtain \eqref{eq:34} (given at the bottom of the page)

\begin{figure*}[b]
\hrulefill{}
    \begin{flalign}\label{eq:34}
        {P_{noise}} &= \sigma _{th}^2 + {\alpha ^2}\left[ {G_2^2\sigma _{sh,Tx}^2 + G_2^2{\beta _A}\sigma _{Opt}^2 + 2G_2^2{P_A}{\beta _A}} \right.+ G_1^2\sigma _{sh,LO}^2 + G_1^2{\beta _B}\sigma _{Opt,LO}^2 + 2G_1^2{P_B}{\beta _B}+ \sigma _{sh,LO}^2\sigma _{sh,Tx}^2&&\\\nonumber
&  + \sigma _{Opt}^2\sigma _{sh,LO}^2{\beta _A}+ 2{P_A}\sigma _{sh,LO}^2{\beta _A} + \sigma _{Opt,LO}^2\sigma _{sh,Tx}^2{\beta _B}+ 2{P_B}\sigma _{sh,Tx}^2{\beta _B} + {\beta _A}{\beta _B}\sigma _{Opt}^2\sigma _{Opt,LO}^2&&\\\nonumber
&+ 2{\beta _A}{\beta _B}{P_A}\sigma _{Opt,LO}^2 + 2{\beta _A}{\beta _B}{P_B}\sigma _{Opt}^2\left. { + 4{\beta _A}{\beta _B}{P_A}{P_B}} \right]&&
    \end{flalign}
    \begin{multline}\label{eq:35}
SN{R_{Rx}} = \frac{{{{\left( {\alpha {G_1}{G_2}} \right)}^2}}}{\begin{array}{c}
\sigma _{th}^2 + {\alpha ^2}\left[ {G_2^2\sigma _{sh,Tx}^2 + G_2^2{\beta _A}\sigma _{Opt}^2 + 2G_2^2{P_A}{\beta _A} + G_1^2\sigma _{sh,LO}^2 + G_1^2{\beta _B}\sigma _{Opt,LO}^2 + 2G_1^2{P_B}{\beta _B}} \right.\\
 + \sigma _{sh,LO}^2\sigma _{sh,Tx}^2 + \sigma _{Opt}^2\sigma _{sh,LO}^2{\beta _A} + 2{P_A}\sigma _{sh,LO}^2{\beta _A} + \sigma _{Opt,LO}^2\sigma _{sh,Tx}^2{\beta _B} + 2{P_B}\sigma _{sh,Tx}^2{\beta _B}\\
\left. { + {\beta _A}{\beta _B}\sigma _{Opt}^2\sigma _{Opt,LO}^2 + 2{\beta _A}{\beta _B}{P_A}\sigma _{Opt,LO}^2 + 2{\beta _A}{\beta _B}{P_B}\sigma _{Opt}^2 + 4{\beta _A}{\beta _B}{P_A}{P_B}} \right]
\end{array}}    
\end{multline}    
%
\begin{flalign}\label{eq:SIN-photon}    
\sigma _{SIN}^2 &:= \sigma _{th}^2 + {\alpha ^2}\left[ {G_2^2\sigma _{sh,Tx}^2 + G_2^2{\beta _A}\sigma _{Opt}^2 + 2G_2^2{P_2}{\beta _A}} \right. + \sigma _{sh,LO}^2\sigma _{sh,Tx}^2 + \sigma _{Opt}^2\sigma _{sh,LO}^2{\beta _A} + 2{P_2}\sigma _{sh,LO}^2{\beta _A}&&\\\nonumber
 &+ \sigma _{Opt,LO}^2\sigma _{sh,Tx}^2{\beta _B} + 2{P_B}\sigma _{sh,Tx}^2{\beta _B} + {\beta _A}{\beta _B}\sigma _{Opt}^2\sigma _{Opt,LO}^2\left. { + 2{\beta _A}{\beta _B}{P_2}\sigma _{Opt,LO}^2 + 2{\beta _A}{\beta _B}{P_B}\sigma _{Opt}^2 + 4{\beta _A}{\beta _B}{P_2}{P_B}} \right]&&
\end{flalign}
\begin{flalign}\label{eq:SDN-photon}
\sigma _{SDN}^2 &:= {\alpha ^2}\left[ {2G_2^2{P_1}{\beta _A} + G_1^2\sigma _{sh,LO}^2 + G_1^2{\beta _B}\sigma _{Opt,LO}^2 + 2G_1^2{P_B}{\beta _B} + 2{P_1}\sigma _{sh,LO}^2{\beta _A} + 2{\beta _A}{\beta _B}{P_1}\sigma _{Opt,LO}^2} \right.&&\\\nonumber
 &+ \left. {4{\beta _A}{\beta _B}{P_1}{P_B}} \right]&&
\end{flalign}
    
\end{figure*}
where we define ${P_A} = {P_1} + {P_2}$, ${P_B} = {P_3} + {P_4}$, ${\beta _A} = {{{R^2}\sigma _{Opt}^2} \mathord{\left/
 {\vphantom {{{R^2}\sigma _{Opt}^2} 4}} \right.
 \kern-\nulldelimiterspace} 4}$, and ${\beta _B} = {{{R^2}\sigma _{Opt,LO}^2} \mathord{\left/
 {\vphantom {{{R^2}\sigma _{Opt,LO}^2} 4}} \right.
 \kern-\nulldelimiterspace} 4}$.

Replacing \eqref{eq:31} and \eqref{eq:34} in \eqref{eq:30}, the received SNR is given by \eqref{eq:35} which is given at the bottom of the page. From \eqref{eq:35}, we observe that in addition to the thermal noise added at the receiver, the overall noise power contains several different other contributions due to the shot noise at transmitter and receiver as well as the noise added by optical components both at transmitter and receiver. Apart from the thermal noise component, the other components are scaled by propagation gain ${\alpha ^2}$; which means that their contribution will vary with the propagation gain. To gain further insights, we categorize the constituent noise components into two categories: 1) Signal-independent noise (SIN) and 2) Signal-dependent noise (SDN). As the name suggests, SIN contributes to noise power independent of the underlying modulation format while SDN contributes to noise which is scaled with the amplitude of modulation symbols. Accordingly, the $SNR_{Rx}$ expression can be written in a simplified form as

\begin{flalign}\label{eq:snr-rx-opt-simple}
SN{R_{Rx}} = \frac{{{{\left( {\alpha {G_1}{G_2}} \right)}^2}}}{{\sigma _{SIN}^2 + \sigma _{SDN}^2}}&&    
\end{flalign}
where $\sigma _{SIN}^2$ and $\sigma _{SDN}^2$ represent the SIN and the SDN, respectively, which are defined in \eqref{eq:SIN-photon} and \eqref{eq:SDN-photon} at the bottom of the page. Further discussion on the impact of different noise components is presented in the numerical results section. 

\textbf{Characterization of noise sources:} In the following, we present the calculations of optical and electronic noise terms required for the calculation of total noise power in \eqref{eq:35}.

\noindent\textbf{Amplified spontaneous emission (ASE) noise due to optical amplifier: } The ASE is an optical noise which is added by the optical amplifier during the amplification process. For an EDFA, the power of ASE is given by \cite{qiaoASEAnalysisCorrection2007} $\sigma _{ASE}^2 = 2{n_{sp}}\left( {G - 1} \right)h\upsilon  \cdot B_{opt} $ where ${n_{sp}}$ is the spontaneous emission factor, $G$ is the gain, $h$ is the Planck’s constant, $\upsilon $ is the optical frequency, and $B_{opt}$ is the optical bandwidth. For a typical operational condition of an EDFA, we have $G \gg 1$, and ${n_{sp}} \approx 1$. Hence, $\sigma _{ASE}^2$ reduces to $\sigma _{ASE}^2 = 2G \cdot h\upsilon  \cdot B_{opt} $.

In this work, no optical bandpass filter is assumed in the photonic transmitter chain. Introducing a narrow optical filter would restrict the tunability of the photonic THz source, introduce insertion loss, increase system cost, and add system complexity. Therefore, the optical noise bandwidth $B_{opt}$ corresponds to the optical bandwidth of the EDFA used in the system. ASE and RIN noise powers are obtained by integrating their respective spectral densities over this bandwidth.

\noindent\textbf{Relative intensity noise (RIN) of a laser:} The RIN is defined as a ratio of random power fluctuations of laser to its mean value. Mathematically, it is expressed as $RIN = {{\left\langle {\delta P{{\left( t \right)}^2}} \right\rangle } \mathord{\left/
 {\vphantom {{\left\langle {\delta P{{\left( t \right)}^2}} \right\rangle } {{{\bar P}^2}}}} \right.
 \kern-\nulldelimiterspace} {{{\bar P}^2}}}$ where $\delta P\left( t \right)$ is the time varying power fluctuation and $\bar P$ is the mean laser power. The noise power due to RIN is then calculated as ${P_{RIN}} = {\bar P^2}RIN \cdot B_{opt} $ where $B_{opt} $ is the optical bandwidth. Due to the presence of an optical amplifier at the transmitter, the noise power due to RIN will be further enhanced by factor $G$. i.e., $\sigma _{RIN}^2 = G \cdot {\bar P^2}RIN \cdot B_{opt}$. 

 \noindent\textbf{Shot noise due to photomixers at the transmitter and receiver:} The shot noise is related to the optical power incident on the photomixer. Its variance can be calculated as (see Eq. 10 of \cite{brattDevelopmentPINHgCdTe1987})
 \begin{flalign}\label{eq:36}
 \sigma _{shot}^2 = E\left[ {i_{shot}^2} \right] = {P_{shot}} = 2qR{\bar P_{in}}B&&
 \end{flalign}
where $q$ is the elementary charge and $B$ is the receiver bandwidth. 
The receiver employs an ideal rectangular filter with bandwidth $B = \left( {1 + \beta } \right){R_s}$, where ${R_s}$ is the symbol rate and $\beta$ is the roll-off factor of the pulse-shaping filter. This electrical bandwidth is used consistently when evaluating shot noise and thermal noise terms.
Here, ${\bar P_{in}}$ is the average incident power and is given by
\begin{flalign}\label{eq:37}
{\bar P_{in}} = E\left[ {{{\left| {{E_{pm}}} \right|}^2}} \right]&&    
\end{flalign}

Replacing ${\left| {{E_{pm}}} \right|^2}$ given by \eqref{eq:3} in \eqref{eq:37}, we obtain
\begin{flalign}\label{eq:38}
{{\bar P}_{in}} &= \frac{1}{2}E\left[ {{P_1} \cdot {{\left| {x\left( t \right)} \right|}^2} + {P_2} + {{\left| {{{\tilde i}_{Opt}}\left( t \right)} \right|}^2}} \right.&&\\\nonumber
& + 2\left[ {\sqrt {{P_2}} \sqrt {{P_1}} {x_I}\left( t \right)\cos \left( {2\pi {f_c}t + {\phi _c}\left( t \right)} \right)} \right.&&\\\nonumber
& + \sqrt {{P_2}} \sqrt {{P_1}} {x_Q}\left( t \right)\sin \left( {2\pi {f_c}t + {\phi _c}\left( t \right)} \right)&&\\\nonumber
& + \sqrt {{P_2}}  \cdot \Re \left\{ {{{\mathord{\buildrel{\lower3pt\hbox{$\scriptscriptstyle\frown$}} 
\over i} }_{Opt}}\left( t \right)} \right\} + \sqrt {{P_1}} \left( {{x_I}\left( t \right) \cdot \Re \left\{ {{{\hat i}_{Opt}}\left( t \right)} \right\}} \right.&&\\\nonumber
&\left. {\left. {\left. { - {x_Q}\left( t \right) \cdot \Im \left\{ {{{\hat i}_{Opt}}\left( t \right)} \right\}} \right)} \right]} \right]&&
\end{flalign}

As defined earlier, $E\left[ {{{\left| {x\left( t \right)} \right|}^2}} \right] = 1$ and by definition $E\left[ {{{\left| {{{\tilde i}_{Opt}}\left( t \right)} \right|}^2}} \right] = \sigma _{Opt}^2$. Additionally, $x\left( t \right)$, ${P_2}$, ${\tilde i_{Opt}}\left( t \right)$, and the carrier are all independent from each other. Hence, we expand \eqref{eq:38} as
\begin{flalign}\label{eq:39}
{{\bar P}_{in}} &= \frac{1}{2}\left( {{P_1} + {P_2} + \sigma _{Opt}^2} \right)&&\\\nonumber
& + \sqrt {{P_2}} \sqrt {{P_1}}  \cdot E\left[ {{x_I}\left( t \right)} \right] \cdot E\left[ {\cos \left( {2\pi {f_c}t + {\phi _c}\left( t \right)} \right)} \right]&&\\\nonumber
& + \sqrt {{P_2}} \sqrt {{P_1}}  \cdot E\left[ {{x_Q}\left( t \right)} \right] \cdot E\left[ {\sin \left( {2\pi {f_c}t + {\phi _c}\left( t \right)} \right)} \right]&&\\\nonumber
& + \sqrt {{P_2}}  \cdot E\left[ {\Re \left\{ {{{\mathord{\buildrel{\lower3pt\hbox{$\scriptscriptstyle\frown$}} 
\over i} }_{Opt}}\left( t \right)} \right\}} \right]&&\\\nonumber
& + \sqrt {{P_1}} \left( {E\left[ {{x_I}\left( t \right)} \right] \cdot E\left[ {\Re \left\{ {{{\hat i}_{Opt}}\left( t \right)} \right\}} \right]} \right.&&\\\nonumber
&\left. { - E\left[ {{x_Q}\left( t \right)} \right] \cdot E\left[ {\Im \left\{ {{{\hat i}_{Opt}}\left( t \right)} \right\}} \right]} \right)&&
\end{flalign}

For symmetric QAM constellations, we have $E\left[ {{x_I}\left( t \right)} \right] = E\left[ {{x_Q}\left( t \right)} \right] = 0$. Furthermore, note that $E\left[ {\cos \left( \theta  \right)} \right] = E\left[ {\sin \left( \theta  \right)} \right] = 0$. Hence, \eqref{eq:39} reduces to 
\begin{flalign}\label{eq:40}
    {\bar P_{in}} = \frac{1}{2}\left( {{P_1} + {P_2} + \sigma _{Opt}^2} \right)&&
\end{flalign}

Replacing \eqref{eq:40} in \eqref{eq:36}, we obtain
\begin{flalign}\label{eq:41}
\sigma _{sh,Tx}^2 &= E\left[ {i_{sh,Tx}^2\left( t \right)} \right]&&\\\nonumber
 &= {P_{sh,Tx}} = qRB\left( {{P_1} + {P_2} + \sigma _{Opt}^2} \right)&&
\end{flalign}

The shot noise power generated by the LO photodiode is calculated in a similar way. As noted earlier, the optical input to the LO photodiode are unmodulated lasers with amplitudes ${P_3}$ and ${P_4}$. Following the same steps as before, the corresponding shot noise power is given as
\begin{flalign}\label{eq:42}
\sigma _{sh,LO}^2 &= E\left[ {i_{sh,LO}^2\left( t \right)} \right]&&\\\nonumber
& = {P_{sh,LO}} = qRB\left( {{P_3} + {P_4} + \sigma _{Opt,LO}^2} \right)&&
\end{flalign}

\noindent\textbf{Thermal noise of the RF mixer:} The variance of thermal noise can be calculated as \cite{friisNoiseFiguresRadio1944}, Eq. 6
\begin{flalign}\label{eq:43}
\sigma _{th}^2 = E\left[ {i_{th}^2\left( t \right)} \right] = {P_{th}} = kTBF{G_e}&&
\end{flalign}
where $k$ is the Boltzmann’s constant, $T$ is the temperature, $F$ is the aggregate noise factor of the mixer and the following amplifiers, and ${G_e}$ is the gain of electrical amplifier. Noise factor $NF$ is the linear equivalent of the noise figure   specified in logarithmic scale in the datasheets of active electronic components e.g., amplifiers and mixers. 

\subsubsection{Phase Noise}\hfill

The phase noise terms of carrier signal and LO are respectively denoted by   ${\phi _c}$ and ${\phi _{LO}}$. Consequently, the phase noise at receiver, $\Delta \phi $, is the result of phase noises from four independent laser sources; two at the transmitter and two at the receiver. The resultant phase noise at the receiver is expressed as
\begin{flalign}\label{eq:48}
\Delta \phi \left( t \right) = {\phi _1}\left( t \right) - {\phi _2}\left( t \right) - {\phi _3}\left( t \right) + {\phi _4}\left( t \right)&&
\end{flalign}

The phase noise of a laser is related to its linewidth $\Delta \nu $ and is typically modeled as a Wiener process, resulting in the Lorentzian shaped spectrum \cite{barryPerformanceCoherentOptical1990}. Mathematically speaking, we can express the laser phase noise as 
\begin{flalign}\label{eq:49}
\phi \left( t \right) = \int_0^t {\psi \left( z \right) \cdot dz}&&
\end{flalign}

Here, $\psi \left( z \right)$ is modelled as zero-mean Gaussian random process (see Eq. 107 and 108 of \cite{barryPerformanceCoherentOptical1990} and related discussion) with a variance of ${\sigma ^2}\left( \tau  \right) = 2\pi \Delta \nu \tau $ where $\tau $ is the observation interval which is typically set equal to the symbol period, i.e., one sample per symbol processing. Using \eqref{eq:49}, we can write \eqref{eq:48} as $\Delta \phi \left( t \right) = \int_0^t {\left( {{\psi _1}\left( z \right) - {\psi _2}\left( z \right) - {\psi _3}\left( z \right) + {\psi _4}\left( z \right)} \right) \cdot dz}$ where ${\psi _i}\left( z \right)$ represents the zero-mean Gaussian random process associated with the ${i^{th}}$ laser with a variance of $\sigma _i^2$. Since ${\psi _i}\left( z \right),\quad i = 1,2,3,4$ are independent Gaussian random processes, they can be combined in an equivalent Gaussian random process with variance equal to the sum of the variances of individual processes. For this purpose, let us define
\begin{flalign}\label{eq:50}
    \Psi \left( z \right) = {\psi _1}\left( z \right) - {\psi _2}\left( z \right) - {\psi _3}\left( z \right) + {\psi _4}\left( z \right)&&
\end{flalign}

This Gaussian process has zero mean and its variance is given by $\sigma _\Psi ^2 = \sigma _1^2 + \sigma _2^2 + \sigma _3^2 + \sigma _4^2$. Under the assumption that each laser has the same linewidth, $\Delta \nu $, the variance of each process is given by $\sigma _i^2\left( \tau  \right) = 2\pi \Delta \upsilon \tau ,{\rm{ }}\quad i = 1,2,3,4$. The resultant variance of $\Psi \left( z \right)$ is therefore $\sigma _\Psi ^2\left( \tau  \right) = 4 \cdot \left( {2\pi \Delta \upsilon \tau } \right) = 8\pi \Delta \upsilon \tau $. 

It should be further emphasized that the lasers used for THz generation at the transmitter and receiver are modeled as independent free-running sources, and their phase noises are therefore treated as independent Wiener processes. In some practical implementations, the optical tones used for THz generation may be derived from a shared optical source or from a phase-locked optical frequency comb, which introduces correlation between the generated optical frequencies and can reduce the relative phase noise between the carrier and the local oscillator. However, even in such configurations the transmitter and receiver units remain independent, and therefore the THz link still exhibits relative phase noise between the oscillators at the two ends of the link. Consequently, the effective phase noise observed at the receiver is governed by the relative phase evolution between the transmitter and receiver systems. The independent-laser assumption adopted here therefore represents a general and conservative modeling scenario. The proposed analytical framework can be readily extended to correlated-source architectures by introducing correlation between the corresponding phase noise processes.
In addition to phase noise, the frequency of different lasers slowly drift back and forth around a nominal value. As a result, a finite $\Delta f$ is experienced by the received signals (see \eqref{eq:22} and \eqref{eq:24}). The process of frequency drift is much slower than the typical symbol period in coherent optical communication systems. Consequently, $\Delta f$ is considered fixed over several thousands of data symbols and not explicitly modeled here.
\subsubsection{IQ imbalance}
So far, we assumed perfectly balanced I/Q branches, which represents an ideal case. In practice, however, the direct-conversion (zero-IF) architecture considered here might introduce mismatches between the I and Q channels. At the photonics transmitter, the I and Q arms of the optical modulator may exhibit differences in amplitude and phase responses, resulting in I/Q imbalance. Similarly, at the receiver, the THz mixer may introduce mismatches in the amplitude response of the I and Q branches connecting the photomixer output to the subsequent stages. In addition, residual phase offsets between these branches can deviate from the ideal $90^\circ$ phase relationship, leading to further imbalance.

The presence of I/Q imbalance introduces in-band image interference, which can be incorporated into the received signal in \eqref{eq:17} as \cite{valkamaAdvancedSignalProcessing2001,anttilaFrequencySelectiveMismatchCalibration2008}
\begin{flalign}\label{eq:iq-imbalance-photon}
    y'\left( t \right) = &\alpha G_1 G_2 \cdot \left[g_1 x\left( t \right) + g_2 x^*\left( t \right)\right] &&\\ \nonumber
    &\cdot \exp \left( -j\left(2\pi \Delta f t + \Delta \phi(t)\right) \right)
    + \tilde{n}(t) &&
\end{flalign}
where $g_1$ and $g_2$ are scalar coefficients determined by the amplitude imbalance $A$ and phase imbalance $\theta$, given by
\begin{equation}
g_1 = \frac{1 + A e^{-j\theta}}{2}, \qquad
g_2 = \frac{1 - A e^{j\theta}}{2}.
\end{equation}
In the ideal case with no imbalance ($A = 1$, $\theta = 0$), we obtain $g_1 = 1$ and $g_2 = 0$, and \eqref{eq:iq-imbalance-photon} reduces to \eqref{eq:17}, i.e., $y'(t) = y(t)$.

\section{Electronics-based THz systems}\label{sec:electronics-basedTHz}
\subsection{Signal Models for Transmission and Reception}

\begin{figure}  
    \centering
    \subfloat[]{\includegraphics[trim={0.4cm 0 0 0}, clip, scale=0.85]{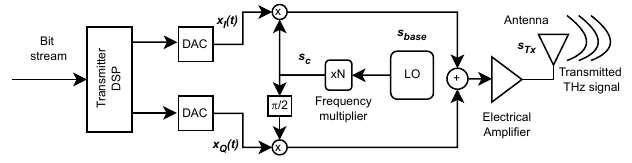}}\label{fig:elecTxBlkDiagram}\\    
    \subfloat[]{\includegraphics[trim={0.1cm 0 0 0}, clip, scale=0.85]{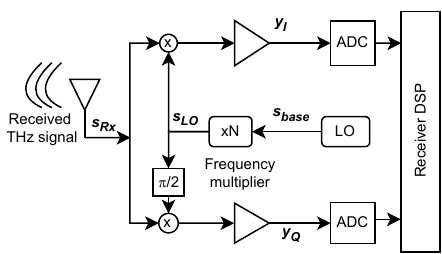}}\label{fig:elecRxBlkDiagram}
    \caption{Electronics based THz system, (a) Transmitter, (b) Receiver.}\label{fig:elecTxRx}
    \vspace{-10pt}
\end{figure}

The block diagram for electronics-based THz signal generation is shown in Fig. \ref{fig:elecTxRx}. The transmitter DSP steps remain the same as those in photonics-based approach.

\noindent\textbf{Transmitted Signal Model:} For the electronics-based THz signal generation, the baseband waveform is upconverted to the carrier frequency using a THz-class electronic IQ mixer. The carrier frequency in THz regime is generated using a lower frequency base oscillator followed by a frequency multiplier. Let ${s_{base}}\left( t \right)$ denote the oscillator signal. It is given by
\begin{flalign}\label{eq:51}
{s_{base}} = \cos \left( {2\pi {f_{base}}t + {\phi _{base}}\left( t \right)} \right) + {i_{base}}\left( t \right)&&
\end{flalign}
where ${f_{base}}$ is the frequency, and ${\phi _{base}}\left( t \right)$ is the phase noise of the base oscillator. ${i_{base}}\left( t \right)$ is the additive white noise floor of the base oscillator and is modeled as ${i_{base}}\left( t \right) \sim N(0;\sigma _{base}^2)$. The output of frequency multiplier is given as ${s_{mult}} = {\left[ {{s_{base}}} \right]^N}$ where $N$ is the multiplication factor. Employing the value of \eqref{eq:51}, the multiplier output is given as
\begin{flalign}\label{eq:52}
    {s_{mult}} = {\left[ {\cos \left( {2\pi {f_{base}}t + {\phi _{base}}\left( t \right)} \right) + {i_{base}}\left( t \right)} \right]^N}&&
\end{flalign}

Using the Binomial expansion formula \cite{weissteinTrigonometricPowerFormulas,beyerCRCStandardMathematical1987}, we can express \eqref{eq:52} as
\begin{flalign}\label{eq:53}
{s_{mult}} &= {\cos ^N}\left( {2\pi {f_{base}}t + {\phi _{base}}\left( t \right)} \right)&&\\\nonumber
& + N \cdot {\cos ^{N - 1}}\left( {2\pi {f_{base}}t + {\phi _{base}}\left( t \right)} \right) \cdot {i_{base}}\left( t \right)&&\\\nonumber
& + \frac{{N\left( {N - 1} \right)}}{{2!}}{\cos ^{N - 2}}\left( {2\pi {f_{base}}t + {\phi _{base}}\left( t \right)} \right) \cdot i_{base}^2\left( t \right)&&\\\nonumber
& +  \ldots  + i_{base}^N\left( t \right)&&
\end{flalign}

Note that the white noise current ${i_{base}}\left( t \right)$ is much smaller in magnitude compared to the oscillator signal, hence, terms with higher powers of ${i_{base}}\left( t \right)$ can be neglected. The multiplier output then becomes
\begin{flalign}\label{eq:54}
{s_{mult}} = {\cos ^N}\left( {2\pi {f_{base}}t + {\phi _{base}}\left( t \right)} \right) + N \cdot {i'_{base}}\left( t \right)&&
\end{flalign}
where ${i'_{base}}\left( t \right) = {\cos ^{N - 1}}\left( {2\pi {f_{base}}t + {\phi _{base}}\left( t \right)} \right) \cdot {i_{base}}\left( t \right) \sim N(0;\sigma _{base}^2)$. From \eqref{eq:53}, it is clear that in addition to the desired upconverted frequency term, a large number of harmonics will be generated at the frequency multiplier. These must be filtered out using a bandpass filter centered around the desired frequency. After passing through a bandpass filter centered around $N{f_{base}}$, the resulting signal is given by
\begin{flalign}\label{eq:55}
{s_c} = \cos \left( {2\pi {f_c}t + {\phi _c}\left( t \right)} \right) + {i_c}\left( t \right)&&  
\end{flalign}
where we define ${f_c} = N{f_{base}}$, ${\phi _c}\left( t \right) = N \cdot {\phi _{base}}\left( t \right)$, and ${i_c}\left( t \right) = N \cdot {i'_{base}}\left( t \right)\sim N(0;{N^2}\sigma _{base}^2)$. Similarly, the carrier signal in the Q branch can be written as ${s'_c} = \sin \left( {2\pi {f_c}t + {\phi _c}\left( t \right)} \right) + {i'_c}\left( t \right)$ where ${i'_c}\left( t \right)$ represents the phase shifted version of ${i_c}\left( t \right)$. After the multiplication of baseband signal with carrier signal, the output is given as
\begin{flalign}\label{eq:56}
{s_{Tx}} = {s_c} \cdot \sqrt {\frac{{{P_s}}}{2}}  \cdot {x_I}\left( t \right) + {s'_c} \cdot \sqrt {\frac{{{P_s}}}{2}}  \cdot {x_Q}\left( t \right) + {i_{th,Tx}}&&
\end{flalign}
where ${P_s}$ is the signal power and ${i_{th,Tx}}\sim N\left( {0;\sigma _{th,Tx}^2} \right)$ is the thermal noise added by the mixer and the following amplifier. Replacing \eqref{eq:55} in \eqref{eq:56}, we obtain
\begin{flalign}\label{eq:57}
{s_{Tx}} &= \sqrt {\frac{{{P_s}}}{2}} {x_I}\left( t \right)\cos \left( {2\pi {f_c}t + {\phi _c}\left( t \right)} \right)&&\\\nonumber
& + \sqrt {\frac{{{P_s}}}{2}} {x_Q}\left( t \right)\sin \left( {2\pi {f_c}t + {\phi _c}\left( t \right)} \right)&&\\\nonumber
& + \sqrt {\frac{{{P_s}}}{2}} {x_I}\left( t \right){i_c}\left( t \right) + \sqrt {\frac{{{P_s}}}{2}} {x_Q}\left( t \right){{i'}_c}\left( t \right) + {i_{th,Tx}}\left( t \right)&&
\end{flalign}

In the form of complex representation, \eqref{eq:57} is written as
\begin{flalign}\label{eq:58}
{{\tilde s}_{Tx}} &= \sqrt {{P_s}} x\left( t \right)\exp \left( { - j\left( {2\pi {f_c}t + {\phi _c}\left( t \right)} \right)} \right)&&\\\nonumber
& + \sqrt {{P_s}} x\left( t \right) \cdot {{\tilde i}_c}\left( t \right) + {{\tilde i}_{th,Tx}}\left( t \right)&&
\end{flalign}
where ${\tilde i_c}\left( t \right)\sim CN(0;{N^2}\sigma _{base}^2)$ and ${\tilde i_{th,Tx}}\sim CN\left( {0;\sigma _{th,Tx}^2} \right)$ are the complex representations of ${i_c}$ and ${i_{th,Tx}}$, respectively.

\noindent\textbf{Received signal model:} The received signal at the antenna can then be written as
\begin{flalign}\label{eq:59}
{{\tilde s}_{Rx}} &= \alpha \left[ {\sqrt {{P_s}} x\left( t \right)\exp \left( { - j\left( {2\pi {f_c}t + {\phi _c}\left( t \right)} \right)} \right)} \right.&&\\\nonumber
&\left. { + \sqrt {{P_s}} x\left( t \right) \cdot {{\tilde i}_c}\left( t \right) + {{\tilde i}_{th,Tx}}\left( t \right)} \right]&&
\end{flalign}

The received signal is multiplied with the LO, ideally with the same frequency as the carrier frequency of ${f_c}$. The LO signal in THz regime is generated in the same way as the carrier signal was generated at the transmitter; employing a base oscillator at lower frequency followed by a frequency multiplier block. The frequency multiplier multiplies the input carrier frequency by a factor of $N$. During the frequency multiplication process, the noise of input oscillator is also enhanced by a factor of $N$\cite{weberWBandX12Frequency2011}. The noise from an oscillator consists of two parts; the near-carrier phase noise (so called to highlight the PSD component close to the carrier frequency) and the white noise floor \cite{chenDoesNoiseFloor2017}. Both noises are enhanced by the frequency multiplier and each distorts the signal in a different way \cite{chenInfluenceWhiteNoise2018}. Here, we assume that the base oscillator at the receiver has the same center frequency as the one at the transmitter and consequently the same multiplication factor $N$ is required to generate LO matching the transmit carrier frequency. In practice, this may not be always the case; the base oscillators at transmitter and receiver may have different center frequency requiring a different multiplication factor to match the carrier and LO. Accordingly, the phase noise and white noise in the multiplier block would be differently scaled. 

After the frequency multiplier block at the receiver, the LO output can be written in complex form as
\begin{flalign}\label{eq:60}
{{\tilde s}_{LO}} &= \sqrt {{P_{LO}}} \exp \left( {j\left( {2\pi \left( {N{f_{base,Rx}}} \right)t} \right.} \right.&&\\\nonumber
&\left. {\left. { + N \cdot {\phi _{base,Rx}}\left( t \right)} \right)} \right) + N \cdot {{\tilde i}_{base,Rx}}\left( t \right)&&
\end{flalign}
where ${P_{LO}}$ is the power of local oscillator, ${f_{base,Rx}}$ is the frequency of the base oscillator at the receiver, ${\phi _{base,Rx}}\left( t \right)$ represents the phase noise and ${\tilde i_{base,Rx}}\left( t \right)\sim CN\left( {0;\sigma _{base,Rx}^2} \right)$ is the white noise floor of the base oscillator. Replacing ${f_{LO}} = N \cdot {f_{base,Rx}}$, ${\phi _{LO}}\left( t \right) = N \cdot {\phi _{base,Rx}}\left( t \right)$ and $N \cdot {\tilde i_{base,Rx}}\left( t \right) = {\tilde i_{LO}}\left( t \right)\sim CN\left( {0;{N^2}\sigma _{base,Rx}^2} \right)$ in \eqref{eq:60}, we have
\begin{flalign}\label{eq:61}
{\tilde s_{LO}} = \sqrt {{P_{LO}}} \exp \left( {j\left( {2\pi {f_{LO}}t + {\phi _{LO}}\left( t \right)} \right)} \right) + {\tilde i_{LO}}\left( t \right)&&
\end{flalign}

The THz mixer mixes both the signal and LO. The resulting signal is given by
\begin{flalign}\label{eq:62}
y\left( t \right) = {\tilde s_{LO}} \cdot {\tilde s_{Rx}} + {\tilde i_{th,Rx}}\left( t \right)&&
\end{flalign}
where ${\tilde i_{th,Rx}}\left( t \right)\sim CN\left( {0;\sigma _{th,Rx}^2} \right)$ is the complex thermal noise of the THz mixer and following amplifiers. Replacing \eqref{eq:59}, \eqref{eq:60} and \eqref{eq:61} in \eqref{eq:62} and expanding the resulting expression, we obtain
\begin{flalign}\label{eq:63}
y\left( t \right) &= \alpha \sqrt {{P_s}{P_{LO}}} x\left( t \right) \cdot \exp \left( { - j\left( {2\pi \Delta {f_{elec}}t + \Delta {\phi _{elec}}\left( t \right)} \right)} \right)&&\\\nonumber
 &+ \alpha \sqrt {{P_s}{P_{LO}}} x\left( t \right) \cdot {{\mathord{\buildrel{\lower3pt\hbox{$\scriptscriptstyle\frown$}} 
\over i} }_c}\left( t \right) + \alpha \sqrt {{P_{LO}}} {{\mathord{\buildrel{\lower3pt\hbox{$\scriptscriptstyle\frown$}} 
\over i} }_{th,Tx}}\left( t \right)&&\\\nonumber
 &+ \alpha \sqrt {{P_s}} x\left( t \right) \cdot {{\mathord{\buildrel{\lower3pt\hbox{$\scriptscriptstyle\frown$}} 
\over i} }_{LO}}\left( t \right) + \alpha \sqrt {{P_s}} x\left( t \right) \cdot {{\tilde i}_{LO}}\left( t \right) \cdot {{\tilde i}_c}\left( t \right)&&\\\nonumber
 &+ \alpha  \cdot {{\tilde i}_{LO}}\left( t \right) \cdot {{\tilde i}_{th,Tx}}\left( t \right) + {{\tilde i}_{th,Rx}}\left( t \right)&&
\end{flalign}
where we define 
\newline ${\mathord{\buildrel{\lower3pt\hbox{$\scriptscriptstyle\frown$}} 
\over i} _c}\left( t \right) = {\tilde i_c}\left( t \right)\exp \left( {j\left( {2\pi {f_{LO}}t + {\phi _{LO}}\left( t \right)} \right)} \right)$, \hfill 
\newline ${\mathord{\buildrel{\lower3pt\hbox{$\scriptscriptstyle\frown$}} 
\over i} _{th,Tx}}\left( t \right) = {\tilde i_{th,Tx}}\left( t \right)\exp \left( {j\left( {2\pi {f_{LO}}t + {\phi _{LO}}\left( t \right)} \right)} \right)$, \hfill 
\newline ${\mathord{\buildrel{\lower3pt\hbox{$\scriptscriptstyle\frown$}} 
\over i} _{LO}}\left( t \right) = {\tilde i_{LO}}\left( t \right)\exp \left( { - j\left( {2\pi {f_c}t + {\phi _c}\left( t \right)} \right)} \right)$, \hfill 
\newline $\Delta {f_{elec}} = \left| {{f_c} - {f_{LO}}} \right|$ and $\Delta {\phi _{elec}}\left( t \right) = {\phi _c}\left( t \right) - {\phi _{LO}}\left( t \right)$.

Note that the noise currents ${\tilde i_c}\left( t \right)$, ${\tilde i_{th,Tx}}\left( t \right)$ and ${\tilde i_{LO}}\left( t \right)$ are modeled as symmetric complex Gaussian distribution and their multiplication with an exponential does not change the statistics of noise. Therefore, ${\mathord{\buildrel{\lower3pt\hbox{$\scriptscriptstyle\frown$}} 
\over i} _c}\left( t \right)$, ${\mathord{\buildrel{\lower3pt\hbox{$\scriptscriptstyle\frown$}} 
\over i} _{th,Tx}}\left( t \right)$ and ${\mathord{\buildrel{\lower3pt\hbox{$\scriptscriptstyle\frown$}} 
\over i} _{LO}}\left( t \right)$ have the same statistical properties as ${\tilde i_c}\left( t \right)$, ${\tilde i_{th,Tx}}\left( t \right)$ and ${\tilde i_{LO}}\left( t \right)$, respectively., i.e. ${\mathord{\buildrel{\lower3pt\hbox{$\scriptscriptstyle\frown$}} 
\over i} _c}\left( t \right)\sim CN(0;{N^2}\sigma _{base}^2) = CN(0;\sigma _c^2)$, ${\mathord{\buildrel{\lower3pt\hbox{$\scriptscriptstyle\frown$}} 
\over i} _{th,Tx}}\left( t \right)\sim CN\left( {0;\sigma _{th,Tx}^2} \right)$ and ${\mathord{\buildrel{\lower3pt\hbox{$\scriptscriptstyle\frown$}} 
\over i} _{LO}}\left( t \right) \sim CN\left( {0;{N^2}\sigma _{base,Rx}^2} \right) = CN\left( {0;\sigma _{LO}^2} \right)$. Apart from the first term in \eqref{eq:63}, rest of the terms represent noise. The downconverted signal from \eqref{eq:63} can then be written as
\begin{flalign}\label{eq:64}
y\left( t \right) &= \alpha  \cdot \sqrt {{P_s}{P_{LO}}} x\left( t \right) \cdot \exp \left( { - j\left( {2\pi \Delta {f_{elec}}t} \right.} \right.&&\\\nonumber
&\left. {\left. { + \Delta {\phi _{elec}}\left( t \right)} \right)} \right) + {{\tilde n}_{elec}}\left( t \right)&&
\end{flalign}
where ${\tilde n_{elec}}$ is given by
\begin{flalign}\label{eq:65}
{{\tilde n}_{elec}}\left( t \right) &= \alpha \sqrt {{P_{LO}}} {{\mathord{\buildrel{\lower3pt\hbox{$\scriptscriptstyle\frown$}} 
\over i} }_{th,Tx}}\left( t \right) + {{\tilde i}_{th,Rx}}\left( t \right)&&\\\nonumber
& + \alpha {{\tilde i}_{LO}}\left( t \right) \cdot {{\tilde i}_{th,Tx}}\left( t \right)&&\\\nonumber
& + \alpha \sqrt {{P_s}} x\left( t \right)\left( {\sqrt {{P_{LO}}} {{\mathord{\buildrel{\lower3pt\hbox{$\scriptscriptstyle\frown$}} 
\over i} }_c}\left( t \right) + {{\mathord{\buildrel{\lower3pt\hbox{$\scriptscriptstyle\frown$}} 
\over i} }_{LO}}\left( t \right)} \right)&&\\\nonumber
& + \alpha \sqrt {{P_s}} x\left( t \right) \cdot {{\tilde i}_{LO}}\left( t \right) \cdot {{\tilde i}_c}\left( t \right)&&
\end{flalign}
Its statistical description is given by
\begin{flalign}\label{eq:66}
{\tilde n_{elec}}\left( t \right) \sim {n_{e1}}\left( t \right) + {n_{e2}}\left( t \right) + {n_{e3}}\left( t \right) + {n_{e4}}\left( t \right)&&
\end{flalign}
where we define
\newline
${n_{e1}}\left( t \right) \sim CN\left( {0;{\alpha ^2}{P_{LO}}\sigma _{th,Tx}^2 + \sigma _{th,Rx}^2} \right)$, \hfill \newline
${n_{e2}}\left( t \right) \sim CN\left( {0;\sigma _{LO}^2} \right) \cdot CN\left( {0;{\alpha ^2}\sigma _{th,Tx}^2} \right)$, \hfill \newline
${n_{e3}}\left( t \right) \sim \alpha \sqrt {{P_s}} x\left( t \right) \cdot CN\left( {0;{P_{LO}}\sigma _c^2 + \sigma _{LO}^2} \right)$, \hfill \newline
${n_{e4}}\left( t \right) \sim \alpha \sqrt {{P_s}} x\left( t \right) \cdot CN\left( {0;\sigma _c^2} \right)CN\left( {0;\sigma _{LO}^2} \right)$. 
\newline
Here, ${n_{e1}}\left( t \right)$ is the complex Gaussian noise term which represents the thermal noise added by the transmitter and receiver mixers. Other noise terms, ${n_{e2}}\left( t \right)$, ${n_{e3}}\left( t \right)$, and ${n_{e4}}\left( t \right)$ exist because of the presence of noise floor of the carrier and LO at the transmitter and receiver, respectively. Furthermore, the noise terms ${n_{e1}}\left( t \right)$ and ${n_{e2}}\left( t \right)$ are independent of the symbol amplitude and in contrast ${n_{e3}}\left( t \right)$ and ${n_{e4}}\left( t \right)$ are scaled with the symbol amplitude. In a classical wireless system operating at lower frequencies, the noise originating from the white noise floor of transmitter carrier and the LO is assumed to be negligible; i.e., $\sigma _c^2 = \sigma _{LO}^2 = 0$. In that case, only the first term in \eqref{eq:66} remains and all others vanish. However, due to frequency multiplication by a large factor $N$, the noise floors, i.e., ${\mathord{\buildrel{\lower3pt\hbox{$\scriptscriptstyle\frown$}} 
\over i} _c}\left( t \right)$ and ${\mathord{\buildrel{\lower3pt\hbox{$\scriptscriptstyle\frown$}} 
\over i} _{LO}}\left( t \right)$, are enhanced by factor $N$ and are not negligible in our case.

The real and imaginary parts of mixer output are given by
\begin{flalign}\label{eq:67}
{y_I}\left( t \right) &= \alpha \sqrt {\frac{{{P_s}{P_{LO}}}}{2}} {x_I}\left( t \right)\cos \left( {2\pi \Delta {f_{elec}}t + \Delta {\phi _{elec}}\left( t \right)} \right)&&\\\nonumber
& + \alpha \sqrt {\frac{{{P_s}{P_{LO}}}}{2}} {x_Q}\left( t \right)\sin \left( {2\pi \Delta {f_{elec}}t + \Delta {\phi _{elec}}\left( t \right)} \right)&&\\\nonumber
& + {n_{I,elec}}\left( t \right)&&
\end{flalign}
\begin{flalign}\label{eq:68}
{y_Q}\left( t \right) &= \alpha \sqrt {\frac{{{P_s}{P_{LO}}}}{2}} {x_Q}\left( t \right)\cos \left( {2\pi \Delta {f_{elec}}t + \Delta {\phi _{elec}}\left( t \right)} \right)&&\\\nonumber
& + \alpha \sqrt {\frac{{{P_s}{P_{LO}}}}{2}} {x_I}\left( t \right)\sin \left( {2\pi \Delta {f_{elec}}t + \Delta {\phi _{elec}}\left( t \right)} \right)&&\\\nonumber
& + {n_{Q,elec}}\left( t \right)&&
\end{flalign}
where ${n_{I,elec}}\left( t \right) = \Re \left\{ {{{\tilde n}_{elec}}\left( t \right)} \right\}$ and ${n_{Q,elec}}\left( t \right) = \Im \left\{ {{{\tilde n}_{elec}}\left( t \right)} \right\}$. Under the assumption of ideal homodyne detection, ${f_c} = {f_{LO}}$ and $\Delta {f_{elec}} = 0$. If in addition, $\Delta {\phi _{elec}}\left( t \right) = 0$, \eqref{eq:67} and \eqref{eq:68} reduce to ${y_I}\left( t \right) = \alpha \sqrt {{{{P_s}{P_{LO}}} \mathord{\left/
 {\vphantom {{{P_s}{P_{LO}}} 2}} \right.
 \kern-\nulldelimiterspace} 2}} {x_I}\left( t \right) + {n_{I,elec}}\left( t \right)$ and ${y_Q}\left( t \right) = \alpha \sqrt {{P_s}{P_{LO}}} {x_Q}\left( t \right) + {n_{Q,elec}}\left( t \right)$, respectively. Here we observe that in the absence of frequency and phase offset, i.e., $\Delta {f_{elec}} = 0$, and $\Delta {\phi _{elec}}\left( t \right) = 0$, ${y_I}\left( t \right)$ and ${y_Q}\left( t \right)$ are the scaled and noisy version of the transmitted waveforms ${x_I}\left( t \right)$ and ${x_Q}\left( t \right)$, respectively. 

 The I and Q components are then sampled by two ADCs. Similar DSP steps, as explained in Section \ref{sec:photonics-basedTHz}, are performed to retrieve the original bit stream.

\subsection{Effect of Hardware Impairments in Electronics-based THz Systems}
In electronic generation and detection of THz signals, various noise sources inherent to the hardware components play a critical role in limiting performance. Chief among these are the thermal noise contributions from active components such as mixers and electronic amplifiers, as well as the noise floors of the local oscillators used at both the transmitter and receiver ends. While in conventional wireless systems at lower frequencies, the oscillator noise floor is typically negligible and can often be ignored in system design. However, this assumption does not hold for THz systems. The use of frequency multipliers amplifies oscillator phase noise significantly. As a result, the oscillator noise becomes a dominant impairment, introducing additional distortion that cannot be treated as simple additive white noise. One distinguishing characteristic of oscillator-induced noise in THz systems is that it scales with the signal amplitude. This means the resulting noise power increases proportionally with the signal power, thereby capping the potential improvement in SNR that can be achieved by merely increasing transmit power. This behavior stands in contrast to additive thermal noise, which remains constant irrespective of signal strength.

Furthermore, similar to photonics-based THz transmission systems, elecronic-based THz systems are subject to frequency shift and phase noise. Electronic oscillators exhibit frequency instability over time. This manifests as a slow drift of the oscillator frequency around a nominal central frequency, introducing a residual frequency offset in the received signal. In addition to this frequency drift, random phase fluctuations also affect the signal. While both electronic and photonic systems experience phase noise, the underlying statistical models describing phase evolution in electronic systems differ due to the nature of electronic oscillators and the specific architecture of frequency synthesis chains.
\subsubsection{Effect of Noise}
\textbf{Signal to noise ratio at the transmitter:} The SNR at the transmitter is defined by
\begin{flalign}\label{eq:69}
SN{R_{elec,Tx}} = \frac{{{P_s}}}{{{P_{n\_elec,Tx}}}}&&    
\end{flalign}
where ${P_s}$ is the signal power and ${P_{n\_elec,Tx}}$ denotes the noise power and is given as 
\begin{flalign}\label{eq:70}
{P_{n\_elec,Tx}} &= E\left[ {{{\left| {\sqrt {{P_s}} x\left( t \right) \cdot {{\tilde i}_c}\left( t \right)} \right|}^2}} \right] + E\left[ {{{\left| {{{\tilde i}_{th,Tx}}\left( t \right)} \right|}^2}} \right]&&\\\nonumber
& = {P_s} \cdot E\left[ {{{\left| {{{\tilde i}_c}\left( t \right)} \right|}^2}} \right] + \sigma _{th,Tx}^2&&\\\nonumber
& = {P_s} \cdot {N^2}\sigma _{base}^2 + \sigma _{th,Tx}^2&&
\end{flalign}
Replacing \eqref{eq:70} in \eqref{eq:69}, we obtain
\begin{flalign}\label{eq:71}
    SN{R_{elec,Tx}} = \frac{{{P_s}}}{{{P_s}{N^2}\sigma _{base}^2 + \sigma _{th,Tx}^2}}&&
\end{flalign}

It can be seen from \eqref{eq:71} that, in addition to the standard additive thermal noise, there exists an additional noise term that is proportional to the baseband signal power itself. As a result, the noise power becomes signal-dependent, i.e., symbols with higher amplitude experience greater noise distortion, and lower-amplitude symbols experience less \cite{chenInfluenceWhiteNoise2018}. Such an effect is not observed at lower carrier frequencies, where oscillator noise contributions are negligible. However, in the THz regime, due to the use of frequency multipliers with noise scaling factor of ${N^2}$, this previously negligible effect becomes significant and must be explicitly considered in system modeling and performance analysis. 

\textbf{Signal to noise ratio at the receiver:} The SNR at the receiver is defined by
\begin{flalign}\label{eq:72}
SN{R_{elec,Rx}} = \frac{{{P_{sig,elec}}}}{{{P_{n\_elec}}}}&&
\end{flalign}
\begin{figure*}[b]
\hrulefill{}
    \begin{flalign}\label{eq:73}
{P_{n\_elec}} &= E\left[ {{{\left| {{{\tilde n}_{elec}}\left( t \right)} \right|}^2}} \right]&&\\\nonumber
& = {\alpha ^2}{P_{LO}}E\left[ {{{\left| {{{\mathord{\buildrel{\lower3pt\hbox{$\scriptscriptstyle\frown$}} 
\over i} }_{th,Tx}}\left( t \right)} \right|}^2}} \right] + E\left[ {{{\left| {{{\tilde i}_{th,Rx}}\left( t \right)} \right|}^2}} \right] + {\alpha ^2}E\left[ {{{\left| {{{\tilde i}_{LO}}\left( t \right)} \right|}^2}} \right]E\left[ {{{\left| {{{\tilde i}_{th,Tx}}\left( t \right)} \right|}^2}} \right] &&\\\nonumber
&+ {\alpha ^2}{P_s}E\left[ {{{\left| {x\left( t \right)} \right|}^2}} \right]\left( {{P_{LO}}E\left[ {{{\left| {{{\mathord{\buildrel{\lower3pt\hbox{$\scriptscriptstyle\frown$}} 
\over i} }_c}\left( t \right)} \right|}^2}} \right]} \right.\left. { + E\left[ {{{\left| {{{\mathord{\buildrel{\lower3pt\hbox{$\scriptscriptstyle\frown$}} 
\over i} }_{LO}}\left( t \right)} \right|}^2}} \right]} \right)+ {\alpha ^2}{P_s}E\left[ {{{\left| {x\left( t \right)} \right|}^2}} \right]E\left[ {{{\left| {{{\tilde i}_{LO}}\left( t \right)} \right|}^2}} \right]E\left[ {{{\left| {{{\tilde i}_c}\left( t \right)} \right|}^2}} \right]&&
\end{flalign}
\newline
\begin{flalign}\label{eq:74}
{P_{n\_elec}} &= {\alpha ^2}{P_{LO}}\sigma _{th,Tx}^2 + \sigma _{th,Rx}^2 + {\alpha ^2}\sigma _{LO}^2\sigma _{th,Tx}^2+ {\alpha ^2}{P_s}\left( {{P_{LO}}\sigma _c^2 + \sigma _{LO}^2} \right) + {\alpha ^2}{P_s}\sigma _{LO}^2\sigma _c^2&&\\\nonumber
& = \sigma _{th,Rx}^2 + {\alpha ^2}\left[ {{P_{LO}}\sigma _{th,Tx}^2 + \sigma _{LO}^2\sigma _{th,Tx}^2} \right.\left. { + {P_s}\left( {{P_{LO}}\sigma _c^2 + \sigma _{LO}^2 + \sigma _{LO}^2\sigma _c^2} \right)} \right]&&
\end{flalign}
\newline
\begin{flalign}\label{eq:75}
SN{R_{elec,Rx}} = \frac{{{\alpha ^2}{P_s}{P_{LO}}}}{
\sigma _{th,Rx}^2 + {\alpha ^2}\left[ {{P_{LO}}\sigma _{th,Tx}^2 + \sigma _{LO}^2\sigma _{th,Tx}^2} \right.\left. { + {P_s}\left( {{P_{LO}}\sigma _c^2 + \sigma _{LO}^2 + \sigma _{LO}^2\sigma _c^2} \right)} \right]}&&
\end{flalign}
\end{figure*}
where ${P_{sig,elec}}$ is the received signal power and ${P_{n\_elec}}$ is the noise power. The power of noise is calculated from \eqref{eq:65} as shown in \eqref{eq:73} at the bottom. 

Note that $E\left[ {{{\left| {x\left( t \right)} \right|}^2}} \right] = 1$, $E\left[ {{{\left| {{{\mathord{\buildrel{\lower3pt\hbox{$\scriptscriptstyle\frown$}} 
\over i} }_{th,Tx}}\left( t \right)} \right|}^2}} \right] = \sigma _{th,Tx}^2$,
$E\left[ {{{\left| {{{\tilde i}_{th,Rx}}\left( t \right)} \right|}^2}} \right] = \sigma _{th,Rx}^2$, $E\left[ {{{\left| {{{\tilde i}_{LO}}\left( t \right)} \right|}^2}} \right] = \sigma _{LO}^2$, and $E\left[ {{{\left| {{{\mathord{\buildrel{\lower3pt\hbox{$\scriptscriptstyle\frown$}} 
\over i} }_c}\left( t \right)} \right|}^2}} \right] = \sigma _c^2$. Replacing these in \eqref{eq:73}, we obtain \eqref{eq:74}, given at the bottom of the page.

From \eqref{eq:64}, we observe that the effective signal power after downconversion is given as ${P_{sig,elec}} = {\alpha ^2}{P_s}{P_{LO}}$. Replacing this and \eqref{eq:74} in \eqref{eq:72}, we calculate the received SNR as \eqref{eq:75} given at the bottom of the page. 

It is observed from \eqref{eq:75} that the received signal SNR is not only impacted by the receiver thermal noise power, but also other noise components due to the noise floors of carrier and LO oscillators. It is observed that there are six terms in the denominator of \eqref{eq:75} which can also be characterized in to SIN and SDN components. The simplified SNR expression can be accordingly written as
\begin{flalign}
    SN{R_{elec,Rx}} = \frac{{{\alpha ^2}{P_s}{P_{LO}}}}{{\sigma _{elec,SIN}^2 + \sigma _{elec,SDN}^2}}&&
\end{flalign} where  $\sigma _{elec,SDN}^2: = {\alpha ^2}{P_s}\left( {{P_{LO}}\sigma _c^2 + \sigma _{LO}^2 + \sigma _{LO}^2\sigma _c^2} \right)$ and $\sigma _{elec,SIN}^2: = \sigma _{th,Rx}^2 + {\alpha ^2}\left( {{P_{LO}}\sigma _{th,Tx}^2 + \sigma _{LO}^2\sigma _{th,Tx}^2} \right)$. 
\newline
\textbf{Characterization of noise sources:} In the following, we present the calculations of noise power in \eqref{eq:71} and \eqref{eq:75}.
\newline
\textbf{Thermal noise of the RF mixer:  }Thermal noise is introduced by active components (such as amplifiers and mixers) at both the transmitter and receiver. Unlike photonics-based systems—where the transmit signal is generated via photomixing (avoiding thermal noise at the transmitter)—the RF architecture adds thermal noise at both ends of the communication link. The corresponding thermal noise power can be calculated using \eqref{eq:43} and is not repeated here to avoid redundancy.

\textbf{Noise floor of RF oscillators:} The noise floor of RF oscillators arises primarily from thermal noise and shot noise generated within the oscillator circuitry. Its power level depends on both the oscillator’s architecture and the underlying semiconductor technology used. The phase noise characteristics of an RF oscillator are typically provided in the datasheet as a function of frequency offset from the carrier. To quantify the additive white noise floor, the value at a large offset frequency (e.g., several MHz) is commonly used, as this region reflects the flat, white noise portion of the spectrum. 

\subsubsection{Phase Noise}\hfill

The phase noise of RF oscillator consists of near carrier colored noise and the white noise floor \cite{khanzadiCalculationPerformanceCommunication2014}. The near carrier phase noise is highly correlated, and for a free-running RF oscillator, its PSD strongly decays with frequency. The phase noise spectra of RF oscillators initially decays proportional to ${1 \mathord{\left/
 {\vphantom {1 {{f^3}}}} \right.
 \kern-\nulldelimiterspace} {{f^3}}}$ followed by a decay proportional to ${1 \mathord{\left/
 {\vphantom {1 {{f^2}}}} \right.
 \kern-\nulldelimiterspace} {{f^2}}}$ eventually reaching a white noise floor at high frequency offset from the carrier \cite{khanzadiCalculationPerformanceCommunication2014}. In our derivation of the effective additive noise in \eqref{eq:65}, we have considered the white noise of the carrier and LO which is the additive noise component of RF oscillator noise. In addition, there is a multiplicative component of white noise floor which originates from the noise floor of oscillator phase noise \cite{chenInfluenceWhiteNoise2018}. 

 The near carrier phase noise and the white noise floor are modeled as a combination of three independent processes following the methodology of  \cite{khanzadiCalculationPerformanceCommunication2014}. Two independent filtered white noise processes are employed to generate phase noise components ${\phi _2}\left( t \right)$ and ${\phi _3}\left( t \right)$ corresponding to ${1 \mathord{\left/
 {\vphantom {1 {{f^2}}}} \right.
 \kern-\nulldelimiterspace} {{f^2}}}$ and ${1 \mathord{\left/
 {\vphantom {1 {{f^3}}}} \right.
 \kern-\nulldelimiterspace} {{f^3}}}$ parts of the phase noise, respectively. The coefficients of the finite impulse response (FIR) filters used for the generation of phase noise are given by \cite{kasdinDiscreteSimulationColored1995}
\begin{multline}\label{eq:79}
    h\left[ k \right] = \left\{ \begin{array}{l}
1, \qquad \qquad k = 0\\
\\
\left( {\frac{\alpha }{2} + k - 1} \right) \cdot \frac{{h\left[ {k - 1} \right]}}{k},{\rm{         }}k = 1,2...N - 1
\end{array} \right.
\end{multline}

\noindent where $\alpha$ is the exponent for phase noise decaying with respect to ${1 \mathord{\left/
 {\vphantom {1 {{f^\alpha }}}} \right.
 \kern-\nulldelimiterspace} {{f^\alpha }}}$ and $N$ denotes the number of FIR filter coefficients. For the white noise floor component (denoted by ${\phi _0}(t)$), an unfiltered white noise process is employed. The total phase noise is given by$
\Delta {\phi _{elec}}(t) = {\phi _0}(t) + {\phi _2}(t) + {\phi _3}(t).$

The variances of the white noise processes, denoted by $\sigma _{w0}^2$, $\sigma _{w2}^2$, and $\sigma _{w3}^2$ for the phase noise terms ${\phi _0}(t)$, ${\phi _2}(t)$, and ${\phi _3}(t)$, respectively, are given by \cite{khanzadiCalculationPerformanceCommunication2014} $
\sigma _{w0}^2 = {{{K_0}} \mathord{\left/
 {\vphantom {{{K_0}} \tau }} \right.
 \kern-\nulldelimiterspace} \tau }, \qquad
\sigma _{w2}^2 = 4{K_2}\tau {\pi ^2}, \qquad
\sigma _{w3}^2 = 8{K_3}{\tau ^2}{\pi ^3},
$
where $\tau$ denotes the observation interval. The constants $K_0$, $K_2$, and $K_3$ control the magnitude of the corresponding phase-noise components ${\phi _0}(t)$, ${\phi _2}(t)$, and ${\phi _3}(t)$, respectively. These parameters determine the power spectral density of the oscillator phase noise, which can be expressed using the commonly adopted multi-slope model
\begin{flalign}
S_{\phi}(f) &= K_0 + \frac{K_2}{f^2} + \frac{K_3}{f^3} &&   \label{eq:phase-noise-psd} 
\end{flalign}

The single-sideband phase noise typically measured for RF oscillators, denoted by $L(f)$ and expressed in dBc/Hz, is related to the phase-noise power spectral density by
\begin{flalign}
L(f) &= 10\log_{10}\left(\frac{S_{\phi}(f)}{2}\right) &&    
\end{flalign}
Therefore, the parameters $K_0$, $K_2$, and $K_3$ can be obtained from measurements of the RF oscillator phase-noise spectrum under investigation. In this work, the chosen values are guided by the measurement-based discussion in \cite{khanzadiCalculationPerformanceCommunication2014}, where the single-sideband phase-noise spectra of a GaN HEMT MMIC oscillator are reported under different bias conditions. The measured spectra exhibit the characteristic multi-slope behavior typically observed in RF oscillators, including near-carrier regions with approximately $30$~dB/decade and $20$~dB/decade slopes followed by a white phase-noise floor. These measurements motivate the use of the multi-slope phase-noise model in \eqref{eq:phase-noise-psd}, where the parameters $K_3$, $K_2$, and $K_0$ control the magnitude of the corresponding $1/f^3$, $1/f^2$, and white phase-noise components, respectively.
 
 In addition to the phase noise, the frequency of the RF oscillators also drift around a nominal value. Nonetheless, the frequency drift of RF oscillators is much slower than the symbol rate considered here. Consequently, it is assumed fixed over thousands of data symbols and not explicitly modeled here.
 
\subsubsection{IQ imbalance}
Similar to the photonics-based THz case, I/Q amplitude and phase imbalances are also inherent in electronics-based THz systems. In particular, mismatches in the amplitude response of the branches connecting the multiplier output to the I- and Q-channel mixers, as well as deviations from the ideal $90^\circ$ phase difference between these branches, introduce I/Q distortion.

Following the same formulation as in the photonics case, \eqref{eq:64} can be modified to incorporate I/Q imbalance as
\begin{flalign}\label{eq:iq-imbalance-electron}
y'\left( t \right) = &\alpha \cdot \sqrt{P_s P_{LO}} \left[g_1 x\left( t \right) + g_2 x^*\left( t \right)\right] &&\\ \nonumber
&\cdot \exp \left( -j\left(2\pi \Delta f_{\mathrm{elec}} t + \Delta \phi_{\mathrm{elec}}(t)\right) \right)
+ \tilde{n}_{\mathrm{elec}}(t) &&
\end{flalign}
In Section~\ref{sec:numericalResults}, we evaluate the impact of I/Q imbalance on the performance of electronics-based THz systems across different QAM formats.

\begin{table}[htb]
\centering
\caption{Simulation Parameters and System Settings}
\label{tab:photon-analytical-params}
\begin{tabularx}{0.5\textwidth}{X X X}
\toprule
\textbf{Parameter} & \textbf{Value / Range} & \textbf{Notes} \\
\midrule
Optical Frequency ($\upsilon $) & 193.4 THz & Common to both transmitter and receiver \\
Optical Bandwidth ($B_{opt} $) & 2 THz & Applicable to transmitter \\
Relative Intensity Noise (RIN) & –140, –145, –150, –200 dB/Hz & Applies to both transmitter and receiver \\
Electrical Bandwidth ($B$) & 40 GHz & Applies to receiver \\
Laser Power Input to EDFA (${{{P_2}} \mathord{\left/
 {\vphantom {{{P_2}} G}} \right.
 \kern-\nulldelimiterspace} G}$) & –13 dBm to 16 dBm & Transmitter \\
Signal Power Input to EDFA (${{{P_1}} \mathord{\left/
 {\vphantom {{{P_1}} G}} \right.
 \kern-\nulldelimiterspace} G}$) & –13 dBm to 16 dBm & Transmitter \\
 EDFA Gain ($G$) &  Case I: Fixed at 20 dB; Case II: Variable (output fixed at 23 dBm)  & Transmitter; case-dependent setup \\
Optical Output of EDFA (${P_1} + {P_2}$) & 23 dBm & Transmitter; fixed in Case II \\
Photomixer Responsivity ($R$) & 0.7 A/W & Transmitter and receiver \\
Laser Power at Receiver ($P_3$, $P_4$) & –10 dBm to 20 dBm & Receiver \\
Electrical Amplifier Gain ($G_\text{e}$) & 5 dB & Receiver \\
Noise Figure ($NF$) & 5 dB & Receiver \\
Temperature ($T$) & 290 K & Receiver \\
\bottomrule
\end{tabularx}
\vspace{-10pt}
\end{table}

\section{Numerical Results and Discussion}\label{sec:numericalResults}
In this section, we present simulation results for the photonics and electronics-based THz transmission schemes.

\subsection{Photonics-based THz system}
In this section, based on the derived transmit and received SNR expressions, we first present numerical results to have insight into the effect of hardware impairments. The expressions for the transmitter and receiver SNR in a photonics-based THz transmission system are provided in \eqref{eq:29} and \eqref{eq:35}, respectively. Using representative values for system parameters (listed in Table \ref{tab:photon-analytical-params}), we calculate the corresponding SNR values. We then carry out Monte Carlo simulations for various square QAM modulation formats to assess system performance in terms of bit error rate (BER). For benchmarking, we also compare the results to the classical AWGN channel. 

In Fig.4, we investigate how optical power levels influence the transmitter-side SNR. We consider the dependence of ASE and RIN noise on the EDFA gain, as described in Section \ref{sec:photon-hardwareImpairments}. Two practical system scenarios are examined:
\begin{itemize}
\item \textbf{Case 1:} Fixed EDFA Gain — This reflects systems where the amplifier operates at a constant gain, such as in low-complexity or cost-sensitive designs where dynamic gain control is not implemented.

\item \textbf{Case 2:} Fixed Photomixer Input Power — In this case, we maintain a constant optical power at the photomixer input while varying the EDFA gain. This scenario is representative of systems that employ automatic gain control (AGC) or link optimization strategies to ensure a stable input level at the photodiode, such as in high-performance coherent optical systems.
\end{itemize}

These two cases allow us to analyze how changes in amplifier gain and system configuration influence the resulting noise levels and SNR at the transmitter.
\begin{figure*}  
    \centering
    \subfloat[]{\includegraphics[scale=0.7]{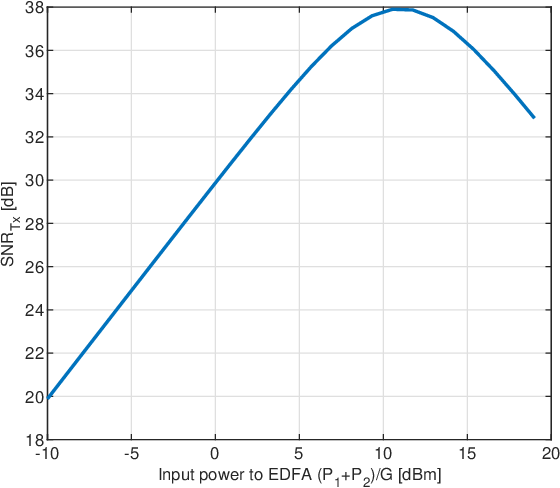}\label{fig:case1-snrTx}}\qquad\qquad    
    \subfloat[]{\includegraphics[scale=0.7]{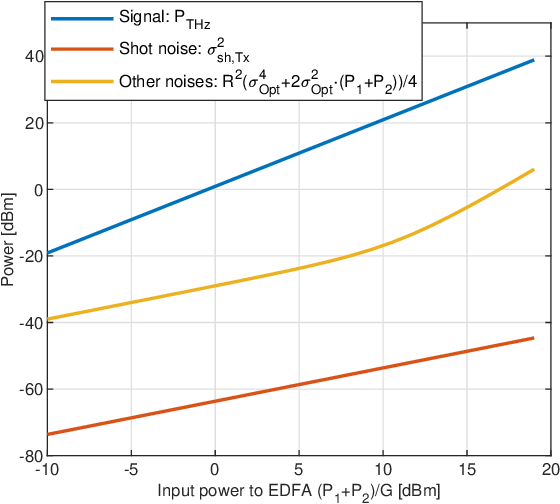}\label{fig:case1-Pnoise}}
    \caption{Case I: fixed EDFA gain $G = 20$dB (a) Transmit SNR versus optical input (b) Signal power and noise powers (see \eqref{eq:29}).}\label{fig:photonSNRTx}
\end{figure*}
\begin{figure*}
    \centering
    \subfloat[]{\includegraphics[scale=0.7]{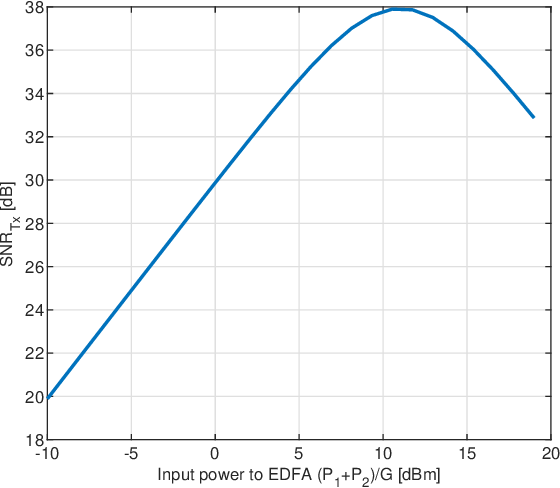}\label{fig:case2-snrTx}}\qquad\qquad    
    \subfloat[]{\includegraphics[scale=0.7]{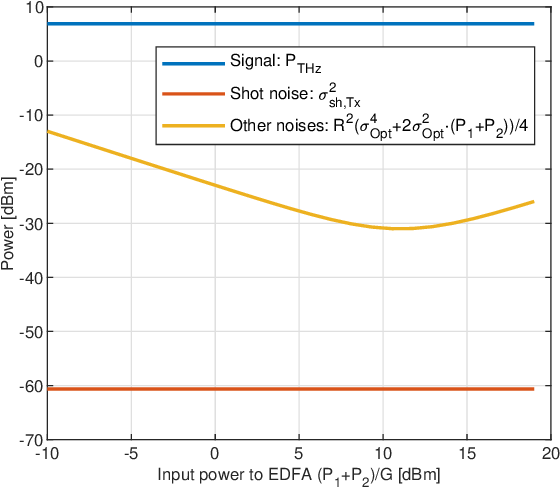}\label{fig:case2-Pnoise}}
    \caption{Case II: fixed output power (${P_1} + {P_2} = 23$ dBm) (a) Transmit SNR versus optical input (b) Signal power and noise powers (see \eqref{eq:29}).}\label{fig:photonSNRTx-case2}
\end{figure*}
In Fig.  \ref{fig:case1-snrTx}, we plot the SNR at the transmitter for a fixed EDFA gain of 20dB. The value of RIN is set to -145dB/Hz; which is a typical value of DFB lasers commercially available. We observe that SNR improves linearly with the input optical power (${{\left( {{P_1} + {P_2}} \right)} \mathord{\left/
 {\vphantom {{\left( {{P_1} + {P_2}} \right)} G}} \right.
 \kern-\nulldelimiterspace} G}$) up to a value of around 8dBm, however, beyond this value, we observe a saturation followed by a degradation in SNR beyond 11dBm. From this curve we can conclude that the optimal input power to EDFA is in range of 10 to 12dBm; which results in the maximum SNR of around 38dB. 

 To better understand the contribution of individual terms in the SNR expression from \eqref{eq:29}, we plot them in Fig.  \ref{fig:case1-Pnoise}. It is observed that both the signal power (${P_{THz}}$) and the shot noise power ($\sigma _{sh,Tx}^2$) increase linearly with the optical input power. However, the signal power increases at a steeper rate because it scales with the product of the signal and laser powers (denoted by ${P_1}$ and ${P_2}$, respectively), whereas the shot noise scales with their sum. This means that if only shot noise were present, the SNR would continue to improve indefinitely as the optical power increases. In practice, however, this is not the case due to additional noise contributions from RIN and ASE introduced by the EDFA. It is observed that these optical noise terms (plotted in yellow) also increase linearly with optical input power up to about 8 dBm, but beyond that point, they grow exponentially, leading to the SNR degradation seen in Fig. \ref{fig:case1-snrTx}. This nonlinear increase in noise power results from the increase of $\sigma _{Opt}^2$ as well as both the signal and unmodulated laser powers. It can be noted that the output power of commercial EDFAs is typically limited to around 23 dBm \cite{ErbiumDopedFiberAmplifiers}. With a fixed amplifier gain of 20 dB, this implies that 0 dBm input power for both the signal and the unmodulated laser (a total of 3 dBm into the EDFA) is sufficient and still falls within the linear operating region shown in Fig. \ref{fig:case1-snrTx}. Under these conditions, the resulting SNR is approximately 33 dB, though this value depends on the specific RIN and ASE levels. 

 The above analysis demonstrates that the dominant limiting factors for transmitter-side SNR in a photonics-based THz system are RIN and ASE noise. The impact of shot noise is relatively minor in comparison. Enhancing SNR would therefore require reducing RIN—by using lasers with better noise performance—and lowering ASE, potentially by reducing EDFA gain. However, these improvements come at the cost of more expensive components.

In Fig. \ref{fig:photonSNRTx-case2} , we consider the second case with fixed optical output power of EDFA with varying value of gain $G$. The value of optical input varies in range of ${{{P_1}} \mathord{\left/
 {\vphantom {{{P_1}} {G = }}} \right.
 \kern-\nulldelimiterspace} {G = }}{{{P_2}} \mathord{\left/
 {\vphantom {{{P_2}} {G = }}} \right.
 \kern-\nulldelimiterspace} {G = }}$-13dBm to 16dBm, as before, and the value of $G$ is accordingly varied in range of $G = $33dB to 4dB; in order to have a fixed output power of ${P_1} + {P_2} = $23dBm. 

It is observed that the transmitter-side SNR for Case II, where the EDFA output power is fixed, is nearly identical to Case I (see Fig. \ref{fig:case1-snrTx}), where the EDFA operates in fixed-gain mode. This implies that, from an SNR standpoint, both operating modes can achieve similar performance at the transmitter. However, the underlying signal and noise components differ significantly, as illustrated in Fig. \ref{fig:case2-Pnoise}. Since the EDFA output power is held constant in Case II, both the signal power and shot noise power remain constant regardless of the input power level. In contrast, the EDFA gain varies with the optical input power, which in turn causes the ASE noise and RIN-induced noise to vary accordingly. We observe that the total optical noise reaches a minimum when the EDFA input power is around 11dBm. Below this point, ASE noise dominates, due to high gain requirements. Above this point, RIN becomes the dominant noise contributor, as the input signal and laser powers increase, leading to SNR degradation. In summary, both Case I (fixed EDFA gain) and Case II (fixed EDFA output) achieve similar transmitter SNR values and the dominant noise contributions come from ASE and RIN. These observations highlight the importance of optical noise sources in photonics-based THz system to optimize system performance.
\begin{figure*}
    \centering
    \subfloat[]{\includegraphics[scale=0.7]{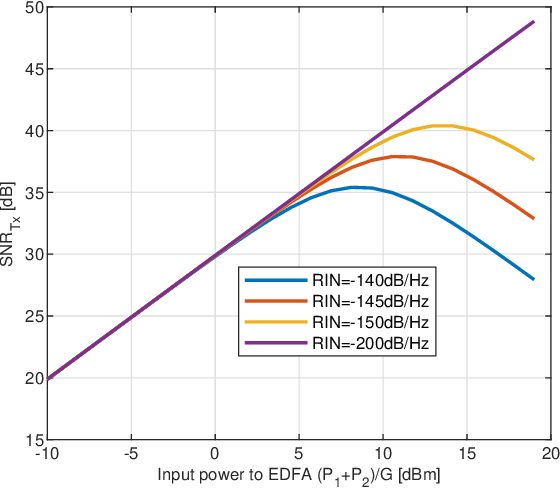}\label{fig:snrTx-varyRIN}}\qquad\qquad    
    \subfloat[]{\includegraphics[scale=0.7]{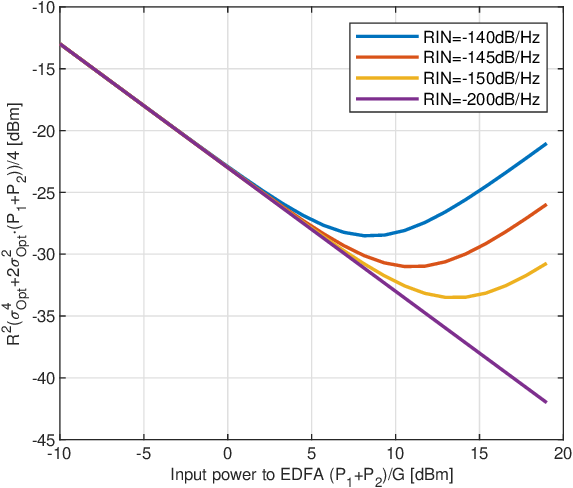}\label{fig:Pnoise-varyRIN}}
    \caption{Impact of RIN on (a) transmitter SNR and (b) power of noise due to ASE and RIN.}\label{fig:photonSNRTx-varyRIN}
\end{figure*}

In Fig. \ref{fig:photonSNRTx-varyRIN}, we further explore the impact of different values of RIN on the optical noise as well as the transmitter SNR. For this evaluation, we assume that the output power of EDFA is fixed to a value of ${P_1} + {P_2} = $23~dBm; same as in Case II. We evaluate the transmitter SNR for different values of RIN.  It is observed from Fig. \ref{fig:snrTx-varyRIN} that, for every 5~dB reduction in RIN, the optimum EDFA input power shifts higher by approximately 2.5~dB, and the corresponding SNR also increases by about 2.5~dB. Fig. \ref{fig:Pnoise-varyRIN} shows the noise power from ASE and RIN, which are the dominant noise sources in a photonics-based THz transmission system and primarily limit the achievable transmitter SNR. For very low RIN values (e.g., –200 dB/Hz), the SNR increases linearly with input power, without any noticeable saturation or inflection point in the region of interest. It is also important to highlight that for EDFA input powers ${{\left( {{P_1} + {P_2}} \right)} \mathord{\left/
 {\vphantom {{\left( {{P_1} + {P_2}} \right)} G}} \right.
 \kern-\nulldelimiterspace} G} \le $5 dBm, the transmitter SNR becomes independent of RIN. This is because, in this region, ASE noise dominates, making the RIN contribution negligible. Conversely, for input powers above ${{\left( {{P_1} + {P_2}} \right)} \mathord{\left/
 {\vphantom {{\left( {{P_1} + {P_2}} \right)} G}} \right.
 \kern-\nulldelimiterspace} G} \ge $5 dBm, RIN begins to dominate, leading to ultimate decay in SNR with increasing optical input to EDFA.
\begin{figure}
    \centering
    \includegraphics[scale=0.7]{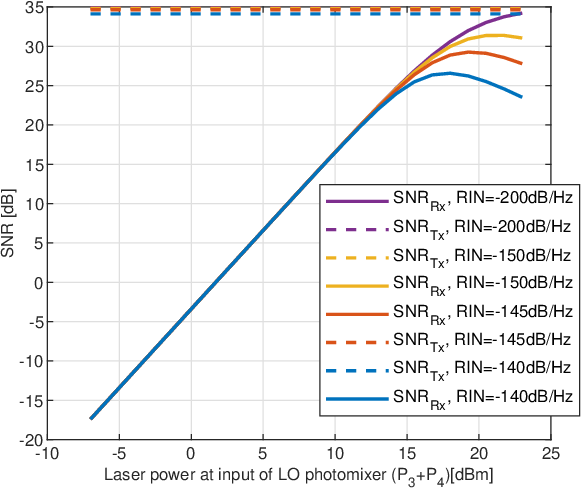}
    \caption{Received SNR versus optical input to the LO photomixer.}
    \label{fig:snrRx-vs-input-varyRIN}
\end{figure}

In Fig. \ref{fig:snrRx-vs-input-varyRIN}, we investigate the received SNR, as described by \eqref{eq:35}. The RIN for the lasers used at both the transmitter and receiver is varied over a range of $RIN = $–140~dB/Hz to –200~dB/Hz, with the assumption that all lasers in the system have identical RIN value. These RIN values are considered to reflect both high- and low-quality commercial lasers. For the analysis, we fixed the optical input to the EDFA at the transmitter to a value of ${{\left( {{P_1} + {P_2}} \right)} \mathord{\left/
 {\vphantom {{\left( {{P_1} + {P_2}} \right)} G}} \right.
 \kern-\nulldelimiterspace} G} = 5$~dBm, and set the EDFA gain to $G = $18~dB, resulting in a total optical power of ${P_1} + {P_2} = 23$~dBm impinging on the photomixer. These values were chosen based on typical specifications of commercial EDFAs and to ensure operation below the transmit SNR inflection points as observed from Fig. \ref{fig:snrTx-varyRIN}. Under these conditions, depending on the RIN level, the transmitter SNR varies between 34.1~dB and 34.9~dB. This range confirms the marginal yet noticeable impact of RIN on transmitter-side SNR when the system is operating in this regime.

In Fig. \ref{fig:snrRx-vs-input-varyRIN}, we also illustrate the received SNR versus the laser power at the input of LO photomixer (${P_3} + {P_4}$). We observe that for each value of RIN (except for -200~dB/Hz), there is an optimum laser power for the LO photomixer; both below and beyond that value, the received SNR rapidly decays. We also note that for a higher RIN value, the difference between the transmit SNR and the received SNR is larger and vice versa. This is expected because the receiver SNR is affected by the optical noise from both the transmitter and receiver lasers. As a limiting case, for a very low RIN value of $RIN = - 200$~dB/Hz, the received SNR approaches the transmit SNR for lasers power ${P_3} + {P_4} \ge 23$~dBm. However, for realistic RIN values of $RIN = -150$~dB/Hz, $ - 145$~dB/Hz, and $- 140$~dB/Hz, the optimum laser power input to the photomixer is observed to be ${P_3} + {P_4} = $~22dBm, 19.25~dBm, and 18~dBm, respectively. At the optimum point, difference between transmitter and receiver SNR are 3.4~dB, 5.4~dB, and 7.6~dB for RIN of $RIN = -150$~dB/Hz, $-145$~dB/Hz, and $-140$~dB/Hz, respectively. It is worth highlighting that for laser powers below the optimum value, the noise from the transmitter limits the SNR while beyond the optimum optical power, the noise from the receiver degrades the SNR. It is also worth highlighting that for lower RIN values, the range of optimum optical powers is larger and vice versa. 
 \begin{figure}
     \centering
     \subfloat[][]{\includegraphics[scale=0.7]{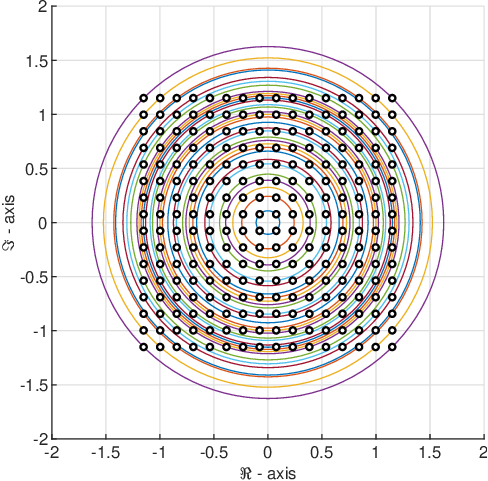}\label{fig:cstll-256QAM}}
     \\
     \subfloat[][]{\includegraphics[scale=0.7]{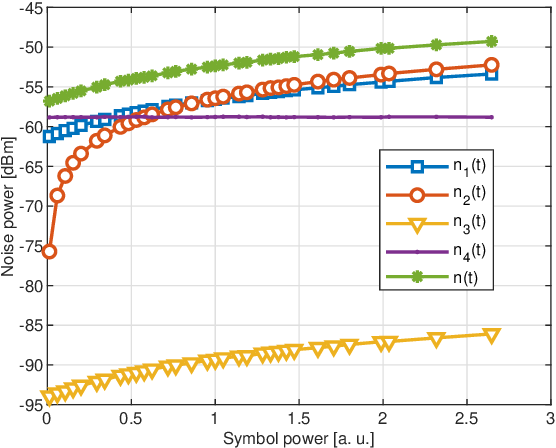}\label{fig:photon-Noise-vs-symPwr}}   
     \caption{(a) Constellation diagram of 256QAM. Circles represent the constellation points with same power. In total there are 32 circles, i.e., 32 groups of symbols with same power. (b) Noise power versus symbol power. ([a. u.] means “arbitrary units”)}
     \label{fig:placeholder}
 \end{figure}

In the following, we analyze the statistics of different noise components for the photonics-based THz transmission case. We consider the modulation format of 256QAM for this analysis. The values of different system parameters considered for this analysis are the same as already listed in Table \ref{tab:photon-analytical-params}. The optical input to the EDFA at the transmitter is assumed to be ${{\left( {{P_1} + {P_2}} \right)} \mathord{\left/
 {\vphantom {{\left( {{P_1} + {P_2}} \right)} G}} \right.
 \kern-\nulldelimiterspace} G} = 5$~dBm and EDFA gain is fixed at $G = 18$~dB; resulting in optical input to the transmitter photomixer of ${P_1} + {P_2} = 23$~dBm. Optimum value of the laser power input to the receiver photomixer are selected according to the RIN value (see Fig. \ref{fig:snrRx-vs-input-varyRIN}), i.e., $19.25$~dBm for a RIN of $-145$~dB/Hz. 

 As it is reflected in \eqref{eq:21}, the aggregate noise consists of four constituent terms. Some of the terms are modulation symbol dependent e.g., ${n_2}\left( t \right)$ and therefore, for the following analysis, it is essential to define different groups of symbols in a 256QAM format. 

 In Fig. \ref{fig:cstll-256QAM}, we plot the constellation diagram of the 256QAM as well as the circles highlighting the symbols which are equidistant from the origin and hence have the same power. We categorize the 256QAM symbols in 32 different groups based on their power level. The power of a 256QAM symbol is given as ${\left| {{a_i}} \right|^2}$ where ${a_i}$ is a constellation point from 256QAM and $i = 1,2,...256$. Group of symbols with the same power value are equidistant from the origin and hence lie on a circle.
\begin{figure}
    \centering
    \includegraphics[scale=0.7]{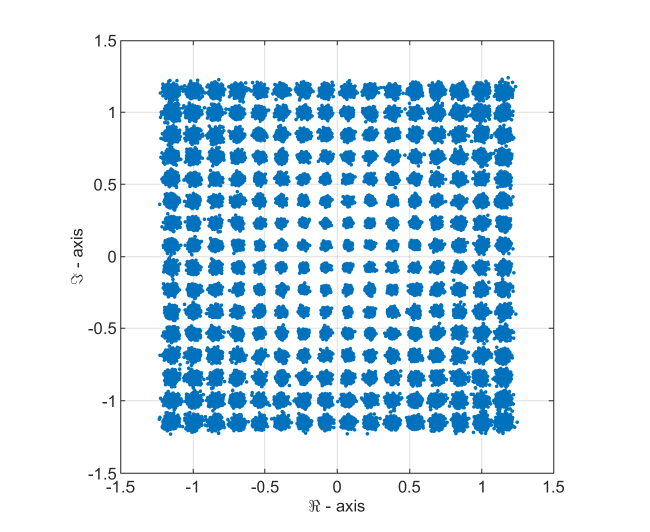}
    \caption{Received signal constellation.}
    \label{fig:cstll-256QAM-photonNoise}
\end{figure}
 To determine the noise contribution depending on the symbol power, we perform a simulation study with 1 million symbols randomly selected from the 256QAM constellation set. Correspondingly, 1~million noise samples are generated employing the statistical definition of different noise terms described by \eqref{eq:21}. The noise samples are then categorized in to 32 groups. The power of different noise components is evaluated and plotted against the symbol power in Fig. \ref{fig:photon-Noise-vs-symPwr}. It is immediately evident that the ${n_3}\left( t \right)$ is negligibly small as compared to the other noise components. 
  We also observe that, apart from ${n_4}(t)$, which represents the thermal noise at the receiver, all other noise terms scale with the symbol power; consequently, symbol groups with higher power experience larger noise variance, while lower-power symbols are less affected. This behavior differs from the conventional AWGN assumption, where the noise variance is constant and independent of the transmitted symbol.

The individual noise components in the proposed model—such as thermal noise, shot noise, ASE, and RIN—are approximated as Gaussian processes following common practice in communication system analysis\cite{loudonTheoryNoiseAccumulation1985, laxQuantumNoiseVII1967}. However, the aggregate noise term obtained from their combination is not strictly Gaussian when considered across all transmitted symbols. 
Instead, the noise can be more accurately described as heteroscedastic Gaussian, where the variance depends on the transmitted symbol amplitude. Consequently, the aggregate noise is modeled as a conditionally Gaussian process, where the noise variance depends on the corresponding symbol amplitude. This representation enables tractable SNR analysis while preserving the dependence of the noise variance on the symbol level.

To illustrate this effect, Fig.~\ref{fig:cstll-256QAM-photonNoise} shows the constellation diagram of the received signal after the addition of the equivalent noise $\tilde{n}(t)$. The impact of heteroscedastic Gaussian noise (or simply symbol-dependent noise) is clearly visible, as higher-power symbols exhibit larger spread compared to lower-power symbols.
\begin{figure}
    \centering
    \subfloat[][]{\includegraphics[scale=0.7]{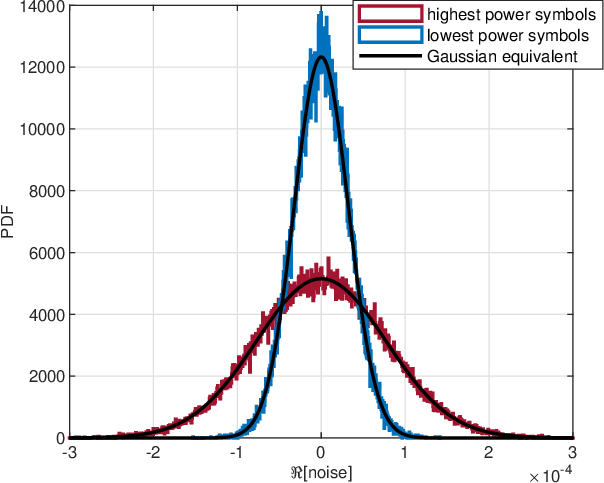}\label{fig:photon-PDF-Noise-eqGauss}}
    \newline
    \subfloat[][]{\includegraphics[scale=0.7]{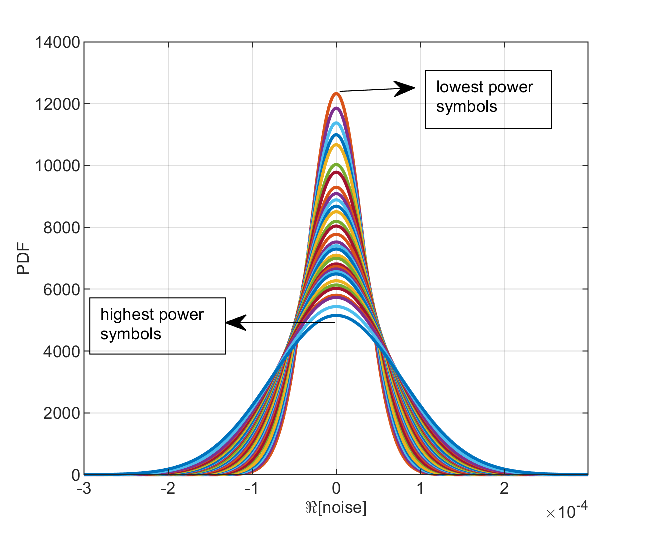}\label{fig:photon-PDF-eqGauss}}
    \caption{PDFs of aggregate noise in photonics-based THz system. (a) corresponding to lowest and highest power symbols group only along with Gaussian equivalent PDF. (b) Gaussian equivalent of noise PDFs for all symbol groups of 256QAM constellation. 32 plots corresponding to 32 symbol groups as shown in Fig. \ref{fig:cstll-256QAM}.}
    \label{fig:placeholder}
\end{figure}

In order to improve the SNR, sources of optical noise at the transmitter and receiver i.e., RIN of lasers and ASE of EDFA, must be properly handled. One straightforward approach to reduce ASE is by filtering out of the band optical noise by employing an optical bandpass filter before the photomixer; thus, reducing the optical bandwidth. Furthermore, lasers with lower RIN values should be employed. However, both of these approaches increase the overall system cost. In addition, the insertion of optical bandpass filter induces optical power loss; which may not be desired for energy efficiency perspective. 
\begin{figure}
    \centering
    \subfloat[][]{\includegraphics[scale=0.7]{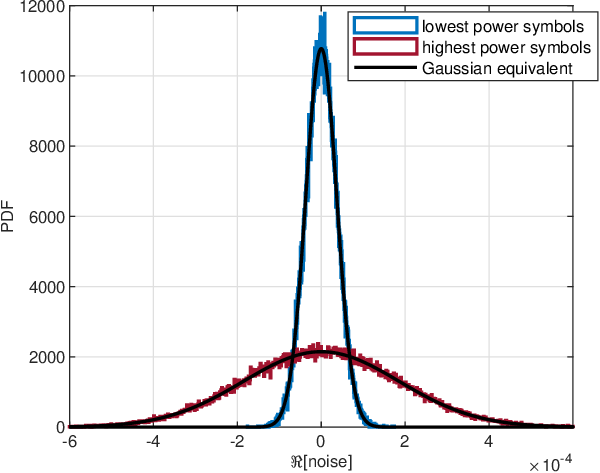}\label{fig:photon-PDF-Noise-eqGauss-135}}
    \newline
    \subfloat[][]{\includegraphics[scale=0.7]{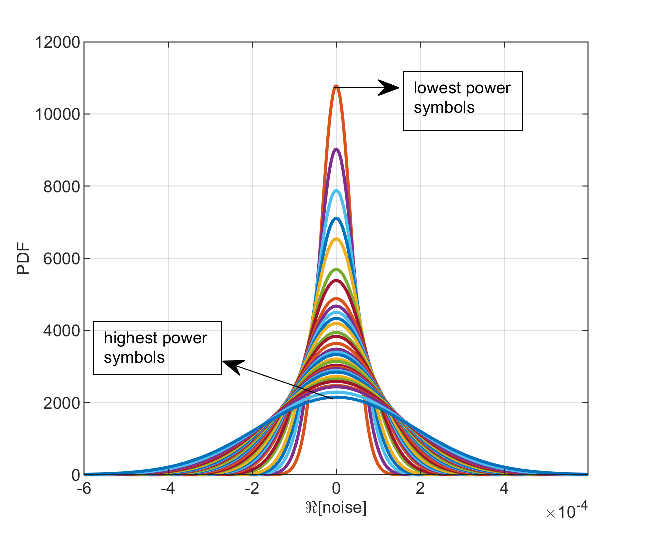}\label{fig:photon-PDF-eqGauss-135}}    
    \caption{PDF of noise for $RIN =  - 135$~dB/Hz (a) corresponding to lowest and highest power symbols only (b) Gaussian equivalent of noise PDFs for all symbol groups of 256QAM constellation.}
    \label{fig:placeholder}
\end{figure}

Next, we analyze the statistical distribution of the noise across different symbol groups of a 256QAM constellation, using the system parameters listed in Table \ref{tab:photon-analytical-params}. Fig. \ref{fig:photon-PDF-Noise-eqGauss} shows the PDF of the real part of the aggregate noise for the symbol groups with the lowest and highest power. For reference, Gaussian PDFs with the same variances as those obtained from the simulated data are overlaid. The results indicate that the noise distribution is well approximated by a Gaussian PDF; however, the variance of the noise is significantly larger for the highest-power symbols compared to the lowest-power ones. This difference in noise variance across symbol groups is further illustrated in Fig. \ref{fig:photon-PDF-eqGauss}, which compares the corresponding equivalent Gaussian PDFs. The difference in noise variance originates from optical noise sources introduced at both the transmitter and receiver, particularly ASE and RIN. In this analysis, we assume a typical RIN value of $RIN =-145$~dB/Hz to evaluate the noise statistics and corresponding PDFs for demonstration. It is worth noting that practical devices such as vertical-cavity surface-emitting lasers (VCSELs) and Fabry–Perot lasers, which are often cheaper alternatives to distributed feedback (DFB) lasers, can exhibit RIN values as high as $RIN =-130$~dB/Hz\cite{1550nmSMVCSEL}. To illustrate the effect of such higher RIN levels on the aggregate noise distribution, we also evaluate the noise PDF with an increased RIN value of $RIN =-135$~dB/Hz as will be discussed next.

Fig. \ref{fig:photon-PDF-Noise-eqGauss-135} shows the PDF of the aggregate noise corresponding to symbol groups with lowest and highest power levels; for an elevated RIN of $RIN=-135$~dB/Hz. Gaussian approximations of the noise PDFs are also overlaid for a visual comparison. We observe that, compared to the previous plot in Fig. \ref{fig:photon-PDF-Noise-eqGauss} for $RIN=-145$~dB/Hz, there is a significantly larger difference between noise variances corresponding to the two symbol groups. This deviation is directly influenced by the optical noise at both the transmitter and receiver. The cost-effective VCSELs exhibiting higher RIN (e.g., $RIN \ge  - 135$~dB/Hz) leads to a larger difference in noise variance among different symbol groups. This distinction is critical for the efficient design of receiver DSP algorithms which typically incorporate a constant noise variance in their design. Depending on the difference on noise variance, relevant parameters in DSP blocks at transmitter and receiver must be optimized for the best performance. Fig. \ref{fig:photon-PDF-eqGauss-135} further illustrates the difference in noise variance corresponding to each symbol group of 256QAM constellation. 

In the following, end-to-end simulation is conducted and BER performance is presented for photonics-based THz transmission systems. The simulation parameters are listed in Table \ref{tab:photon-MCsim-params}.
\begin{table}[t]
\centering
\caption{Parameter values for simulation analysis of THz transmission system}
\begin{tabularx}{0.5\textwidth}{p{0.45\linewidth} X}
\toprule
\textbf{Parameter} & \textbf{Value} \\
\midrule
Modulation & QPSK, 16QAM, 64QAM, 256QAM \\ 
Pulse shape / Matched filter, $p\left( t \right)$ & Root Raised Cosine (roll-off = 0.2) \\ 
Symbol rate ($R_s$) & 32~GBd \\ 
Samples per symbol & 4 \\ 
Number of symbols & 1~million \\ 
Laser frequency drift ($\Delta f$) & $\pm 1$~GHz \\ 
Laser linewidth ($\Delta\nu$) for each laser & 0~kHz, 10~kHz, 100~kHz, 1~MHz \\ 
Relative Intensity Noise (RIN) & $-145$~dB/Hz \\ 
Tx modulated signal power ($P_1/G$) & 2~dBm \\ 
Tx unmodulated laser power ($P_2/G$) & 2~dBm \\ 
Receiver laser power ($P_3, P_4$) & 16.25~dBm \\ 
EDFA gain (at Tx only), $G$ & 18~dB \\ 
Photodiode responsivity, $R$ & 0.7~A/W \\ 
Propagation gain, $\alpha$ & 0.01 to 2.6 \\ 
Noise figure, $NF$ & 5~dB \\ 
Electrical amplifier gain, $G_e$ & 5~dB \\ 
I/Q amplitude imbalance (Tx and Rx), $A$ & 0.25~dB \\
I/Q phase imbalance (Tx and Rx), $\theta$ & 1$^{\circ}$ \\

\bottomrule
\end{tabularx}
\label{tab:photon-MCsim-params}
\end{table}
\subsubsection{BER performance in the absence of phase noise}
We conduct a baseband Monte Carlo simulation and randomly generate 1~million symbols of the desired modulation format. The QAM symbols are then upsampled by a factor of 4 and a root-raised cosine filter with a roll-off factor of $0.2$ is subsequently applied for pulse shaping. The aggregate noise term in \eqref{eq:34} is generated according to the parameters listed in Table \ref{tab:photon-MCsim-params} and subsequently added to the pulse shaped signal. A matched filter is then applied to the noisy signal. The matched filter is another root-raised cosine filter with the same roll-off factor as the pulse shaping filter. After matched filtering, the resulting signal is down-sampled to $1$~sample per symbol and a minimum Euclidean-distance based decision of QAM symbols then followed. A QAM symbol-to-bit de-mapper is then employed to convert decided symbols to a bit-stream. The same de-mapper is utilized to convert transmitted QAM symbols to a bitstream. The retrieved bit stream is compared with the transmitted bit stream and the resulting BER is subsequently calculated. In our simulations, we keep the power of transmitted THz signal fixed. Therefore, to simulate the BER performance at different values of received power, we vary the propagation coefficient $\alpha$ which effectively changes the power of the received THz signal.
\begin{figure}
    \centering
    \includegraphics[scale=0.8]{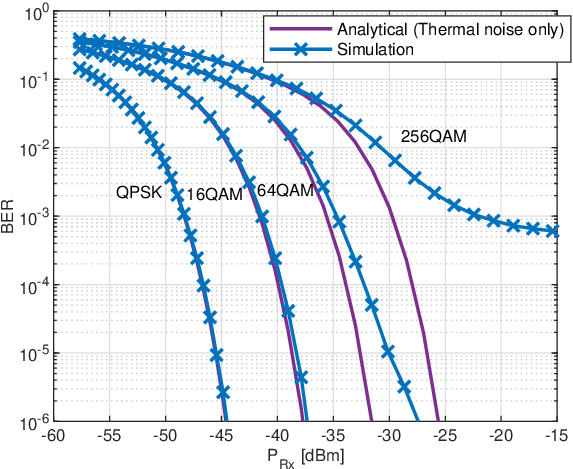}
    \caption{BER versus received THz power for photonics-based THz transmission}
    \label{fig:photon-BER-vs-Prx}
\end{figure}

In Fig. \ref{fig:photon-BER-vs-Prx}, we plot the simulated BER versus received THz signal power. For comparison, we include an analytical curve that shows the BER performance over the classical AWGN channel considering the thermal noise only. For QPSK, the analytical and simulated curves coincide, indicating that the performance is limited solely due to thermal noise of the receiver and impact of optical noises is negligible at such low received THz powers. For 16QAM, we observe that the simulated curve and the analytical curve shows a slight deviation for power values $\ge -41$~dBm. This indicates that the impact of noise sources other than thermal noise in the system has started to take effect. The performance gap at a BER of $10^{-3}$ is about $0.2$~dB in terms of received THz power. For higher-order modulations, deviations from the AWGN  curve become more significant. Starting from lower received power values, the optical noise term becomes dominant for both 64 QAM and 256 QAM. The BER for 256 QAM experiences an error floor eventually due to the dominance of signal power dependent noise terms, see \eqref{eq:35}; leading to a saturation in received SNR and consequently a BER floor. It should be noted that the BER floor can be mitigated by mitigating the signal-dependent noise components which are related to the RIN of lasers and ASE of EDFA. For example, a high quality laser with low RIN can be employed to reduce the effect. Furthermore, ASE noise can be mitigated by including an optical bandpass filter prior to photomixing.
\subsubsection{Contribution of constituent noise terms}
For photonics-based THz systems, the SNR expression at the receiver is presented in \eqref{eq:35}. Note that apart from the receiver thermal noise $\sigma_{th}^2$, all the other SIN and SDN components are scaled by $\alpha^2$. Hence, depending on its value, either the SIN or the SDN will dominate. In case the SIN dominates, the simulated BER performance becomes closer to the analytical BER curve. On the other hand, if SDN increases in magnitude, we observe deviation between the simulated and analytical BER curves. To highlight the contributions of SIN and SDN we plot a bar graph showing each component for several different $P_{Rx}$ values; employing parameter values listed in Table \ref{tab:photon-MCsim-params}. 
\begin{figure*}[t]
    \centering
    \subfloat[][]{\includegraphics[scale=0.8]{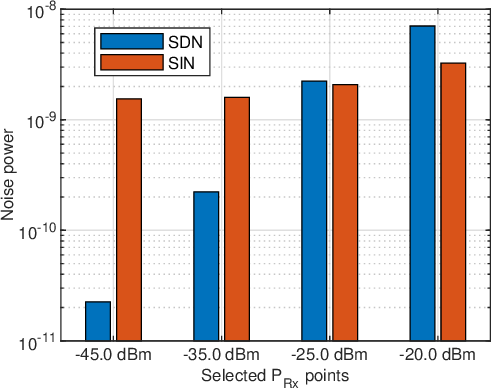}}\qquad\qquad
    \subfloat[][]{\includegraphics[scale=0.8]{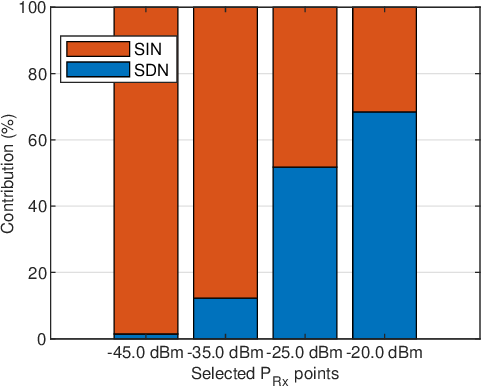}}
    \caption{(a) Magnitude of SIN and SDN terms and (b) Percentage contribution of SDN and SIN to the total noise for several $P_{Rx}$ values.}\label{fig:NoiseContributions_photonics}
\end{figure*}

Fig. \ref{fig:NoiseContributions_photonics}a shows the magnitude of SIN and SDN for several ${P_{Rx}}$ values while their percentage contribution to the total noise power is shown in Fig. \ref{fig:NoiseContributions_photonics}b. For lower ${P_{Rx}}$ values, SIN becomes dominant while for high ${P_{Rx}}$ values, SDN is the dominant noise contributor. In particular, the contribution of SDN for $P_{Rx}=-45$~dBm is only about $1.5\%$. However, for higher power values, it increases gradually; for $P_{Rx}=-20$~dBm the SDN contribution reaches about $68\%$. In summary, the dominant noise term is dependent on the received power; for high received power, the SDN dominates while for the low received power, SIN dominates. Hence, for high received power values, we observe a BER floor (Fig. \ref{fig:photon-BER-vs-Prx}).

\subsubsection{BER performance including phase noise}
In the previous section, we evaluated BER performance under the assumption of ideal lasers with no frequency offset or phase noise. In this section, we extend the analysis to include these two effects. The simulation process follows the same steps as before, with two key additions: after introducing aggregate noise to the pulse-shaped signal, we apply a fixed frequency offset of $\Delta f$ and phase noise of $\Delta \phi (t)$. Both are calculated using the parameters listed in Table \ref{tab:photon-MCsim-params}. 

\begin{figure*}[t]
    \centering
    \subfloat[][]{\includegraphics[scale=0.6]{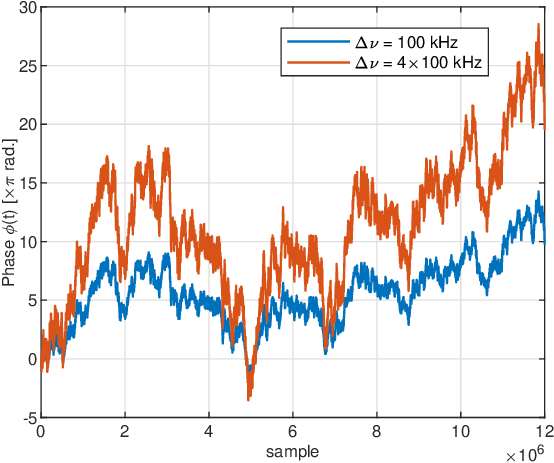}}
    \subfloat[][]{\includegraphics[scale=0.6]{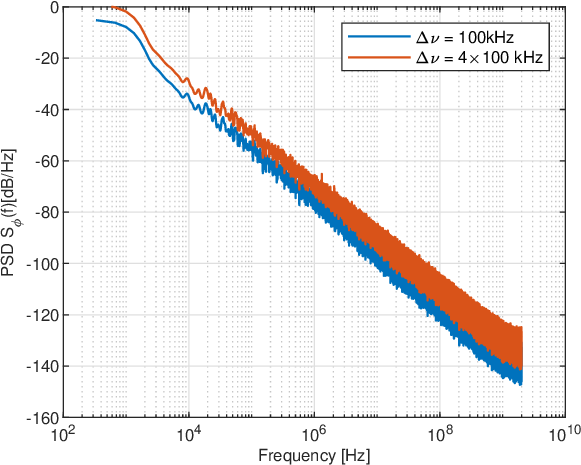}}
    \subfloat[][]{\includegraphics[scale=0.7]{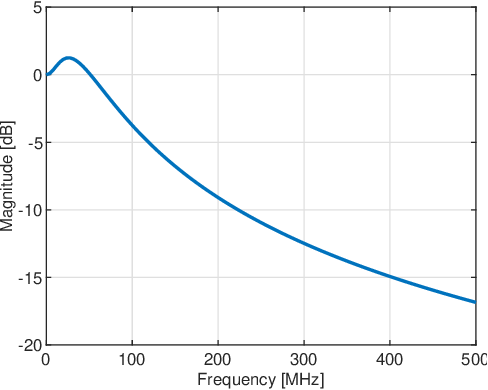}}
    \caption{Phase noise of a laser with linewidth $100$~kHz and $400$~kHz (a) Time domain phase noise evolution (sampling period $\tau=250$~ps) (b) PSD of phase noise (c) Magnitude response of the PLL used for carrier recovery.}\label{fig:laser-phaseNoise-tdfd}
\end{figure*}

Fig. \ref{fig:laser-phaseNoise-tdfd}a and \ref{fig:laser-phaseNoise-tdfd}b shows an example of phase noise PSD for a laser with linewidth $\Delta \upsilon  = 100$~kHz. If independent lasers with same linewidth are used to generate the THz carrier and the LO, the effective phase noise is enhanced by a factor of $4$ whose PSD is also shown in the same figure for a visual comparison. 
If the frequency offset and phase noise are left unmitigated, it leads to a random rotation of QAM constellation points on the cartesian plane which leads to a BER of $\sim0.5$ after detection; regardless of received power. Hence, frequency offset and phase noise must be compensated using suitable algorithms for a proper symbol detection. 

Next, we redo the simulations with laser phase noise and employ a PLL based phase noise mitigation scheme at the receiver DSP. For this purpose, we use a second-order digital data-aided phase-locked loop (PLL) operating at one update per received symbol. The loop employs a cross-product phase detector and a proportional-integral controller. The phase detector gain is $K_P=\frac{1}{15}$ and the VCO gain is $K_O=1$. The proportional and integral gains of the loop filter are determined from the damping factor $\zeta=1$ and the normalized loop noise bandwidth $B_n=0.0045$. For the 32~GBd system considered in this work, this corresponds to a loop bandwidth that is a small fraction of the symbol rate, enabling the PLL to track slow carrier phase variations while suppressing high-frequency noise. The resulting PLL magnitude response is given in Fig. \ref{fig:laser-phaseNoise-tdfd}c.

The laser linewidth values used in the simulations are selected based on specifications of commercially available optical sources commonly used in photonic THz transmitters. For example, high-quality external cavity lasers typically exhibit linewidths on the order of $\Delta\nu=10$~kHz, while distributed feedback (DFB) lasers commonly achieve linewidths around $\Delta\nu=1$~MHz. These ranges are consistent with optical sources employed in many photonics-based THz communication experiments.

\begin{figure*}
    \centering
    \subfloat[][]{\includegraphics[scale=0.7]{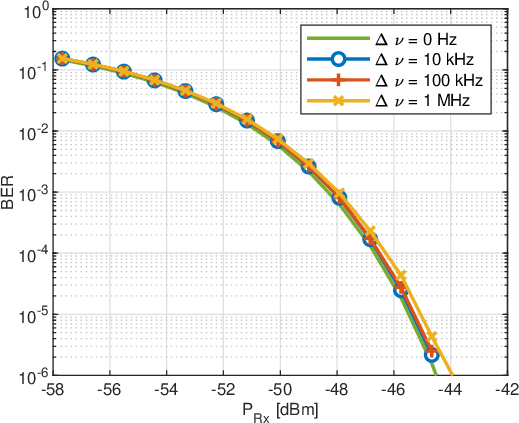}\label{fig:photon-qpsk-BER-vs-Prx-PN-PLL}}\qquad\qquad
    \subfloat[][]{\includegraphics[scale=0.7]{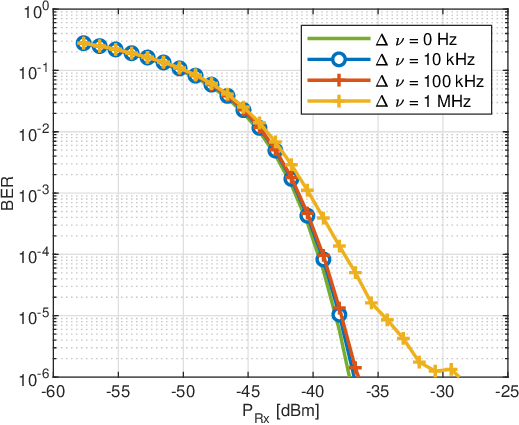}\label{fig:photon-qam16-BER-vs-Prx-PN-PLL}}
    \\
    \subfloat[][]{\includegraphics[scale=0.7]{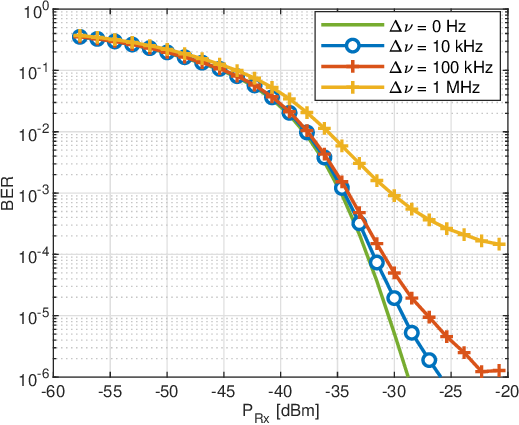}\label{fig:photon-qam64-BER-vs-Prx-PN-PLL}}\qquad\qquad
    \subfloat[][]{\includegraphics[scale=0.7]{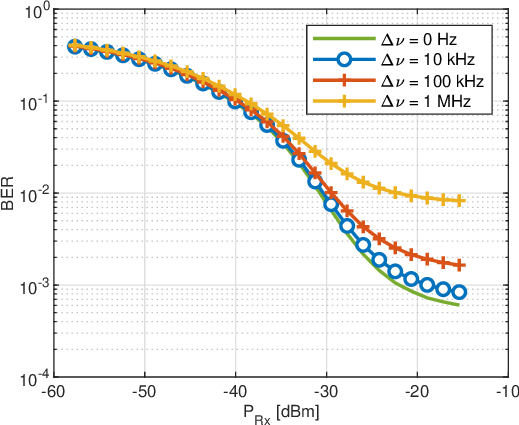}\label{fig:photon-qam256-BER-vs-Prx-PN-PLL}}
    \caption{Impact of laser phase noise on performance of square QAM formats (a) QPSK, (b) 16QAM, (c) 64QAM, (d) 256QAM.}\label{fig:BER-vs-Prx-PhaseNoise-PLL}
\end{figure*}

In Fig.\ref{fig:BER-vs-Prx-PhaseNoise-PLL}, we plot the results of different square QAM formats for different values of laser linewidth. The BER performance significantly improves after phase noise mitigation. In particular, we observe that QPSK is robust to phase noise for laser linewidths up to $1$~MHz; and there is no significant penalty induced on the QPSK signal. This is because of the wide angular separation between constellation points of QPSK modulation. A significant penalty in performance is however observed for $1$~MHz laser for 16QAM, 64QAM, and 256QAM formats. In case of 16QAM, the BER reaches below $1 \times {10^{ - 6}}$ for $P_{Rx} \ge -29$~dBm. However, for 64QAM and 256QAM the minimum BER is $1.5 \times {10^{ - 4}}$ and $9\times 10^{-3}$, respectively, at the maximum $P_{Rx}$. It is worth highlighting that the residual phase error after the PLL has a Gaussian distribution whose variance depends on the value of its loop bandwidth. A larger loop bandwidth is necessary to track faster phase fluctuations due to larger linewidth. On the other hand, a larger loop bandwidth leads to a larger residual phase noise after PLL which can degrade performance of higher order QAM formats because of a smaller angular separation between neighboring points. Table \ref{tab:linewidth_penalty} summarizes the power penalty of different considered modulation formats at a BER of $1 \times {10^{ - 2}}$. 

\begin{table}[h]
\centering
\caption{Power penalty of different QAM formats for typical linewidths of lasers at BER of $1\times 10 ^{-2}$}
\begin{tabular}{l c c c }
\toprule
\textbf{Modulation} & $\Delta\nu=$\textbf{10 kHz} & $\Delta\nu=$\textbf{100 kHz} & $\Delta\nu=$\textbf{1 MHz} \\ \midrule
QPSK & $0.1$ dB & $0.1$ dB & $0.3$ dB \\ 
16QAM & $0.2$ dB & $0.2$ dB & $0.6$ dB \\
64QAM & $0.2$ dB & $0.3$ dB & $2.2$ dB \\ 
256QAM & $0.3$ dB & $1.2$ dB & $8.3$ dB \\ 
\bottomrule
\end{tabular}
\label{tab:linewidth_penalty}
\end{table}

\subsubsection{Impact of I/Q imbalance on the system performance}

To model this effect, we adopt the methodology proposed in \cite{anttilaFrequencySelectiveMismatchCalibration2008}. Since the frequency responses of the I and Q channels are assumed to be flat over the signal bandwidth, the FIR filters $g_1(t)$ and $g_2(t)$ in \cite{anttilaFrequencySelectiveMismatchCalibration2008} reduce to single coefficients. The amplitude and phase imbalance values used in the analysis are listed in Table~\ref{tab:photon-MCsim-params}. With these parameter values, the resulting in-band image interference level is $-35.5$~dB, computed using (2) in \cite{anttilaFrequencySelectiveMismatchCalibration2008}. The same I/Q imbalance parameters are assumed for both the transmitter and receiver.

In this analysis, the laser linewidth is fixed at $\Delta \nu=10$~kHz. As in the previous section, phase noise mitigation is performed in the receiver DSP using a data-aided PLL.

\begin{figure*}
    \centering
    \subfloat[][]{\includegraphics[scale=0.8]{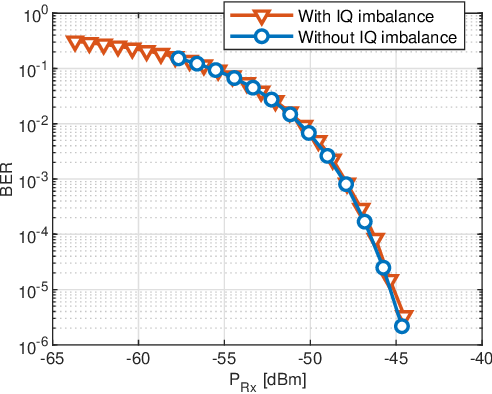}\label{fig:photon-qpsk-BER-vs-Prx-IQimb}}\qquad\qquad
    \subfloat[][]{\includegraphics[scale=0.8]{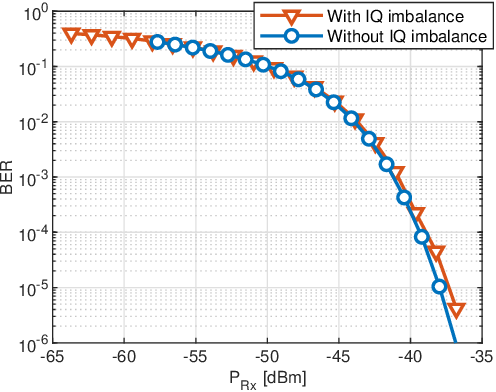}\label{fig:photon-qam16-BER-vs-Prx-IQimb}}
    \\
    \subfloat[][]{\includegraphics[scale=0.8]{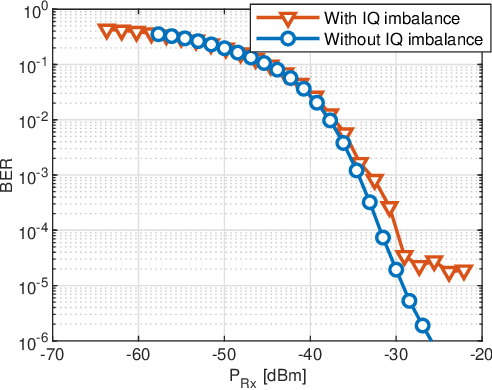}\label{fig:photon-qam64-BER-vs-Prx-IQimb}}\qquad\qquad
    \subfloat[][]{\includegraphics[scale=0.8]{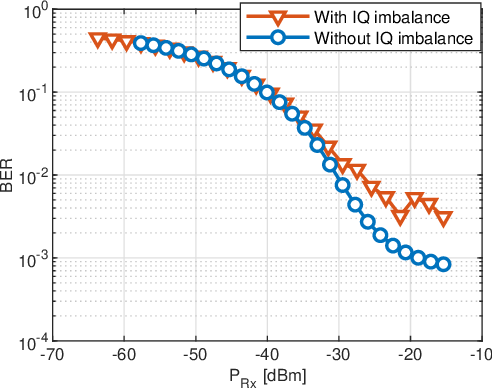}\label{fig:photon-qam256-BER-vs-Prx-IQimb}}
    \caption{Impact of I/Q imbalance in photonics-based THz system on the performance of square QAM formats (a) QPSK, (b) 16QAM, (c) 64QAM, (d) 256QAM.}\label{fig:BER-vs-Prx-IQimb}
\end{figure*}

Fig.~\ref{fig:BER-vs-Prx-IQimb} illustrates the impact of I/Q imbalance on different QAM modulation formats. As shown in Fig.~\ref{fig:BER-vs-Prx-IQimb}a and Fig.~\ref{fig:BER-vs-Prx-IQimb}b, the performance degradation for QPSK and 16QAM is negligible, and no error floor is observed. However, for higher-order modulation formats such as 64QAM and 256QAM (Fig.~\ref{fig:BER-vs-Prx-IQimb}c and Fig.~\ref{fig:BER-vs-Prx-IQimb}d), the impact becomes more pronounced. In addition to an increased power penalty, an error floor emerges for 64QAM due to the image interference introduced by the I/Q imbalance. This error floor becomes progressively more significant as the modulation order increases, indicating that higher-order QAM formats are considerably more sensitive to I/Q mismatch. These results highlight that while lower-order constellations remain relatively robust, I/Q imbalance can impose a fundamental performance limitation for spectrally efficient modulation schemes.

\color{black}

\subsection{Electronics-based THz system}

The expressions for SNR at the transmitter and receiver for an electronics-based THz system are given are given in \eqref{eq:71} and \eqref{eq:75}, respectively. Here, we consider some typical values of different system parameters (see Table \ref{tab:elec-params-analytical}) and evaluate the corresponding SNR. Similar to previous section, we assume $\alpha  = 1$ for simplicity. We consider a base oscillator frequency of ${f_{base}} = 15$~GHz and the target carrier frequency of ${f_c} = 300$~GHz. Hence, the required multiplication factor is $N = 20$. Furthermore, we consider that a bandpass filter is present around the carrier frequency; which suppresses the noise floor. This way, the noise bandwidth of oscillator noise floor is smaller than the signal bandwidth. Here, we consider the bandwidth of oscillator filter to be $2\%$ of the carrier frequency; i.e., for a carrier frequency of ${f_c} = 300$~GHz, the filter bandwidth is $6$~GHz. We fix the signal power ${P_s} = 6.9$~dBm; same as in the photonics-based case when the EDFA output power was set to $23$~dBm. 

The noise floor of the carrier oscillator is varied in the range 
${S_{{i_{base}}}}\left( f \right) = -200$~dBc/Hz to $-100$~dBc/Hz. 
This range is selected to represent typical phase-noise characteristics 
reported for oscillators used in multiplier-based THz electronic transmitters, 
as discussed in the literature (e.g., 
\cite{danSuperheterodyne300GHz2020,weberWBandX12Frequency2011,chenDoesNoiseFloor2017}). 
For reference, commercial microwave signal generators such as the 
Keysight N5183B exhibit measured phase-noise levels of approximately 
$-143$~dBc/Hz at an oscillator frequency of $20$~GHz 
\cite{keysightN5183BMXGXSeries}. These values are therefore representative 
of practical RF oscillator performance used in experimental THz communication systems.

\begin{table}[t]
\centering
\caption{Parameters for SNR evaluation of electronics-based THz system}
\begin{tabularx}{0.5\textwidth}{p{0.4\linewidth} p{0.2\linewidth} X}
\toprule
\textbf{Parameter} & \textbf{Value} & \textbf{Applies To} \\ 
\midrule
Carrier and LO frequency ($f_c$, $f_{LO}$) & $300$~GHz & Transmitter, Receiver \\ 
Base oscillator frequency ($f_{base}$, $f_{base,LO}$) & $15$~GHz & Transmitter, Receiver \\ 
Frequency multiplication factor ($N$) & $20$ & Transmitter, Receiver \\ 
Noise floor of base oscillator (${S_{{i_{base}}}}\left( f \right)$, ${S_{{i_{base,Rx}}}}\left( f \right)$) & $–200$ to $–100$~dBc/Hz & Transmitter, Receiver \\ 
Specific oscillator noise floor value ${S_{{i_{base}}}}\left( f \right)$ & $–135.4$~dBc/Hz & Used for calculation of $SN{R_{Rx}}$ \\ 
Oscillator noise bandwidth & $6$~GHz & Transmitter, Receiver \\ 
Signal power ($P_s$) & $6.9$~dBm (same as Photonics case) & Transmitter, Receiver \\
Local oscillator power ($P_{LO}$) & $20$~dBm & Receiver \\
Amplifier gain ($G_e$) & $5$~dB & Transmitter, Receiver \\
Noise figure ($NF$) & $5$~dB & Transmitter, Receiver \\
Bandwidth ($B$) (thermal noise) & $40$~GHz & Transmitter, Receiver \\
Temperature ($T$) & $290$~K & Transmitter, Receiver \\
\bottomrule
\end{tabularx}
\label{tab:elec-params-analytical}
\end{table}

\begin{figure}[!tb]
    \centering
    \subfloat[][]{\includegraphics[scale=0.7]{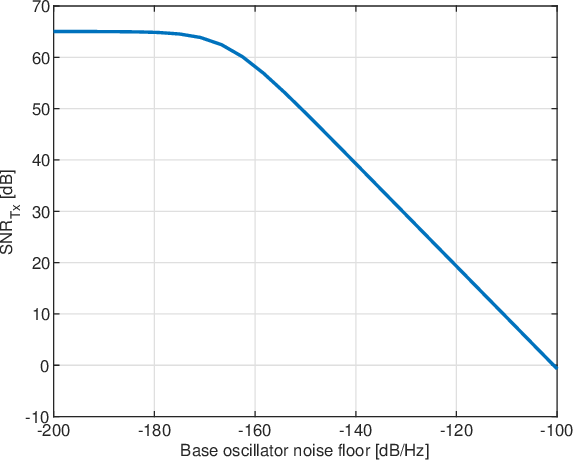}\label{fig:elec-snrTx-analytical}}\\
    \subfloat[][]{\includegraphics[scale=0.7]{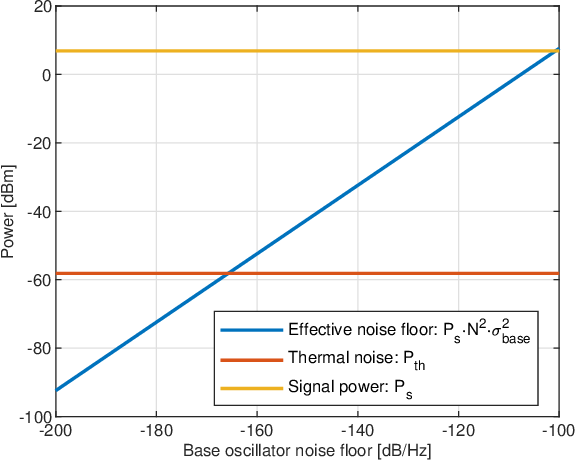}\label{fig:elec-NoisePwr-analytical}}
    \caption{(a) Transmit SNR versus noise floor of base oscillator (b) Signal power (${P_s}$) and noise power ( ${P_s}{N^2}\sigma _{base}^2$ and $\sigma _{th,Tx}^2$) versus the noise floor of base oscillator (${S_{{i_{base}}}}\left( f \right)$).}
\end{figure}

First, based on \eqref{eq:71}, we evaluate the SNR at the transmitter for different values of base oscillator noise floor.
In Fig. \ref{fig:elec-snrTx-analytical}, we plot the transmitter SNR versus the noise floor of base oscillator at the transmitter side, i.e., ${S_{{i_{base}}}}\left( f \right)$. We observe that the maximum SNR is saturated to a value of $66$dB for oscillator noise floor PSD of ${S_{{i_{base}}}}\left( f \right) \le  - 170$~dB/Hz. In that case, contribution of noise due to oscillator noise floor is negligibly small; i.e., the thermal noise of transmitter plays a dominating role. However, as the noise floor PSD increases, we observe a linear decay in SNR. The contribution of oscillator noise floor becomes dominant already for ${S_{{i_{base}}}}\left( f \right) \ge  - 160$~dB/Hz. In order to achieve a transmitter $SN{R_{Tx}} \ge 34.6$~dB, same as for the photonics-based system with ${P_s} = 6.9$~dBm and $RIN =  - 145$~dB/Hz, a base oscillator noise floor of ${S_{{i_{base}}}}\left( f \right) \le  - 135.4$~dBc/Hz is required. It is worth highlighting that the required value of base oscillator noise floor for a specific $SN{R_{Tx}}$ value is dependent on the frequency multiplication factor $N$. For a higher value of $N$ we will need further reduction in oscillator noise floor ${S_{{i_{base}}}}\left( f \right)$ and vice versa. Hence, to achieve the highest SNR at the transmitter side, it is desirable to use a lower multiplication factor and a higher base oscillator frequency with a lower phase noise. 

\begin{figure}
    \centering
    \includegraphics[scale=0.7]{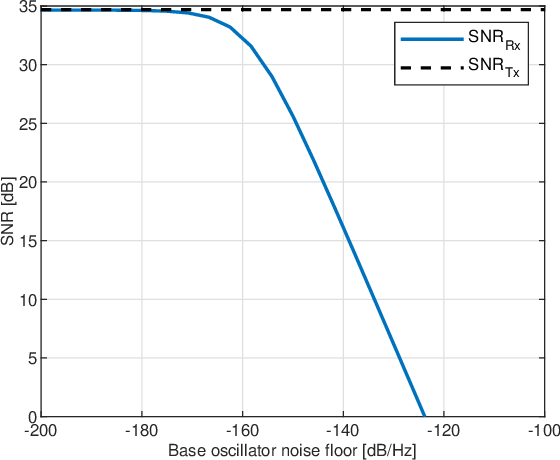}
    \caption{Receive SNR for electronics-based THz transmission system.}
    \label{fig:elec-snrRx-analytical}
\end{figure}

Now we focus on the SNR at the receiver end; whose expression is given in \eqref{eq:75}. We choose the noise floor of the transmitter base oscillator to be $ - 135.4$~dBc/Hz; which results in transmitter SNR of $34.6$~dB. All the parameters remain the same as in Table \ref{tab:elec-params-analytical}. We assume that the noise figure $NF$ and the electrical amplifier gain ${G_e}$ are the same for both transmitter and receiver. In addition, the bandwidth of transmitter and receiver are also the same. The signal power is ${P_s} = 6.9$~dBm and the LO power is ${P_{LO}} = 20$~dBm. The frequency of base oscillator at transmitter and receiver are also assumed to be the same, i.e., ${f_{base}} = {f_{base,LO}} = 15$~GHz. The noise floor of the LO base oscillator is varied in range ${S_{{i_{base,Rx}}}}\left( f \right) =  - 200$~dBc/Hz to $-100$~dBc/Hz and the oscillator noise bandwidth is $6$~GHz. In Fig. \ref{fig:elec-snrRx-analytical}, we plot the receive SNR versus noise floor of the base oscillator ${S_{{i_{base,Rx}}}}\left( f \right)$. We observe a negligible offset between transmitter and receiver SNRs towards the left side of the graph in Fig. \ref{fig:elec-snrRx-analytical}. The impact of oscillator noise floor becomes dominating already at a value of ${S_{{i_{base,Rx}}}}\left( f \right) \ge  - 165$~dB/Hz; a value which is about $30$~dB lower than that of the transmitter. This means that the noise floor of receiver oscillator is much more critical than that of the transmitter.

\begin{figure}
    \centering
    \includegraphics[scale=0.8]{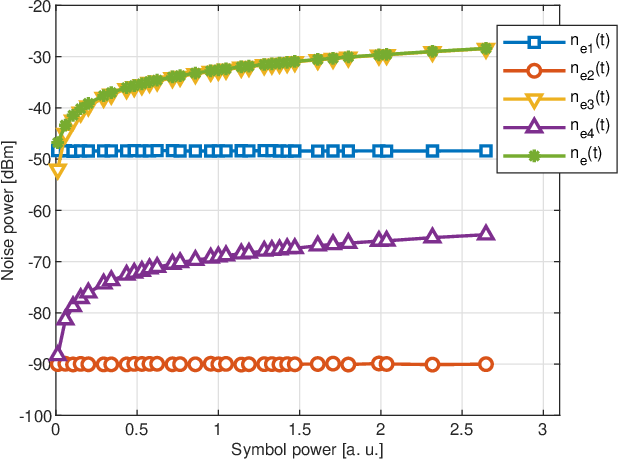}
    \caption{Noise power versus symbol power.}
    \label{fig:elec-NoisePwr-vs-symbolPwr}
\end{figure}

We now turn our attention to analyzing the statistics of the effective noise and its individual components in the context of an electronics-based THz transmission system. For this analysis, we focus on 256QAM modulation, whose constellation diagram is shown in Fig. \ref{fig:cstll-256QAM}. The simulation procedure follows the same steps used for the photonics-based system. From \eqref{eq:66}, we observe that the noise components ${n_{e3}}\left( t \right)$ and ${n_{e4}}\left( t \right)$ are scaled by the transmit signal $x\left( t \right)$. To evaluate their impact, we group the 256QAM constellation points into 32 symbol-power levels (as done previously) and calculate the noise power for each group. Fig. \ref{fig:elec-NoisePwr-vs-symbolPwr} plots the noise power versus symbol power for the electronics-based system. It is evident that the ${n_{e3}}\left( t \right)$ is the dominant component, followed by the thermal ${n_{e1}}\left( t \right)$. Since ${n_{e3}}\left( t \right)$ is directly proportional to the symbol power, constellation points with higher amplitude experience greater noise, while lower-power symbols are less affected. This trend is clearly visible in the plot of individual noise contributions in Fig. \ref{fig:elec-NoisePwr-vs-symbolPwr}. In contrast, the thermal noise ${n_{e1}}\left( t \right)$  remains constant across all symbols, as it is independent of signal amplitude. However, because ${n_{e3}}\left( t \right)$ increases with symbol power, the relative impact of thermal noise becomes negligible for higher-power constellation points. It only plays a significant role for low-power symbols—those clustered near the center of the constellation, as shown in Fig. \ref{fig:cstll-256QAM}. In summary, for high-power symbols, oscillator noise is the sole significant contributor to total noise. This behavior mirrors that of the photonics-based system and is further illustrated by the received signal constellation after aggregate noise ${n_e}\left( t \right)$ is added.

\begin{figure}
    \centering
    \includegraphics[scale=0.7]{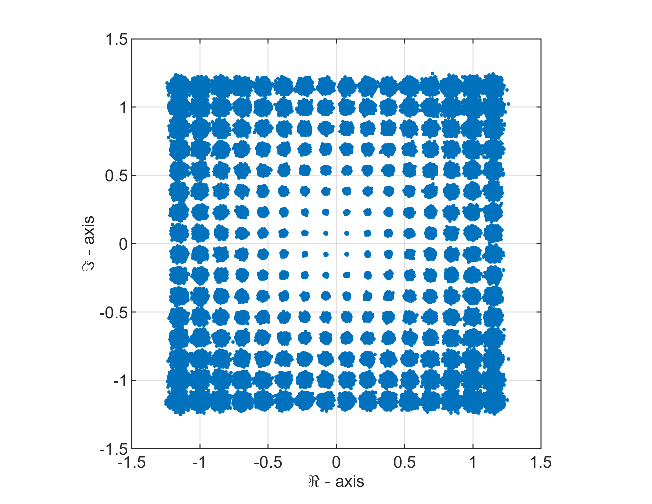}
    \caption{Constellation diagram of 256QAM after addition of equivalent electrical noise, ${n_e}\left( t \right)$.}
    \label{fig:cstll-256QAM-ElecNoise}
\end{figure}

In Fig.~\ref{fig:cstll-256QAM-ElecNoise}, we show the 256QAM constellation after the addition of the equivalent electrical noise ${n_e}(t)$. The impact of signal-dependent noise (heteroscedastic Gaussian noise) is clearly observed, where higher-power symbols exhibit larger dispersion compared to lower-power symbols. This behavior is more pronounced in the electronics-based system due to stronger scaling of noise with symbol power.
In order to further analyze the statistics of equivalent electrical noise, we calculate the PDF of real component of aggregate noise term corresponding to different groups of symbols. The PDFs are then plotted on the same graph for comparison. For the reference, Gaussian equivalent PDFs are also shown in addition to PDFs of the simulated noise samples. 

\begin{figure}
    \centering
    \subfloat[][]{\includegraphics[scale=0.7]{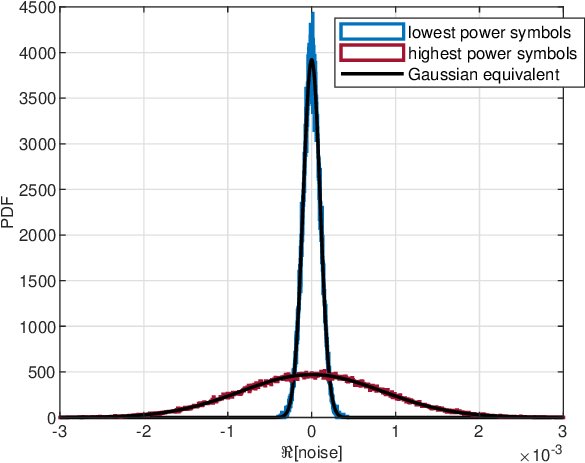}\label{fig:elec-PDF-Noise-eqGauss}}
    \\
    \subfloat[][]{\includegraphics[scale=0.7]{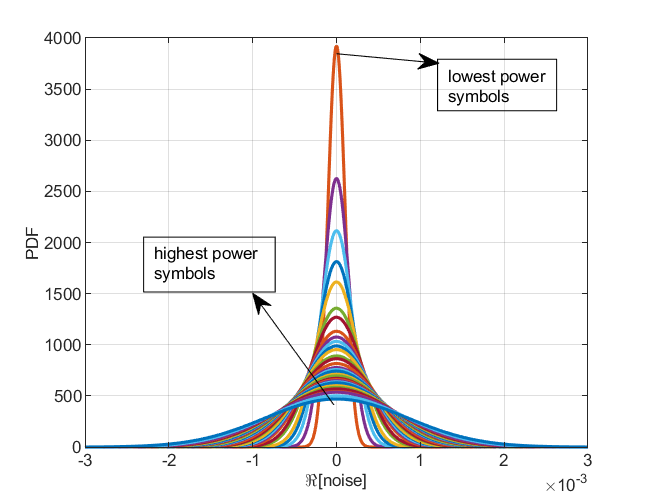}\label{fig:elec-PDF-eqGauss}}    
    \caption{PDF of noise in electronics-based THz system.}
    \label{fig:placeholder}
    \vspace{-10pt}
\end{figure}

In Fig. \ref{fig:elec-PDF-Noise-eqGauss}, the noise PDFs for the lowest and highest power symbol groups are plotted along with Gaussian equivalent PDFs. Similar to the photonics-based system, here also we observe a large difference between the variance of noise for different symbol groups. As we also explored during the signal modeling, the signal dependent noise originates from the noise floor of base oscillators at the transmitter and receiver end; which is further amplified by the frequency multiplication blocks. 
Despite the difference in variance, the shape of the noise PDF is well approximated by a Gaussian distribution. 

Fig.~\ref{fig:elec-PDF-eqGauss} further illustrates the Gaussian equivalent PDFs for all 32 symbol groups of the 256QAM constellation, confirming that the aggregate noise follows a heteroscedastic Gaussian (signal-dependent) model, consistent with the behavior observed in the photonics-based case.  

In the following, simulation performance of electronics-based THz transmission system is evaluated. The parameters of simulation are listed in Table \ref{tab:elec-MCsim-params}.

\begin{table}[h]
\centering
\caption{Parameters for simulation of Electronics-based THz transmission system.}
\begin{tabularx}{0.5\textwidth}{p{0.5\linewidth} X}
\toprule
\textbf{Parameter} & \textbf{Value} \\ 
\midrule
Modulation & QPSK, 16QAM, 64QAM, 256QAM \\ 
Pulse shape / Matched filter $p\left( t \right)$ & Root raised cosine (roll-off factor 0.2) \\ 
Symbol rate & $32$~GBd \\ 
Samples per symbol & 4 \\ 
Number of symbols & $1 $~million \\ 
Signal power ($P_s$) & $6.9$~dBm \\ 
LO power ($P_{LO}$) & $20$~dBm \\ 
Propagation gain ($\alpha$) & $0.002$ to $0.6$ \\ 
Electrical bandwidth ($B$) & $40$~GHz \\ 
Temperature ($T$) & $290$~K \\ 
Noise bandwidth of RF oscillator & $6$~GHz \\ 
Noise figure ($NF$) & $5$~dB \\ 
Electrical amplifier gain ($G_{e}$) & $5$~dB \\ 
$K_0$ (After multiplier) & $–145$, $–140$, $–130$, $–120$~dB/Hz \\ 
$K_2$ (After multiplier) & $10$, $100$, $1000$ \\ 
$K_3$ (After multiplier) & $10^4$ \\ 
Noise floor of transmitter base oscillator ${S_{{i_{base}}}}\left( f \right)$ & $–135.4$~dBc/Hz \\ 
Noise floor of receiver base oscillator ${S_{{i_{base,Rx}}}}\left( f \right)$ & $–155$~dBc/Hz \\ 
Frequency multiplication factor ($N$) & $20$ \\ 
I/Q amplitude imbalance (Tx and Rx), $A$ & 0.25~dB \\
I/Q phase imbalance (Tx and Rx), $\theta$ & 1$^{\circ}$ \\
\bottomrule
\end{tabularx}\label{tab:elec-MCsim-params}
\vspace{-5pt}
\end{table}

The system performance is evaluated by conducting a baseband Monte Carlo simulation with 1million randomly selected symbols from a square QAM constellation set. The aggregate noise term is calculated using \eqref{eq:66} and added to the signal and the effective phase noise is applied to it. The received THz power is varied by controlling only the coefficient $\alpha$ in the range of $0.002$ to $0.6$ which varies the $SN{R_{Rx}}$ according to \eqref{eq:75} and the noisy signal is processed by the receiver DSP. The results are then plotted in terms of BER versus received THz power. At first, no frequency offset nor phase noise are assumed in the simulations.
\begin{figure}[t]
    \centering
    \includegraphics[scale=0.8]{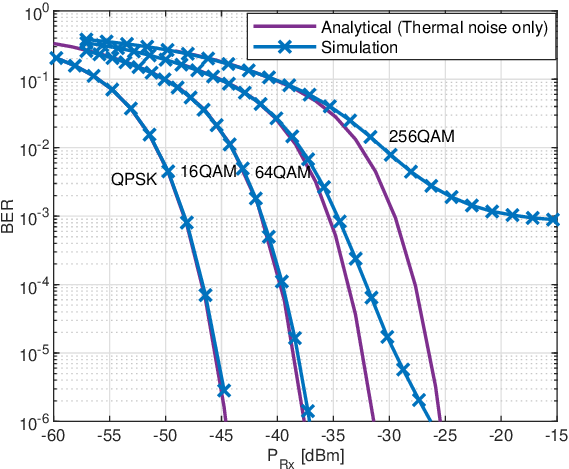}
    \caption{BER versus received THz power for electronics-based THz transmission.}
    \label{fig:elec-BER-vs-Prx}
\end{figure}

\subsubsection{BER performance without phase noise}
In Fig. \ref{fig:elec-BER-vs-Prx}, we plot the BER performance of QAM formats versus received THz power along with the benchmarking AWGN curveWe observe that the simulation results  overlap almost identically to the AWGN performance for QPSK, indicating that the impact of noise sources other than the thermal noise is negligible . However, for higher order modulation formats i.e., 16QAM, 64QAM and 256QAM, we see increasing deviations from the AWGN performance. In particular, for received powers below $-40$~dBm, the impact of thermal noise dominates; leading to a perfect overlap between the corresponding simulated and analytical curves (see BER curves for $P_{Rx} \leq -40$~dBm for all modulation formats). In contrast, for received powers above  $-40$~dBm, the impact of thermal noise gradually reduces and the degrading effects of other noise sources start to take effect. It should be emphasized that although the underlying noise sources differ and consequently the aggregate noise equations are distinct in photonics-based and electronics-based approaches, the overall performance characteristics in Fig. \ref{fig:photon-BER-vs-Prx} and Fig. \ref{fig:elec-BER-vs-Prx} look similar to each other. However, unlike the photonics-based THz system where optical noise sources are the dominant factor in error floors, the noise floor of the base RF oscillators becomes the critical factor in the electronics-based THz system. Similar to the photonics-based THz systems, a BER floor is observed in electronics-based THz systems for 256QAM occurs. This error floor is due to the signal-dependent noise components which saturate the received SNR for high received powers, i.e.see \eqref{eq:75}. In order to mitigate the signal-dependent noise components, and consequently to avoid the error floor, high quality RF sources with lower noise floor must be used at both transmitter and receiver. A low noise, higher frequency base oscillator which would require a smaller frequency multiplication factor $N$ can also mitigate the total signal-dependent noise component in the system. 

It should be further noted that deviations from the analytical curve are much larger than those in the photonics-based approach as the modulation order increases. For example, in 64QAM, the deviation from the analytical curve with thermal noise only shows a penalty of around $0.87$~dB around a BER of $10^{-3}$; which is comparable to the penalty observed in photonics-based case. For 256QAM however, there is a penalty of $5$~dB; which is much larger than the $2$~dB penalty observed in photonics-based case.
\subsubsection{Contribution of constituent noise terms}
For electronics-based THz systems, the receiver SNR expression is given in \eqref{eq:75}. Employing typical values of different parameters for electronics-based THz systems (listed in Table \ref{tab:elec-MCsim-params} of the manuscript), the magnitude of SIN and SDN components is calculated and shown in a bar graph. 

\begin{figure*}[t]
    \centering
    \subfloat[][]{\includegraphics[scale=0.8]{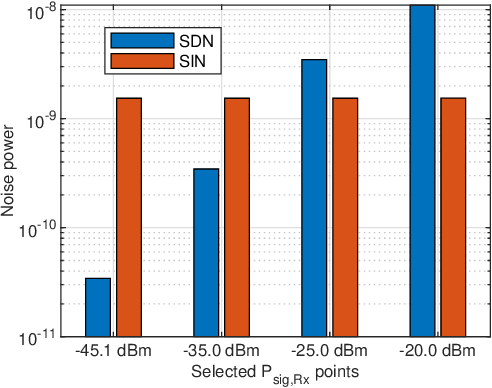}}\qquad\qquad
    \subfloat[][]{\includegraphics[scale=0.8]{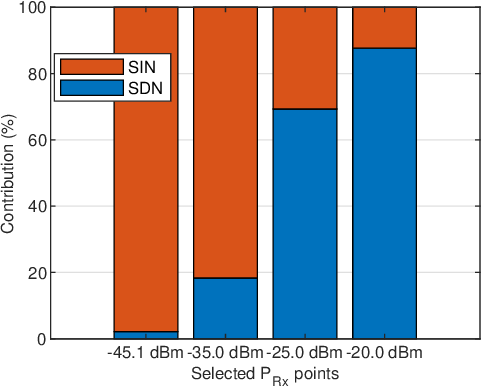}}
    \caption{(a) Magnitude of different noise terms and (b) Contribution of SDN and SIN to the total noise for several $P_{Rx}$ values.}\label{fig:NoiseContributions_Elec}
\end{figure*}

Fig. \ref{fig:NoiseContributions_Elec}a shows the magnitude of SIN and SDN for several values of ${P_{Rx}}$. Similar to the photonics-based case, the contribution of different noise terms varies with the ${P_{Rx}}$. Notice that for low ${P_{Rx}}$ value, the SIN dominates and for high ${P_{Rx}}$ values, SDN become dominant. Although, here we observe a similar trend as for the photonics-based THz case, but the magnitude of SDN and SIN are quite different. 

Fig. \ref{fig:NoiseContributions_Elec}b shows the percentage contribution of SIN and SDN to the total noise power. It shows that for $-45.1$~dBm, the contribution of SDN is only 2\% and SIN is the dominant component. However, for higher received powers, the contribution of SDN drastically increases; at $-20$~dBm, SDN constitutes about 88\% of the total noise power. This ratio is significantly higher than the photonics-based case where SDN component was 68\% for the same $P_{Rx}$ value.

In summary, the dominant noise component in both photonics- and electronics-based THz system depends the received power. In both cases, SIN is dominant for low received powers while other noise components dominate for higher received powers. The dominance of SDN for high received power values results in a BER floor. For electronics-based THz systems, the ratio of SDN is much higher than that for the photonics-based THz systems. 

\subsubsection{BER performance with phase noise}
\begin{figure}
    \centering
    \subfloat[][]{\includegraphics[scale=0.7]{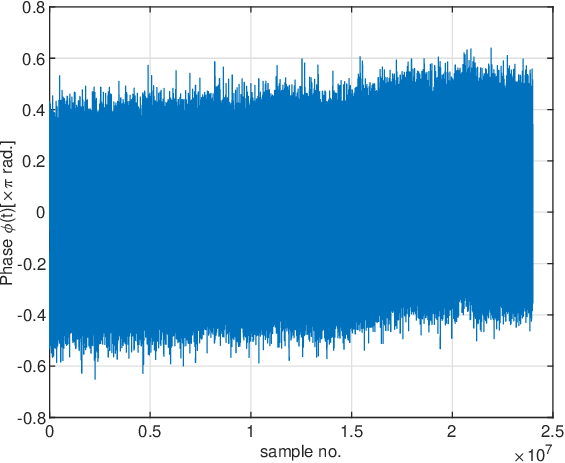}\label{fig:elec-phaseNoise-randomWalk}}\\
    \subfloat[][]{\includegraphics[scale=0.7]{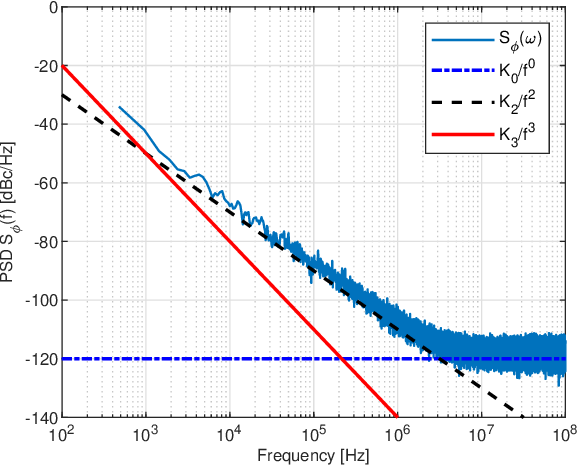}\label{fig:elec-phaseNoise-PSD}}
    \caption{(a) Phase noise evolution in time domain (sampling period $\tau  = 7.8125$~ps) (b) Phase noise PSD of a RF oscillator. $K_0 = -120$~dB/Hz, $K_2 = 10$, $K_3 = 10^4$.}
\end{figure}

In the previous section, we evaluated BER performance under the assumption of ideal oscillators with no frequency offset or phase noise; i.e., $\Delta {f_{elec}} = 0$ and $\Delta {\phi _{elec}}(t) = 0$. Now we include both of those impairments and evaluate their impact on the system performance. First, we provide an example of RF oscillator phase noise and plot its PSD as well as its constituent components for ${K_0} =  - 120$~dB, $K_2 = 10$, and $K_3 = 10^4$ . In Fig. \ref{fig:elec-phaseNoise-PSD}, we present the PSD of RF oscillator phase noise; highlighting some distinct regions in the curve. It is clearly observed that near the carrier frequency, the PSD quickly decays with respect to ${f^3}$ and ${f^2}$ followed eventually by a white noise floor. Fig. \ref{fig:elec-phaseNoise-randomWalk} shows the time domain phase noise samples corresponding to the phase noise PSD shown in Fig. \ref{fig:elec-phaseNoise-PSD}. 

The next step is to evaluate the performance in the presence of phase noise of RF oscillators. The simulation parameters for phase noise of RF oscillators are listed in Table \ref{tab:elec-MCsim-params} which represent the combined value for both carrier and LO sources. First, we evaluate the impact of near carrier phase noise by varying the parameter ${K_2}$, which represents the phase noise components decaying with ${f^2}$ in the PSD of phase noise. The value of phase noise floor, represented by ${K_0}$, is fixed at $-145$~dB/Hz and ${K_3}$, the component decaying with ${f^3}$, is fixed to a value of $10^4$. First, we visualize the impact of varying ${K_2}$ on the phase noise PSD. For different value of ${K_2}$, the phase noise PSD is calculated and shown on a single curve for comparison. In Fig.  24, we show the PSD of phase noise with varying value of ${K_2}$. With increasing value of ${K_2}$, the phase noise PSD accordingly increases. Higher value of ${K_2}$ represents a worse oscillator and vice versa. A summary of phase noise performance for different values of ${K_2}$ is given in Table \ref{tab:elec-phaseNoise-PSD-varyK2}. 

\begin{figure}
    \centering
    \includegraphics[scale=0.65]{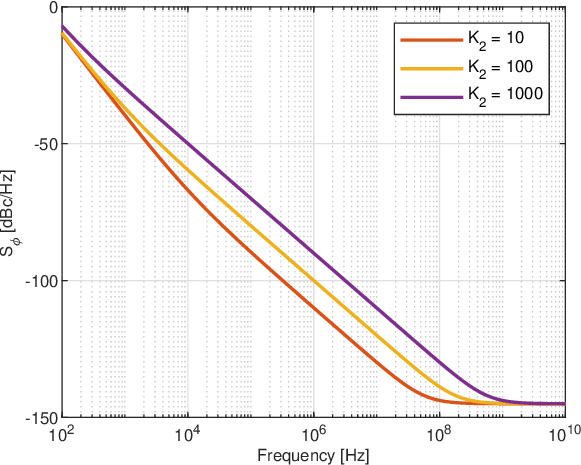}
    \caption{Phase noise PSD with varying $K_2$.}
    \label{fig:placeholder}
\end{figure}

\begin{table}[t]
\centering
\caption{Relation between value of ${K_2}$ and the corresponding value of phase noise PSD.}
\begin{tabular}{cc}
\toprule
\textbf{$K_2$} & \textbf{Phase Noise PSD at $1$~MHz offset} \\
\midrule
10   & $-110$~dBc/Hz \\
100  & $-100$~dBc/Hz \\
1000 & $-90$~dBc/Hz  \\
\bottomrule
\end{tabular}
\label{tab:elec-phaseNoise-PSD-varyK2}
\end{table}

Next, we evaluate the impact of oscillator phase noise on the performance of QAM formats. A simulation was conducted with the same system parameter as listed in Table \ref{tab:elec-MCsim-params}. 
Similar to the performance of photonics-based case without phase noise mitigation, the BER remains around $0.5$ regardless of $P_{Rx}$ value. 

\begin{figure}
    \centering
    \includegraphics[scale=0.8]{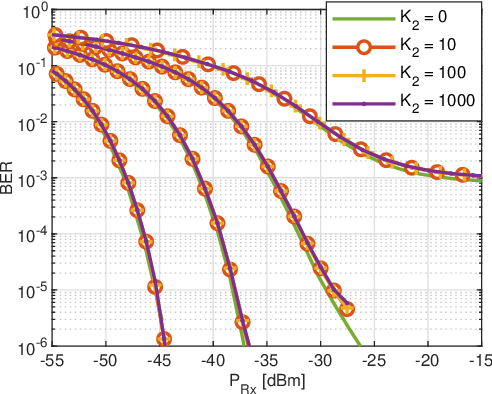}
    \caption{BER performance of square QAM formats due to near carrier phase noise of RF oscillator.}
    \label{fig:elec-BER-vs-Prx-K2phase-PLL}
\end{figure}

As the next step the simulation was redone with the inclusion of PLL based phase noise mitigation at the receiver DSP. In Fig. \ref{fig:elec-BER-vs-Prx-K2phase-PLL}, we show the performance of square QAM formats under different values of the near-carrier phase noise; obtained by varying parameter ${K_2}$. For the case of QPSK and 16QAM, we do not observe any penalty with increasing value of ${K_2}$ up to a value of $1000$. For 64QAM and 256QAM formats, a small penalty is observed with the inclusion of phase noise in simulations. 

Next, we evaluate the performance by simulating the white noise floor of the RF oscillator represented by $K_0$ in the phase noise model. Depending on the noise floor of base oscillator and the frequency multiplication factor, the resulting noise floor of THz source can significantly limit the overall system performance. As an example we refer to \cite{weberWBandX12Frequency2011} where the base oscillator has a noise floor of about $-140$~dBc/Hz and after a $\times12$ multiplier chip it is enhanced to about $-115$~dBc/Hz. Due to its uncorrelated nature, the white noise floor of phase noise (represented by $K_0$) cannot be compensated by the PLL and it contributes to an increased error floor in the BER performance curves as we will show later. Same as the near-carrier phase noise, the white noise floor of phase noise is enhanced by the frequency multiplier block used for THz generation from a low frequency base oscillator. Since it cannot be compensated by the PLL, it can severely limit the transmission performance specially for higher order QAM formats; which are inherently more susceptible to phase noise. Its impact increases linearly with the signal bandwidth, so reducing the bandwidth can help alleviate its effect. However, wide bandwidth is a core advantage of THz communication, meaning that to fully exploit this benefit, the white noise floor must be minimized. For the following simulation, we fix the values of ${K_2} = 10$ and ${K_3} = {10^4}$; while varying ${K_0} =  - 145 \to  - 120$~dB/Hz. 

\begin{figure*}
    \centering
    \subfloat[][]{\includegraphics[scale=0.7]{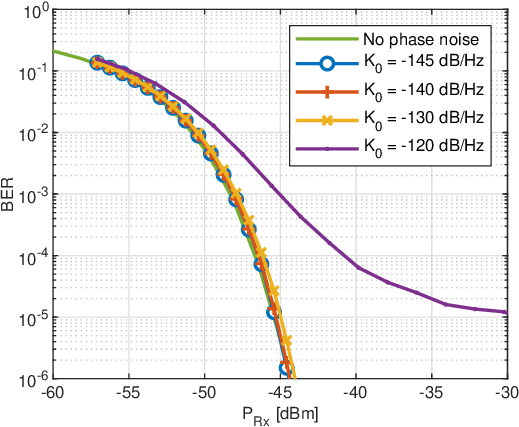}\label{fig:qpsk-BER-vs-Prx-K0-PLL}}\qquad\qquad
    \subfloat[][]{\includegraphics[scale=0.7]{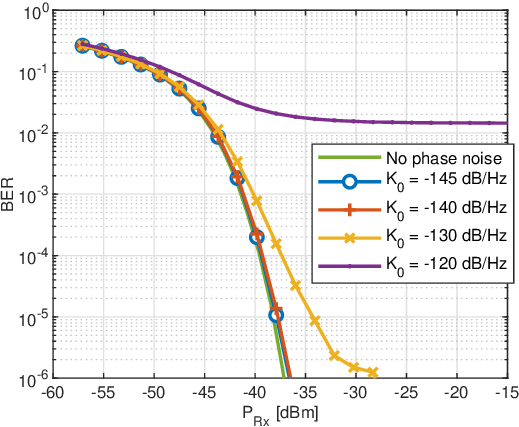}\label{fig:qam16-BER-vs-Prx-K0-PLL}}\\
    \subfloat[][]{\includegraphics[scale=0.7]{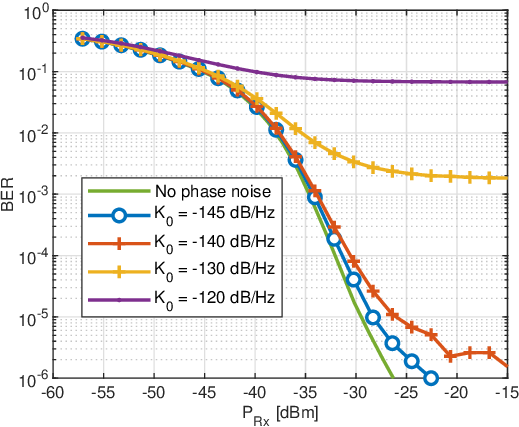}\label{fig:qam64-BER-vs-Prx-K0-PLL}}\qquad\qquad
    \subfloat[][]{\includegraphics[scale=0.7]{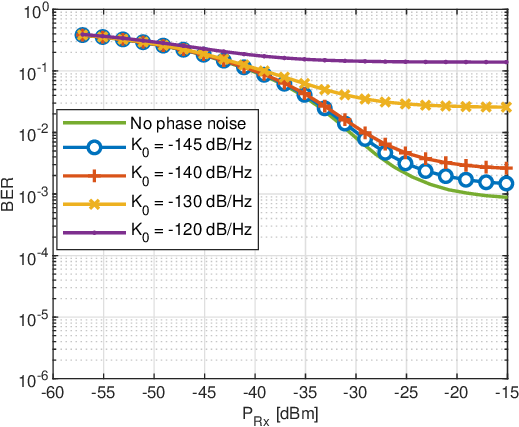}\label{fig:qam256-BER-vs-Prx-K0-PLL}}
    \caption{BER Performance of square QAM formats due to the white noise floor of the oscillator phase noise. (a) QPSK, (b) 16QAM, (c) 64QAM, and (d) 256QAM.}\label{fig:elec-BER-vs-Prx-K0phase-PLL}
\end{figure*}

In Fig. \ref{fig:elec-BER-vs-Prx-K0phase-PLL}, we show BER results for different square QAM formats. As QPSK is the most robust to phase noise among the considered modulation formats, it shows negligible penalty for phase noise floor values up to ${K_0} =  - 130$~dB/Hz. Increasing the phase noise floor further shows a large penalty for QPSK compared to the case without any phase noise. The results for 16 QAM in Fig. \ref{fig:qam16-BER-vs-Prx-K0-PLL} show that there is a considerable penalty already for a value of ${K_0} =  - 130$~dB/Hz and for larger value of ${K_0} =  - 120$~dB/Hz, a BER floor of $>10^{-2}$ is observed. The performance of higher orders of modulation format i.e., 64QAM and 256QAM, in Fig. \ref{fig:qam64-BER-vs-Prx-K0-PLL} and \ref{fig:qam256-BER-vs-Prx-K0-PLL}, respectively, clearly shows increased susceptibility to the white noise floor of the oscillator phase noise. While the penalty for ${K_0} \le  - 140$~dB/Hz is small, for higher values of the phase noise floor a BER floor is observed. For 64 QAM, a BER floor of $2 \times {10^{ - 3}}$ is observed for ${K_0} =  - 130$~dB/Hz. For 256QAM, a BER floor of $2.5 \times {10^{ - 2}}$ is observed for ${K_0} =  - 130$~dB/Hz. 

Earlier, for photonics-based THz transmission system, we noticed that even for very large laser linewidth of $1$~MHz, the BER for 256QAM could reach below $1 \times {10^{ - 2}}$ (see Fig. \ref{fig:photon-qam256-BER-vs-Prx-PN-PLL}). For a laser with linewidth of $100$~kHz, a typical value for commercial lasers, the power penalty was $\le0.7$~dB for all the considered modulation formats (see Table \ref{tab:linewidth_penalty}). In case of electronics-based THz transmission system, we noticed that the near-carrier phase noise is adequately compensated by employing a PLL at the receiver DSP. However, the white noise floor of the phase noise, which is uncorrelated in nature, and cannot be compensated by the receiver DSP, sets the ultimate performance limit. An aggregate phase noise floor of $\le  - 140$~dB/Hz after the multiplier is required to achieve a similar performance as that of the photonics-based THz case with laser linewidth of $100$~kHz. For a multiplication factor of $N = 20$, this means that the base oscillator must have phase noise floor of $\le  - 166$~dB/Hz. Note that this is a combined value for both transmitter and receiver oscillators, i.e., we will require $\le  - 169$~dB/Hz for each base oscillator at transmitter and receiver. Also note that, these simulations were conducted at a fixed symbol rate of $32$~GBd; for smaller symbol rates, requiring smaller bandwidth, the required white noise floor of the phase noise will accordingly scale down.

\begin{figure*}
    \centering
    \subfloat[][]{\includegraphics[scale=0.8]{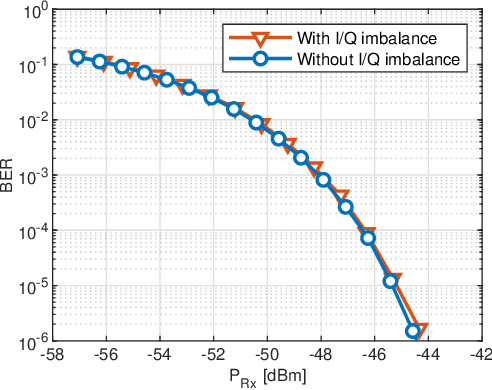}\label{fig:elec-qpsk-BER-vs-Prx-IQimb}}\qquad\qquad
    \subfloat[][]{\includegraphics[scale=0.8]{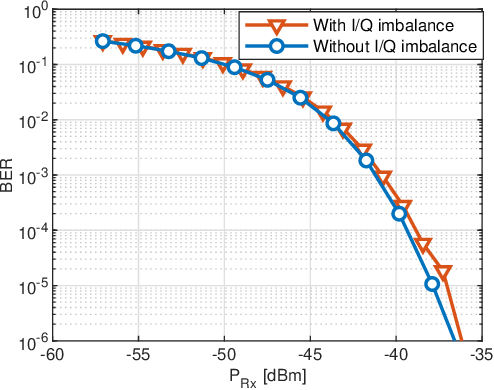}\label{fig:elec-qam16-BER-vs-Prx-IQimb}}
    \\
    \subfloat[][]{\includegraphics[scale=0.8]{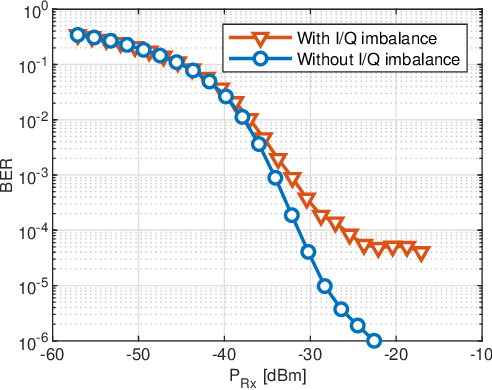}\label{fig:elec-qam64-BER-vs-Prx-IQimb}}\qquad\qquad
    \subfloat[][]{\includegraphics[scale=0.8]{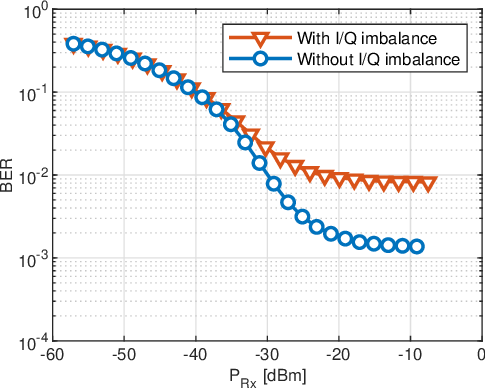}\label{fig:elec-qam256-BER-vs-Prx-IQimb}}
    \caption{Impact of I/Q imbalance in electronics-based THz system on performance of square QAM formats (a) QPSK, (b) 16QAM, (c) 64QAM, (d) 256QAM.}\label{fig:elec-BER-vs-Prx-IQimb}
\end{figure*}
\subsubsection{Impact of I/Q imbalance on the system performance}

 In this section, we evaluate the impact of I/Q imbalance on electronics-based THz systems.The modeling approach follows the same methodology used earlier for the photonics-based THz system. For this analysis, the phase noise parameters are fixed at $K_0 = -145$~dB/Hz, $K_2 = 10$, and $K_3 = 10^4$. As in the previous section, phase noise is mitigated using a data-aided PLL in the receiver DSP, while no additional compensation or predistortion is applied to mitigate the I/Q imbalance. To ensure consistency between the photonics- and electronics-based architectures, the same amplitude and phase imbalance values, listed in Table~\ref{tab:elec-MCsim-params}, are used. With these parameters, the resulting in-band image interference level is $-35.5$~dB.

Fig.~\ref{fig:elec-BER-vs-Prx-IQimb}a and~\ref{fig:elec-BER-vs-Prx-IQimb}b show the performance of QPSK and 16QAM, respectively. Similar to the photonics-based case, the impact of I/Q imbalance on these lower-order modulation formats is negligible, and no noticeable error floor is observed. In contrast, for higher-order modulation schemes such as 64QAM and 256QAM, the effect becomes much more pronounced. In particular, the presence of I/Q imbalance significantly enhances the BER floor, indicating that higher-order constellations are considerably more sensitive to the image interference introduced by I/Q mismatch. Although a similar error-floor trend is observed in the photonics-based system, the degradation is noticeably more severe in the electronics-based implementation. 

\subsection{Model Validation Using Experimental Results}

To validate the proposed SNR expressions, we compare the analytical predictions with representative experimental demonstrations for both photonics-based and electronics-based THz systems selected from Tables~1--3.

For the photonics-based THz system, we consider the experimental study reported in~\cite{maekawa300GHzbandWirelessLink2023a}. The authors report key system parameters including a symbol rate of 20~GBd, laser output power of 150~mW (21.8~dBm), and a roll-off factor of 0.35. The receiver performance is evaluated using the Keysight OMA software, which reports results in terms of error vector magnitude (EVM). Under the assumption of an AWGN channel, EVM is related to SNR as $SN{R_{dB}} =  - 20{\log _{10}}\left( {EVM} \right)$, enabling comparison with the analytical model.

To complete the analysis, we assume typical values for parameters not explicitly provided in~\cite{maekawa300GHzbandWirelessLink2023a}, namely RIN $= -145$~dB/Hz, modulated optical power $P_1 = -10$~dBm, and unmodulated laser power $P_2 = -10$~dBm at the EDFA input, with a gain of $G = 27$~dB. At the receiver, we assume an electrical amplifier noise figure of 8~dB, gain $G_e = 10$~dB, and operating temperature $T = 290$~K. Under these conditions, the predicted receiver SNR is approximately 20.6~dB, which is in close agreement with the SNR inferred from the reported EVM of 11.3\% (corresponding to 18.94~dB). The slight overestimation is attributed to additional distortion sources not captured in the analytical model, such as quantization noise, I/Q imbalance, and nonlinear effects.

For the electronics-based THz system, we consider the experimental demonstration in~\cite{hamada300GHz100GbInPHEMT2018}. The system employs a symbol rate of 20~GBd with 16QAM modulation. The carrier and local oscillator are generated using base oscillators followed by frequency multipliers with a multiplication factor of 18 at both the transmitter and receiver. The oscillator output power is 5~dBm at both ends, and the receiver employs an electrical amplifier with gain $G_e = 21$~dB. We assume an operating temperature of 290~K and a base oscillator phase noise floor of $-147$~dBc/Hz, representative of laboratory-grade RF synthesizers.

Under these assumptions, the predicted receiver SNR is approximately 21.35~dB, which is in reasonable agreement with the experimentally reported value of 19.9~dB. The small discrepancy is again attributed to unmodeled impairments such as quantization noise and nonlinear distortions. 

It is worth noting that the above validation relies on a combination of reported parameters and assumed typical values for missing system details. Therefore, the comparison is intended to demonstrate qualitative agreement and the predictive capability of the proposed modeling framework, rather than an exact replication of the experimental results.
\color{black}
\subsection{Comparison of photonics- and electronics-based THz communication systems}
In this section we compare two implementation approaches of THz communication systems in light of the simulation results presented in the preceding sections. Both photonics- and electronics-based THz communication systems are subject to additive as well as multiplicative noises and accordingly their performance is different from the conventional wireless systems operating at lower frequencies which are mainly limited by AWGN. Our main observations are summarized in the following. 
\begin{itemize}
    \item In photonics-based THz systems, the SNR at the transmitter is limited by optical noise originating from the laser’s RIN and the ASE of the optical amplifier. At the receiver, the RIN of the laser generating the local oscillator reference signal, along with thermal noise from the mixer and subsequent electrical amplifiers, further constrains the SNR. Unlike thermal noise, which is signal-independent, other noise components scale with the signal power. In the photonics-based THz systems, the noise scaling occurs during the photomixing of two optical signals generating the THz signal of interest. Furthermore, the generation of THz LO using another photomixing process and its subsequent product with the received THz signal in a mixer leads to further enhancement of signal-dependent noise. 
    \item In electronics-based THz systems, additive noise originates from the mixers at both the transmitter and receiver, as well as from the intrinsic noise floor of the RF oscillators. Due to this oscillator noise floor, the effective noise scales with the signal power—an effect often negligible at lower frequency bands but significant in the THz range. This is because THz carriers are typically generated by multiplying a low-frequency base oscillator, a process that amplifies the associated noise floor significantly, making it critical to account for in system modeling and design. 
    \item As noted above, in both photonics-based and electronics-based systems, noise scales with the signal power.
    Considering typical values of system parameters for both implementation approaches, we observed from our simulation studay that the difference in noise variance for different symbol groups of QAM constellation is much larger for electronics-based systems as compared to the photonics-based systems. In both THz systems, the aggregate additive noise closely follows a Gaussian probability distribution. However, the variance of noise differs across symbol groups within a QAM constellation. A comparison of noise statistics in both systems show that under typical system parameters, the signal-dependent noise is more pronounced in electronics-based THz systems as compared to photonics-based THz system.
    \item The BER simulation results for both photonics- and electronics-based THz systems show close match with  the conventional AWGN performance only for lower order modulation formats, i.e., QPSK, and 16QAM. For higher order formats, i.e., 64QAM and 256QAM, a clear deviation from the AWGN performance is observed. These results indicate that thermal noise is dominant in low SNR regime while other noise components dominate in the high SNR regime. At higher received powers, the performance becomes limited by signal power-dependent noise sources, i.e., optical noise in photonic systems and the oscillator noise floor in electronic systems. While the origin of signal dependent noise in both systems is quite different, the BER performance trends are remarkably similar, eventually leading to some error floor. 
     
\end{itemize}
\vspace{-2pt}
\quad In addition to the aggregate additive noise, phase noise originating from the instabilities of RF oscillators and lasers, in electronics- and photonics-based THz systems, respectively, distort the received signal and potentially limit the performance of coherent QAM formats. 
\begin{itemize}
    \item For electronics-based THz system, the frequency offset between oscillators of transmitter and receiver results in a linearly increasing phase with respect to time. Frequency offset is compensated inside the receiver using DSP on blocks of symbols. Due to the presence of frequency multipliers at the transmitter and receiver oscillators, the frequency offset between them is enhanced by the multiplication factor   compared to the base oscillators’ frequency offset. Hence, the tolerance range of frequency correction algorithms must be increased by   times. Alternatively, using higher base oscillator frequencies (to reduce the required multiplication factor) or employing oscillators with improved frequency stability can help minimize the aggregate carrier–LO frequency difference, reducing the burden on receiver DSP algorithms.
    \item Similarly, for the photonics-based THz system, multiple infrared lasers at transmitter and receiver are employed leading to frequency offset increase by factor four (two independent lasers generate a THz signal at both transmitter and receiver). The tolerance range of frequency compensation algorithms must be accordingly enhanced. Employing laser combs at transmitter and receiver instead of independent laser sources, the net frequency offset can be reduced by one half. The different wavelengths output of the laser comb are phase-locked i.e., there is no relative phase and frequency difference among them. Hence, employing a laser comb at both transmitter and receiver reduces the frequency as well as phase deviations by one half in magnitude and consequently relaxes the requirements of frequency and phase recovery algorithms \cite{liPhotonicsAidedTerahertzWaveWireless2022,yuTerahertzWaveGenerationBased2020}.
\end{itemize}

In addition to the slow frequency drift, phase noise among the transmitter and receiver carriers distorts the received QAM constellation and it must be removed using DSP. In our simulations for both photonics- and electronics-based THz systems, we have considered a data-aided PLL to track the phase noise.
\begin{itemize}
    \item In photonics-based THz transmission, laser phase noise—modeled as a Wiener process—accumulates from both the transmitter and receiver, leading to significant performance degradation for higher-order QAM formats. While lower-order schemes like QPSK remain relatively unaffected, formats such as 256QAM are highly sensitive, making the choice of low-linewidth lasers or comb-based sources critical for maintaining acceptable BER performance.
    \item In electronics-based THz transmission, phase noise—amplified by frequency multipliers—poses a critical performance challenge. While near-carrier phase noise can be effectively mitigated using PLL, the white noise floor cannot be compensated and becomes the dominant limiting factor, especially for higher-order QAM formats and wider bandwidths. For example, simulations show that a 32-GBd QPSK signal can tolerate a phase noise floor of about –130 dBc/Hz, while 32-GBd 256QAM requires approximately –145 dBc/Hz to avoid significant performance penalties. This highlights the importance of minimizing oscillator noise floors to enable reliable, high-capacity THz links.
\end{itemize}
\section{Practical Implications and Design Considerations for THz Systems}
While the preceding sections have focused on the development of signal models and the evaluation of system performance under hardware impairments, it is equally important to interpret these results from a practical implementation perspective. In particular, the derived models and numerical findings provide valuable insight into how different impairment mechanisms manifest in electronics-based and photonics-based THz systems, and how they can be effectively mitigated in real-world transceivers. In this section, we discuss the role of digital signal processing (DSP) in compensating for THz-specific impairments, as well as the associated scalability and implementation complexity considerations in the context of future 6G systems. These discussions aim to bridge the gap between analytical modeling and practical system design, highlighting key trade-offs that must be addressed for the successful deployment of THz communication technologies.

\subsection{Role of DSP in THz System Design}
As extensively analysed in preceding sections, both photonics-based and electronics-based THz systems experience various impairments which must be compensated digitally. 

In photonics-based THz architectures, multiple free-running lasers are used at the transmitter and receiver. The THz carrier is generated from the frequency difference between optical sources. As demonstrated in Sec.\ref{sec:numericalResults}.A.3, aggregate carrier frequency offset and phase noise can induce severe performance degradations of QAM formats. In particular, the effective frequency offset can reach several gigahertz, exceeding the operating range of conventional coherent optical DSP blocks. This can be mitigated by employing advanced carrier recovery algorithms. For example, a recent work by Rha \textit{et al.} \cite{rhaNovelPhaseCFO2022} has demonstrated CFO estimation capable of handling $\pm5$~GHz offsets for a photonics-based THz communication system. Similarly, advanced phase recovery approaches such as principal component-based phase estimation \cite{ding1248GbitPS256QAMSignal2022} has been shown to outperform blind phase search algorithm to mitigate phase noise in photonic THz systems. In \cite{castroExperimentalValidationCoherent2019}, Castro \textit{et al.}, experimentally evaluated the performance of state-of-the-art optical DSP for THz communication systems employing 16QAM format with symbol rates ranging from 2~GBd to 32~GBd. Their results demonstrate that, by employing coherent optical DSP algorithms, stable BER performance can be maintained in the presence of frequency offsets of up to 4~GHz for a 16~GBd signal. 

In addition to enhanced phase distortions, photonics-based THz systems are also affected by I/Q imbalances; as demonstrated in Sec.\ref{sec:numericalResults}.A.4. To mitigate these impairments, equalization techniques can be used. For example, Zhu \textit{et al.} \cite{zhuPhotonicsAidedTerahertzWaveWireless2022} proposed a 2$\times$2 MIMO-based post-equalizer operating on the real-valued I and Q signal components. Their experimental results demonstrate improved BER performance for 20~GBd 16QAM signals after 1~m wireless transmission; showing the effectiveness of DSP based I/Q imbalance compensation technique. Gram-Schmidt orthogonalization process (GSOP), proposed initially to compensate receiver I/Q imbalance of coherent photonic receivers in \cite{fatadinCompensationQuadratureImbalance2008}, could be also employed to compensate I/Q imbalance of THz receiver, see e.g., the work by Ding \textit{et al.} in \cite{dingHighSpeedLongDistancePhotonicsAided2023}; highlighting the effectiveness of DSP algorithms on THz receivers. 

Electronics-based THz systems are particularly affected by amplified oscillator phase noise arising from frequency multiplier chains as well as the I/Q imbalance at both the transmitter and receiver. The resulting phase noise can be decomposed into a slowly varying (correlated) component and a rapidly varying (uncorrelated) component. As demonstrated by the simulation analysis in Sec.\ref{sec:numericalResults}.B, the correlated phase noise can be effectively mitigated using digital phase-tracking techniques (e.g., PLL), but the residual Gaussian phase noise component is more difficult to track and often limits the performance of estimation algorithms at the receiver DSP. This challenge has motivated the development of advanced channel estimation and equalization techniques. In \cite{shaChannelEstimationEqualization2021}, Sha \textit{et al.} proposed a two-stage framework for joint estimation of I/Q imbalance and channel impulse response in the presence of strong phase noise. In the first stage, channel state information (CSI) is obtained from pilot symbols and corrected using an extended Kalman filter, while in the second stage, I/Q imbalance and channel response are estimated using the phase-corrected pilots. A low-complexity equalization scheme is further employed to jointly compensate for wideband I/Q imbalance, phase noise, and channel distortion.

Under strong residual phase noise, the design of customized waveforms and constellations can further enhance robustness. For example, phase-noise-tolerant modulation formats such as spiral constellations \cite{ugoliniSpiralConstellationsPhase2019} can improve immunity to residual phase fluctuations in electronics-based THz systems. 

Overall, advanced DSP is instrumental to compensate for impairments in both photonics-based and electronics-based THz systems, including large frequency offsets, phase noise, and I/Q imbalance. While DSP techniques can significantly improve performance, they do so at the expense of increased computational complexity, memory requirements, power consumption, as well as an increased thermal load. Consequently, scalable 6G THz transceiver design requires joint optimization of laser/ oscillator quality, multiplier architecture, modulation format, and DSP algorithms to achieve an appropriate trade-off among performance, cost, and energy efficiency.

\subsection{Scalability and Implementation Complexity for 6G}

While our work primarily focuses on link-level performance, this section provides a discussion on the scalability and implementation complexity of both approaches.

The electronics-based architecture largely follows design principles similar to those employed at lower frequencies in terms of overall system structure. In a modern wireless transceiver, baseband processing is performed digitally, while upconversion to the desired carrier frequency is carried out in the analog/RF domain. Silicon-based (CMOS) technologies are typically used to implement high-speed analog-to-digital and digital-to-analog converters, as well as digital signal processors (DSPs) or application-specific integrated circuits (ASICs). CMOS offers high integration density, scalability, and power efficiency, making it well suited for complex baseband processing required in modern wireless systems. 

Silicon technologies however face performance limitations at THz frequencies due to transistor speed ($f_T$ and $f_{max}$ ) constraints and reduced gain and output power at very high frequencies \cite{pfeifferCurrentStatusTerahertz2018}. While sub-THz operation and harmonic-based THz generation are possible in advanced CMOS or SiGe processes, achieving efficient and high-power carrier generation above several hundred gigahertz remains challenging. For this reason, THz analog front-ends are often implemented using III–V semiconductor technologies such as indium phosphide (InP) or gallium arsenide (GaAs), particularly high electron mobility transistor (HEMT) or heterojunction bipolar transistor (HBT) processes \cite{kissingerMillimeterWaveTerahertzTransceivers2021}. These technologies provide superior electron mobility and higher maximum oscillation frequencies, enabling improved THz performance.

The resulting heterogeneous system architecture, however, introduces scalability and implementation challenges. III–V technologies require specialized foundries that are separate from CMOS manufacturing lines, limiting large-scale integration and increasing fabrication cost. Furthermore, integrating CMOS-based DSP with an InP- or GaAs-based THz front end often requires advanced packaging, low-loss high-speed interconnect design, and careful thermal co-design; in architectures that rely on coherent multi-chip operation, tight clock/phase synchronization is also important. These requirements increase implementation complexity and can raise power consumption and packaging cost \cite{fayUltrawideBandwidthInterChip2016,alonso-delpinoMicromachiningAdvancedTerahertz2020,linHeterogeneousIntegrationEnabled2021,dessetInPCMOSCointegration2021}.

Beyond electronics-based implementations, photonics-based architectures provide an alternative paradigm for THz signal generation and processing, relying on fundamentally different physical mechanisms. Photonics-based systems generate THz signals through optical heterodyning or photomixing. In this approach, two optical carriers (typically around 190–200 THz) are combined in a high-speed photodiode, such as a uni-traveling carrier photodiode (UTC-PD), and their frequency difference produces a signal in the 0.1–3 THz range.

A key advantage of this approach is wide and continuous frequency tunability: by adjusting the optical frequency spacing between the lasers, a broad THz range can be covered without hardware modification. Photonic THz systems also offer excellent spectral purity and potentially very large modulation bandwidths, supported by high-speed optical modulators. Nevertheless, photonics-based THz systems face several scalability and implementation challenges that are critical in the context of 6G networks. The photomixing process is inherently inefficient, and the generated THz output power is typically limited to the microwatt to low milliwatt range, restricting link distance and increasing the need for high-gain antennas or beamforming. In addition, the use of narrow-linewidth lasers, optical amplifiers, and precise polarization control increases system complexity. It could be further noted that photonic systems often involve higher component count, larger footprint, and greater energy consumption due to continuous laser operation. Commercial deployment of photonics-based THz transceivers would likely require photonic integrated circuits (PICs) to reduce footprint and power consumption \cite{andrianopoulosPhotonicIntegratedCircuits2023,houPhotonicIntegratedCircuits2020}. Although PIC technology has progressed significantly, it remains less mature and less cost-optimized than CMOS-based electronics in terms of wafer-scale manufacturing, yield, and integration density. 

As a summary, electronic THz systems currently offer stronger prospects for compact integration and higher output power. However, achieving cost-effective mass production at scale remains still challenging when heterogeneous CMOS–III–V integration is required. In contrast, although photonic THz systems provide superior tunability and broader frequency reach, their cost, power efficiency, and integration complexity presently limit their scalability for widespread 6G access deployments. Nevertheless, this conclusion should not be viewed as definitive, since continued advances in PIC technology may substantially improve the manufacturability and integration efficiency of photonic THz transceivers. In particular, progress toward monolithic or hybrid PIC implementations and recent multiband/ultra-wideband demonstrations \cite{taoUltrabroadbandOnchipPhotonics2025} suggest that the scalability gap could narrow over time.

\section{Conclusions and Future Directions}\label{sec:conclusions}

In this work, we presented a comprehensive framework for the analysis of photonics-based and electronics-based THz communication systems under realistic hardware impairments. By developing unified signal and noise models for both architectures, we derived analytical expressions for the transmitter and receiver SNR and evaluated BER performance under practical operating conditions. The results demonstrate that hardware impairments play a fundamental role in determining achievable link performance at THz frequencies. In particular, photonics-based systems are primarily limited by optical noise sources such as laser RIN and ASE, whereas in electronics-based systems, the oscillator phase-noise floor becomes the dominant constraint, especially for wideband transmissions and higher-order QAM formats.

In light of the developed models and numerical findings, we also discussed the role of DSP techniques and the associated scalability and implementation complexity considerations for 6G THz systems. These discussions demonstrate how the identified impairment mechanisms translate into specific signal processing requirements and architectural trade-offs in practical implementations. Overall, the insights provided in this work offer a systematic basis for understanding impairment-limited performance in THz systems and for evaluating design trade-offs between photonics-based and electronics-based architectures in future 6G networks.

Several directions for future research emerge from the developed framework and the presented results. The proposed models were validated against representative experimental demonstrations reported in the literature for both photonics-based and electronics-based THz systems, showing reasonable agreement between analytical predictions and observed performance trends. However, this validation is based on a limited set of studies and some assumed parameters due to incomplete reporting of implementation details. Therefore, further work is required to enable a more comprehensive validation across a wider range of carrier frequencies, bandwidths, and system configurations throughout the THz spectrum.

While the analytical framework developed in this work focuses on the two fundamental implementation approaches, i.e., photonics-based and electronics-based THz systems, it can be naturally extended to hybrid architectures that combine these technologies. In such systems, the overall impairment behavior can be interpreted as the combination of the corresponding noise mechanisms in the photonic and electronic subsystems. For example, a hybrid link employing a photonic transmitter and an electronic receiver would inherit optical noise sources such as RIN and ASE from the transmitter while being affected by oscillator phase noise and electronic front-end impairments at the receiver. A detailed analytical treatment of such hybrid architectures therefore constitutes an important direction for future work.

 In this work, the performance evaluation focuses on single-carrier square QAM formats, which are well suited for typical THz communication scenarios. Due to the highly directional nature of THz links and the resulting negligible multipath propagation, the channel delay spread is typically very small, and the condition for frequency-flat response, $R_s\cdot\tau_{rms}\ll1$ is generally satisfied. Consequently, the channel behaves as frequency-flat, and single-carrier transmission is sufficient and often preferred due to its lower complexity. Nevertheless, multi-carrier waveforms such as OFDM may still be relevant in scenarios where the THz front-end response is not flat over the signal bandwidth or where in-band impairments, such as clock leakage, affect specific frequency components within the signal band. While OFDM can mitigate such frequency-selective distortions, its advantages are less pronounced in typical THz links. Moreover, OFDM is generally more sensitive to oscillator phase noise and peak-to-average power ratio (PAPR), which can exacerbate front-end impairments. Nonetheless, the impairment-aware models developed in this work can also provide insight into the behavior of alternative waveform candidates, while a detailed comparative study is left for future work.

\ifCLASSOPTIONcaptionsoff
  \newpage
\fi
\bibliographystyle{IEEEtran}
\bibliography{IEEEabrv,./bibtex/THz-communication-review.bib}
\end{document}